\documentclass[twocolumn,pra,aps,eqsecnum]{revtex4-1}
\usepackage[T1]{fontenc}
\usepackage[latin9]{inputenc}
\usepackage{color}
\usepackage{bm}
\usepackage{amsmath}
\usepackage{amssymb}
\usepackage{graphicx}
\usepackage{esint}
\PassOptionsToPackage{normalem}{ulem}
\usepackage{ulem}
\usepackage[unicode=true,
 bookmarks=false,
 breaklinks=false,pdfborder={0 0 1},backref=false,colorlinks=true]
 {hyperref}
\hypersetup{
 pdfborderstyle=}

\makeatletter

\usepackage{color}
\usepackage{bm}
\usepackage{bbold}
\usepackage{soul}
\usepackage{epsfig}
\usepackage{verbatim}
\usepackage{array}
\newcommand{\be}{\begin{equation}}
\newcommand{\ee}{\end{equation}}
\newcommand{\bea}{\begin{eqnarray}}
\newcommand{\eea}{\end{eqnarray}}

\definecolor{kyrylo}{rgb}{0.7,0.5,0.2}
\definecolor{rein}{rgb}{0.2,0.5,0.8}
\definecolor{notneed}{rgb}{0.3,0.7,0.9}

\makeatother

\begin{document}
\global\long\def\sp#1#2{\langle#1|#2\rangle}%
\global\long\def\abs#1{\left|#1\right|}%
\global\long\def\avg#1{\langle#1\rangle}%
\global\long\def\ket#1{|#1\rangle}%
\global\long\def\bra#1{\langle#1|}%

\title{Non-Abelian Berry phase for open quantum systems: Topological protection
vs geometric dephasing }
\author{Kyrylo Snizhko,$^{1}$ Reinhold Egger,$^{2}$ and Yuval Gefen$^{1}$}
\affiliation{$^{1}$~Department of Condensed Matter Physics, Weizmann Institute,
Rehovot, Israel ~\\
 $^{2}$~Institut f\"ur Theoretische Physik, Heinrich-Heine-Universit\"at,
D-40225 D\"usseldorf, Germany}
\date{\today}
\begin{abstract}
We provide a detailed theoretical analysis of the adiabatic evolution
of degenerate open quantum systems, where the dynamics is induced
by time-dependent fluctuating loop paths in control parameter space.
For weak system-bath coupling, the fluctuations around a deterministic
base path obey Gaussian statistics, where we assume that the quantum
adiabatic theorem is also satisfied for fluctuating paths. We show
that universal non-Abelian geometric dephasing (NAGD) contributions
are contained in the fluctuation-averaged evolution operator. This
operator plays a key role in all experimental protocols proposed in
this work for the detection of NAGD. In particular, we formulate interference
measurements providing full access to NAGD. Apart from a factor due
to the dynamic phase and its fluctuations, the averaged evolution
operator contains the averaged Berry matrix. A polar decomposition
of this non-unitary matrix into the product of a unitary, $V$, and
a positive semi-definite Hermitian matrix, $R$, reveals the physics
of the NAGD contributions. Unlike the conventional dynamic dephasing
rate, the NAGD eigenrates encoded by $R$ depend on the loop orientation
sense and, in particular, change sign under a reversal of the direction.
A negative rate then implies amplification of coherences as compared to the ones when only dynamic dephasing is present.
The non-Abelian character of geometric dephasing is rooted in the
non-commutativity of $V$ and $R$ and causes smoking-gun signatures
in interference experiments. We also clarify why systems subject to
topological protection do not show geometric dephasing. Without full
protection, however, geometric dephasing can arise and has to be taken
into account. As concrete application, we propose spin-echo NAGD detection
protocols for modified Majorana braiding setups.
\end{abstract}
\maketitle

\section{Introduction}

\label{sec1}

In this paper, we consider the effect of environment-induced fluctuations
on the adiabatic dynamics of a degenerate quantum system. In the absence
of couplings to the environment, the system Hamiltonian, $H=H({\bm{\lambda}})$,
may depend on $d$ classical parameters contained in the vector ${\bm{\lambda}}$,
and one assumes that $H$ has an $N$-fold degenerate level with energy
$E_{1}({\bm{\lambda}})$. This degeneracy has to persist for all relevant
parameter configurations. Under a slow time-dependent loop trajectory,
${\bm{\lambda}}(t)$, of total duration $T$ such that ${\bm{\lambda}}(0)={\bm{\lambda}}(T)$,
the Hamiltonian $H({\bm{\lambda}}(t=0))$ returns to its initial form
at time $t=T$. Nonetheless, the quantum state will change under such
a parameter loop. In the adiabatic limit, apart from a dynamic phase
factor, the state after the round-trip is connected to the initial
state by a unitary $N\times N$ Berry matrix (``non-Abelian Berry
phase''), ${\cal U}_{B}$, which can be written as a path-ordered
Wilson loop amplitude and thus has a purely geometric meaning \citep{Wilczek1984,Zee1988}.
For $N=1$, one recovers the Abelian case with ${\cal U}_{B}=e^{i\varphi_{B}}$,
where $\varphi_{B}$ is the celebrated Berry phase \citep{Berry1984}.

In the presence of system-environment couplings, the Abelian Berry
phase is known to acquire an imaginary contribution due to cross-correlations
of energy and state trajectory fluctuations (``geometric dephasing''),
see Refs.~\citep{Whitney2005,Leek2007,Berger2015}. The widespread
importance of non-Abelian Berry phases and non-Abelian gauge theories
motivated us to investigate whether non-Abelian geometric dephasing
(NAGD) contributions could arise when a weak system-bath coupling
to an environment is present. For a short exposition of our key results,
see Ref.~\citep{ourprl}. In the present paper, we provide a detailed
description of our theoretical formalism, and we also present many
additional results beyond Ref.~\citep{ourprl}. In particular, below
we discuss a variety of experimental detection schemes for accessing
the physics of NAGD, see Sec.~\ref{sec4}, and we relate our results
to known generalizations of the Stokes theorem to non-Abelian systems,
see Sec.~\ref{sec3a}. Moreover, in Sec.~\ref{sec3b}, we show that
in topologically protected systems, NAGD contributions are absent.

In order to highlight the influence of key concepts like the non-Abelian
Berry connection or the corresponding Berry curvature \citep{Nakahara2003},
let us briefly point to several different physical scenarios where
those concepts have been particularly fruitful. Early applications
included the fractional quantum Hall effect \citep{Arovas1984,Wen1991},
while more recent applications concern the fields of topological quantum
computation \citep{Nayak2008} and geometric quantum computation \citep{Pachos1999,Jones2000}.
Other examples come from studies of nuclear quadrupole resonance \citep{Zee1988},
or the physics of topological states of matter \citep{Hasan2010,Yang2014,Wen2017,Nayak2008}.
In condensed matter physics, the degenerate Bloch bands in solids
can be described by non-Abelian Berry connections \citep{Xiao2010}.
For systems of ultracold atoms, synthetic non-Abelian gauge fields
can be designed largely at will \citep{Osterloh2005,Li2016}. Related
ideas have also been put forward for artifical gauge fields in graphene
\citep{Vozmediano2010}. Finally, these concepts also show up in general
relativity through the geometric interpretation of Christoffel symbols
\citep{Carroll}. The question of how system-bath couplings will affect
the physics is of immediate relevance for most of these applications.

When the system is weakly coupled to a quantum bath, the control parameters
will be subject to Gaussian random fluctuations (noise) \citep{WeissBook}.
Throughout, we assume that the coupling of the system to the bath
does not introduce perturbations that lift the $N$-fold degeneracy
of the level of interest. A similar scenario arises if the control
parameters are decorated by classical time-dependent fluctuations.
Due to the system-bath coupling, there are classical fluctuations
around the deterministic loop trajectory, ${\bm{\lambda}}(t)\rightarrow{\bm{\lambda}}(t)+\delta{\bm{\lambda}}(t)$,
with weak fluctuations $\delta{\bm{\lambda}}(t)$. The question then
arises whether the dynamics of the degenerate open quantum system
--- which follows by performing the Gaussian average over the fluctuations
--- can still be characterized by geometric quantities which do not
depend on the detailed time dependence of the protocol but only on
the overall geometry of the reference path in loop space. For the
Abelian case, the authors of Ref.~\citep{Whitney2005} have shown
that on top of the well-known Berry phase, \emph{geometric dephasing}
contributions do exist. Remarkably, the sign of the associated geometric
dephasing rate depends on the direction sense of the loop protocol,
i.e., changes from decay to amplification as one reverses the orientation.
This prediction has been verified in recent experiments using superconducting
nanocircuits with artifically generated (and thus controllable) noise
fluctuations \citep{Berger2015}.

We here formulate the non-Abelian generalization of the theory in
Ref.~\citep{Whitney2005}. This case takes place when studying the
adiabatic dynamics of degenerate open quantum systems. As in the Abelian
case, we find geometric dephasing contributions that allow for a model-independent
general expression, and thus deserve to be called ``universal.''
The key object in our theory is the fluctuation-averaged $N\times N$
Berry matrix $\bar{{\cal U}}_{B}$. This matrix is non-unitary but
always admits a polar decomposition, $\bar{{\cal U}}_{B}=VR$, with
a unitary matrix $V$ and a positive semi-definite Hermitian matrix
$R$. In generic non-Abelian systems, we have the commutator $[V,R]\ne0$,
and the typical non-Abelian features discussed in detail below will
characterize NAGD contributions. On the other hand, we will also study
``poor man's non-Abelian'' systems below. Such systems have an $N$-fold
degenerate subspace but the above commutator vanishes, $[V,R]=0$.
Therefore, such systems do not allow to probe the rich physics associated
with NAGD in ``truly non-Abelian'' systems.

Since the averaged Berry matrix $\bar{{\cal U}}_{B}$ is hard to detect
directly, a large part of this work will be devoted to the formulation
of specific interference protocols that should enable the development
of NAGD detection experiments. As a concrete application of the general
formalism, we study noisy Majorana braiding protocols. In such setups,
braiding is executed by running time-dependent protocols for the tunnel
matrix elements $\Delta_{j}(t)$ between different Majorana pairs.
These tunnel matrix elements play the role of the control parameter
vector ${\bm{\lambda}}(t)$, and by allowing for fluctuations in these
quantities, one can gain access to the physics of NAGD. We believe
that this setup is probably the most elementary example for probing
NAGD. Its simplicity allows us to obtain fully analytical predictions
for different interference protocols. In addition, Majorana braiding
is presently also of considerable experimental interest \citep{Alicea2012,Lutchyn2017},
and we hope that this interest will also be beneficial to the development
of experimental tests for NAGD.

The structure of this article is as follows. In Sec.~\ref{sec2},
we provide a general discussion of the adiabatic quantum dynamics
of degenerate open systems. In particular, we derive a compact expression
for $\bar{{\cal U}}_{B}$, see Eq.~\eqref{avU}, which involves both
the Berry connection and the Berry curvature. We also discuss general
features of NAGD expected from the polar decomposition of $\bar{{\cal U}}_{B}$.
In Sec.~\ref{sec3}, we briefly discuss a generalization of the Stokes
theorem for non-Abelian systems and some implications thereof. In
particular, we show that in topologically protected systems, the vanishing
of the Berry curvature and the absence of NAGD are two sides of the
same coin. In Sec.~\ref{sec4}, we then present general interference
protocols aimed at NAGD detection. These protocols are applied to
a concrete example defined by noisy Majorana braiding setups in Sec.~\ref{sec5}.
Finally, we conclude with an outlook in Sec.~\ref{sec6}. Some technical
details have been delegated to the Appendix. Throughout we consider
the zero-temperature limit and put $\hbar=1$.

\section{Non-Abelian Geometric Dephasing}

\label{sec2}

In this section, we present a general analysis of the non-Abelian
adiabatic dynamics of open quantum systems. We derive and study the
averaged time evolution operator which follows by taking a Gaussian
average over the bath-induced fluctuations around a deterministic
loop path in control parameter space. In Sec.~\ref{sec2a}, we first
summarize known results for closed non-Abelian systems, see Refs.~\citep{Wilczek1984,Zee1988},
thereby also introducing our notation. We prepare the ground for treating
open non-Abelian systems in Sec.~\ref{sec2b}, where we study what
happens for a weakly perturbed parameter trajectory. The Gaussian
average over fluctuating trajectories is then performed in Sec.~\ref{sec2c}.
We show in Sec.~\ref{sec2d} that the resulting averaged Berry matrix,
$\bar{{\cal U}}_{B}$, is gauge covariant. Finally, in Sec.~\ref{sec2e},
we perform a matrix polar decomposition of $\bar{{\cal U}}_{B}$ in
order to illustrate the characteristic physical effects caused by
NAGD. We note that $\bar{{\cal U}}_{B}$ and its properties will directly
determine the outcome of the NAGD detection protocols introduced in
Secs.~\ref{sec4} and \ref{sec5}.

\subsection{Non-Abelian Adiabatic Dynamics}

\label{sec2a}

We start by considering a closed quantum system described by a Hamiltonian
$H({\bm{\lambda}})$ that depends on $d$ classical parameters, $\lambda^{\mu}$.
We combine these parameters to the vector ${\bm{\lambda}}=(\lambda^{1},\ldots,\lambda^{d})$.
For the non-Abelian case studied in this work, one requires that an
$N$-fold degenerate Hilbert subspace ($N\ge2)$ exists for all $\bm{\lambda}$
configurations of interest \citep{Wilczek1984,Zee1988}. We can express
the Hamiltonian as
\begin{equation}
H({\bm{\lambda}})=U({\bm{\lambda}})D({\bm{\lambda}})U^{\dagger}({\bm{\lambda}}),\label{eq:degenerate_Ham_diag}
\end{equation}
where $U({\bm{\lambda}})$ is a parameter-dependent unitary and $D({\bm{\lambda}})$
is a diagonal matrix. It contains several subspaces (``blocks''),
each having an $N_{j}$-fold degeneracy and a parameter-dependent
energy $E_{j}({\bm{\lambda}})$,
\begin{equation}
D({\bm{\lambda}})=\sum_{j=1,2,\ldots}E_{j}({\bm{\lambda}})P_{j},\label{diagmat}
\end{equation}
with the projector $P_{j}$ to the respective Hilbert subspace. The
degenerate blocks need not be of the same size but all $N_{j}$ are
assumed parameter-independent. We define $N_{1}=N$, thus identifying
the $N$-fold degenerate level with block $j=1$. Note that $N_{j\ne1}=1$
is allowed.

For a time-dependent parameter protocol ${\bm{\lambda}}(t)$, we then
have a time-dependent Hamiltonian, $H(t)=H({\bm{\lambda}}(t))=U(t)D(t)U^{\dagger}(t)$.
States evolve according to the Schr\"odinger equation,
\begin{equation}
i\partial_{t}\ket{\psi(t)}=H(t)\ket{\psi(t)},\label{schrodinger}
\end{equation}
where it is convenient to switch to the instantaneous eigenbasis,
\begin{equation}
\ket{\tilde{\psi}(t)}=U^{\dagger}(t)\ket{\psi(t)}.\label{psitilde}
\end{equation}
Using $\partial_{t}(U^{\dagger}U)=0$, Eq.~\eqref{schrodinger} yields
\begin{equation}
i\partial_{t}\ket{\tilde{\psi}(t)}=\tilde{H}(t)\ket{\tilde{\psi}(t)},\quad\tilde{H}(t)=D(t)-iU^{\dagger}(t)\partial_{t}U(t).\label{schr2}
\end{equation}
Throughout this work, we consider closed trajectories of total duration
$T$ such that ${\bm{\lambda}}(0)={\bm{\lambda}}(T)$.

For sufficiently slow ${\bm{\lambda}}(t)$ variation, and taking the
initial state from block 1, $P_{1}\ket{\tilde{\psi}(0)}=\ket{\tilde{\psi}(0)}$,
the quantum adiabatic theorem implies that the state will remain in
block 1 at all times, $P_{1}\ket{\tilde{\psi}(t)}=\ket{\tilde{\psi}(t)}$.
The adiabatic theorem holds when the inter-block transitions induced
by $U^{\dagger}(t)\partial_{t}U(t)$ can be neglected. The matrix
elements of $U^{\dagger}(t)\partial_{t}U(t)$ are typically of size
${\cal O}(1/T)$. Defining the minimal gap between block 1 and all the other
blocks,
\begin{equation}
{\cal E}={\rm min}_{{\bm{\lambda}},j\ne1}\left|E_{j}-E_{1}\right|,\label{defE}
\end{equation}
one can write the adiabatic theorem validity condition as ${\cal E}T\gg1$.
Note that the intra-block components of $U^{\dagger}(t)\partial_{t}U(t)$
may lift the in-block degeneracy by terms of order ${\cal O}(1/T)$ and, therefore,
should be taken into account.

In the adiabatic limit, we thus obtain the final state at
$t=T$ as $\ket{\tilde{\psi}(T)}={\cal U}\ket{\tilde{\psi}(0)}$,
with the unitary time evolution operator
\begin{equation}
{\cal U}=e^{-i\int_{0}^{T}dtE_{1}({\bm{\lambda}}(t))}\mathcal{T}e^{-\int_{0}^{T}dt\dot{\lambda}^{\mu}(t)A_{\mu}({\bm{\lambda}}(t))}.\label{adiabatic_evolution}
\end{equation}
Summation over repeated indices is used throughout, $\mathcal{T}$
denotes time ordering, and with $\partial_{\mu}=\partial_{\lambda^{\mu}}$
the non-Abelian Berry connection $A_{\mu}$ is given by \citep{Nakahara2003}
\begin{equation}
A_{\mu}({\bm{\lambda}})=P_{1}\alpha_{\mu}({\bm{\lambda}})P_{1}=-A_{\mu}^{\dagger},\quad\alpha_{\mu}=U^{\dagger}\partial_{\mu}U.\label{eq:A_mu_def}
\end{equation}
Writing ${\cal U}=e^{-i\int_{0}^{T}dtE_{1}}{\cal U}_{B}$, we arrive
at the $N\times N$ unitary Berry matrix \citep{Wilczek1984,Zee1988}
\begin{equation}
{\cal U}_{B}={\cal P}\exp\left(-\oint d\lambda^{\mu}A_{\mu}\right),\label{udef}
\end{equation}
where ${\cal P}$ denotes path ordering. The path-ordered Wilson loop
formulation of ${\cal U}_{B}$ underlines the geometric nature of
the Berry matrix. The detailed time dependence of the protocol does
not matter as long as the conditions behind the adiabatic theorem
are met.

The Berry connection \eqref{eq:A_mu_def} is defined only up to gauge
transformations, which correspond to changing the basis in the degenerate
block. Writing $U=W\Omega$ with parameter- (and hence time-)dependent
unitaries $W$ and $\Omega$, we demand that the commutator $[\Omega({\bm{\lambda}}),P_{j}]=0$
for all ${\bm{\lambda}}$ and $j$. In that case, $W({\bm{\lambda}})$
is a valid replacement for $U({\bm{\lambda}})$ in Eq.~(\ref{eq:degenerate_Ham_diag})
since $\Omega$ does not mix different blocks: $H({\bm{\lambda}})=W\Omega D\Omega^{\dagger}W^{\dagger}=WDW^{\dagger}.$
At the same time, Eq.~\eqref{eq:A_mu_def} gives $\alpha_{\mu}=\Omega^{\dagger}\left(W^{\dagger}\partial_{\mu}W+\partial_{\mu}\right)\Omega,$
and hence the Berry connection transforms as
\begin{equation}
A_{\mu}\rightarrow\Omega^{\dagger}A_{\mu}\Omega+P_{1}\Omega^{\dagger}\partial_{\mu}\Omega P_{1}.\label{eq:A_gauge_transform}
\end{equation}
States transform according to $\ket{\tilde{\psi}(t)}\rightarrow\Omega^{\dagger}(t)\ket{\tilde{\psi}(t)}$,
and by discretization of the time-ordered exponential in Eq.~\eqref{adiabatic_evolution},
one verifies that the Berry matrix is gauge covariant, cf.~Sec.~\ref{sec2d}
below,
\begin{equation}
{\cal U}_{B}\rightarrow\Omega^{\dagger}({\bm{\lambda}}(T))\ {\cal U}_{B}\ \Omega({\bm{\lambda}}(0)).\label{eq:T-exp_gauge_transform}
\end{equation}

Below we also encounter the Berry curvature (or field strength) tensor
\citep{Nakahara2003},
\begin{equation}
F_{\mu\nu}({\bm{\lambda}})=\partial_{\mu}A_{\nu}-\partial_{\nu}A_{\mu}+\left[A_{\mu},A_{\nu}\right]=-F_{\mu\nu}^{\dagger},\label{eq:F_munu_def}
\end{equation}
which is gauge covariant,
\begin{equation}
F_{\mu\nu}({\bm{\lambda}})\rightarrow\Omega^{\dagger}({\bm{\lambda}})F_{\mu\nu}({\bm{\lambda}})\Omega({\bm{\lambda}}).\label{eq:F_gauge_transform}
\end{equation}
In this context, it is instructive to highlight the role of the projectors
$P_{j}$. When replacing $A_{\mu}\rightarrow\alpha_{\mu}$, see Eq.~\eqref{eq:A_mu_def},
in order to define a tensor $\Phi_{\mu\nu}$ as in Eq.~\eqref{eq:F_munu_def},
one finds easily that $\Phi_{\mu\nu}$ vanishes identically,
\begin{eqnarray}
\Phi_{\mu\nu} & = & \partial_{\mu}\alpha_{\nu}-\partial_{\nu}\alpha_{\mu}+\left[\alpha_{\mu},\alpha_{\nu}\right]\\
 & = & \partial_{[\mu}U^{\dagger}\partial_{\nu]}U+\left[U^{\dagger}\partial_{\mu}U,U^{\dagger}\partial_{\nu}U\right]=0.\nonumber
\end{eqnarray}
Here we use the notation $\partial_{[\mu}U^{\dagger}\partial_{\nu]}U=\partial_{\mu}U^{\dagger}\partial_{\nu}U-(\mu\leftrightarrow\nu)$
and the relation
\begin{equation}
\partial_{\mu}U^{\dagger}=-(U^{\dagger}\partial_{\mu}U)U^{\dagger}.\label{eq:der_U^dag_identity}
\end{equation}
In contrast, because the projector $P_{1}$ appears in the definition
of the Berry connection in Eq.~\eqref{eq:A_mu_def}, the Berry curvature
is given by
\begin{eqnarray}
F_{\mu\nu} & = & P_{1}\partial_{[\mu}U^{\dagger}\partial_{\nu]}UP_{1}+\left[P_{1}U^{\dagger}\partial_{\mu}UP_{1},P_{1}U^{\dagger}\partial_{\nu}UP_{1}\right]\nonumber \\
 & = & -\sum_{j\neq1}P_{1}\alpha_{[\mu}P_{j}\alpha_{\nu]}P_{1},\label{eq:Fmunu_via_U_and_Ps}
\end{eqnarray}
which in general does not vanish.

\subsection{Perturbed parameter trajectory}

\label{sec2b}

In preparation for the averaging over bath-induced fluctuations in
open systems, let us now consider a parameter trajectory ${\bm{\lambda}}(t)+\delta{\bm{\lambda}}(t)$
which contains a slow and weak perturbation $\delta{\bm{\lambda}}(t)$
around the base path ${\bm{\lambda}}(t)$. We assume that the fluctuation
trajectory $\delta{\bm{\lambda}}(t)$ also meets the assumptions behind
the quantum adiabatic theorem. For simplicity, we also impose $\delta{\bm{\lambda}}(t=0,T)=0$
even though this condition could be relaxed  \citep{foot_fluctuations_at_0}.
The noise trajectory $\delta{\bm{\lambda}}(t)$ may be caused by classical
control parameter fluctuations or by the coupling of the system to
a quantum bath, see Sec.~\ref{sec2c}.

To make progress, we expand the Hamiltonian $\tilde{H}(t)$, cf.~Eq.~\eqref{schr2},
for the perturbed trajectory in powers of $\delta{\bm{\lambda}}$
up to second order,
\begin{equation}
\tilde{H}(t)=D({\bm{\lambda}}(t))+\tilde{H}_{1}(t)+\tilde{H}_{2}(t)+{\cal O}(\delta\lambda^{3}).\label{expa1}
\end{equation}
By using Eqs.~\eqref{eq:degenerate_Ham_diag} and \eqref{eq:der_U^dag_identity}
together with $U^{\dagger}\partial_{\mu}\partial_{\nu}U=\partial_{\mu}\alpha_{\nu}+\alpha_{\mu}\alpha_{\nu}$
and $(\partial_{\mu}\partial_{\nu}U^{\dagger})U=-\partial_{\mu}\alpha_{\nu}+\alpha_{\nu}\alpha_{\mu},$
we obtain in a first step
\begin{eqnarray}
\partial_{\mu}H({\bm{\lambda}}) & = & U\left(\partial_{\mu}D+\left[\alpha_{\mu},D\right]\right)U^{\dagger},\\
\partial_{\mu}\partial_{\nu}H({\bm{\lambda}}) & = & U\bigl(\partial_{\mu}\partial_{\nu}D+\left[\partial_{\mu}\alpha_{\nu},D\right]\nonumber \\
 & + & \left[\alpha_{\{\mu},\partial_{\nu\}}D\right]+\left[\alpha_{\mu},\left[\alpha_{\nu},D\right]\right]\bigr)U^{\dagger},\nonumber
\end{eqnarray}
where $\alpha_{\{\mu}\partial_{\nu\}}D=\alpha_{\mu}\partial_{\nu}D+(\mu\leftrightarrow\nu).$
We thus arrive at the first-order contribution
\begin{equation}
\tilde{H}_{1}=-i\dot{\lambda}^{\mu}\alpha_{\mu}+\delta\lambda^{\mu}\left(\partial_{\mu}D+\left[\alpha_{\mu},D\right]\right).\label{dH1}
\end{equation}
With the anticommutator $\{\cdot,\cdot\}$ the second-order term is
given by
\begin{eqnarray}
\tilde{H}_{2} & = & \delta\lambda^{\mu}\delta\lambda^{\nu}\left(\frac{1}{2}\partial_{\mu}\partial_{\nu}D+\frac{1}{2}\left\{ \alpha_{\mu}\alpha_{\nu},D\right\} \right.\label{dH2}\\
 & - & \left.\alpha_{\mu}D\alpha_{\nu}+\left[\alpha_{\mu},\partial_{\nu}D\right]+\frac{1}{2}\left[\partial_{\mu}\alpha_{\nu},D\right]\right).\nonumber
\end{eqnarray}
Separating Eqs.~\eqref{dH1} and \eqref{dH2} into block-diagonal
and off-diagonal terms, it is now straightforward to perform a Schrieffer-Wolff
transformation which projects the full Hamiltonian to the $N$-fold
degenerate subspace. The effective low-energy Hamiltonian acting only
within block $j=1$ follows as
\begin{widetext}
\begin{equation}
\tilde{H}_{\mathrm{eff}}^{11}=P_{1}E_{1}(t)-i\dot{\lambda}^{\mu}A_{\mu}+\sum_{j\neq1}P_{1}\alpha_{\mu}P_{j}\alpha_{\nu}P_{1}\left(\frac{\dot{\lambda}^{\mu}\dot{\lambda}^{\nu}}{E_{j}-E_{1}}+i\left(\dot{\lambda}^{\nu}\delta\lambda^{\mu}-\dot{\lambda}^{\mu}\delta\lambda^{\nu}\right)\right)+{\cal O}\left(\left[\delta\lambda^{2}+\dot{\lambda}^{2}\right]^{3/2}\right).\label{eq:H_eff_11_second_order_calculation}
\end{equation}
\end{widetext}

Here the time dependence of the first term $\sim E_{1}(t)$ is inherited
from ${\bm{\lambda}}(t)+\delta{\bm{\lambda}}(t)$. Equation~\eqref{eq:H_eff_11_second_order_calculation}
reveals a remarkable cancellation of all contributions not containing
$\dot{\lambda}^{\mu}$, apart from the first term $\sim E_{1}(t)$.

At this point we note that $\dot{\lambda}^{\mu}={\cal O}(\omega)$,
where the adiabatic theorem requires $\omega=2\pi/T$ to be small
compared to the minimal energy difference ${\cal E}$ between the
$N$-fold degenerate level and all other states. (The precise condition
for adiabatic evolution also has to account for the matrix elements
between adiabatic states, see, e.g., Refs.~\citep{Lubin1990,Shimshoni1991}.)
The energy scale ${\cal E}$ is here defined by Eq.~\eqref{defE}.
 In the adiabatic limit, we have ${\cal E}T\gg1$ and thus can neglect
the subleading term $\sim\dot{\lambda}^{\mu}\dot{\lambda}^{\nu}={\cal O}\left(T^{-2}\right)$
in Eq.~(\ref{eq:H_eff_11_second_order_calculation}). Furthermore,
by using Eq.~\eqref{eq:Fmunu_via_U_and_Ps}, the last term in Eq.~\eqref{eq:H_eff_11_second_order_calculation}
can be rewritten in terms of the field strength tensor,
\begin{eqnarray}
 &  & i(\dot{\lambda}^{\nu}\delta\lambda^{\mu}-\dot{\lambda}^{\mu}\delta\lambda^{\nu})\sum_{j\neq1}P_{1}\alpha_{\mu}P_{j}\alpha_{\nu}P_{1}=\\
 &  & =-i\dot{\lambda}^{\mu}\delta\lambda^{\nu}\sum_{j\neq1}P_{1}\alpha_{[\mu}P_{j}\alpha_{\nu]}P_{1}=i\dot{\lambda}^{\mu}\delta\lambda^{\nu}F_{\mu\nu}.\nonumber
\end{eqnarray}
With these simplifications, Eq.~\eqref{eq:H_eff_11_second_order_calculation}
takes the form
\begin{equation}
\tilde{H}_{\mathrm{eff}}^{11}=P_{1}E_{1}(t)-i\dot{\lambda}^{\mu}A_{\mu}+i\dot{\lambda}^{\mu}\delta\lambda^{\nu}F_{\mu\nu},\label{eq:H_eff_11_second_order}
\end{equation}
where ${\cal O}(T^{-2},\delta\lambda^{2}/T,\delta\lambda^{3})$ terms
have been dropped.

In summary, the adiabatic time evolution operator follows as
\begin{equation}
{\cal U}=e^{-i\int_{0}^{T}dtE_{1}\left({\bm{\lambda}}(t)+\delta{\bm{\lambda}}(t)\right)}\ {\cal U}_{B},\label{Uperturbed}
\end{equation}
where the Berry matrix for the perturbed trajectory is given by
\begin{equation}
{\cal U}_{B}=\mathcal{P}\exp{\oint d\lambda^{\mu}\left(-A_{\mu}+\delta\lambda^{\nu}F_{\mu\nu}\right)}.\label{eq:corrected_evolution_first_order}
\end{equation}
Here $A_{\mu}$ and $F_{\mu\nu}$ are evaluated for the reference
path with $\delta{\bm{\lambda}}=0$. Equation \eqref{eq:corrected_evolution_first_order}
looks natural in the Abelian case, where the Berry phase can be written
as an area integral over the Berry curvature. Since the area is extended
via $\delta\lambda^{\nu}$ in Eq.~\eqref{eq:corrected_evolution_first_order},
we get a correction term. We show in Sec.~\ref{sec3a} that Eq.~\eqref{eq:corrected_evolution_first_order}
also looks natural in the non-Abelian case when invoking a non-Abelian
variant of the Stokes theorem.

\subsection{Noise-averaged Berry matrix}

\label{sec2c}

In this subsection, we discuss how to average the time evolution operator
${\cal U}$ in Eq.~\eqref{Uperturbed} over different realizations
of the fluctuating trajectories $\delta{\bm{\lambda}}(t)$. The result,
${\cal U}\rightarrow\bar{{\cal U}}$, appears in all experimental
protocols aimed at NAGD detection, see Secs.~\ref{sec4} and \ref{sec5}.
In addition, after discussing the gauge covariance of $\bar{{\cal U}}$
in Sec.~\ref{sec2d}, we show in Sec.~\ref{sec2e} that a matrix
polar decomposition of $\bar{{\cal U}}$ affords a good intuitive
understanding of geometric dephasing in non-Abelian systems.

\subsubsection{Control parameter fluctuations}

We assume that the fluctuations of the control parameters are distributed
according to the measure
\begin{equation}
\sim\int\mathcal{D}\delta{\bm{\lambda}}\ e^{-\frac{1}{2}\int_{0}^{T}dtdt'[\tilde{\sigma}^{-1}]_{\mu\nu}(t-t')\delta\lambda^{\mu}(t)\delta\lambda^{\nu}(t')},\label{measure}
\end{equation}
implying Gaussian statistics with vanishing mean, $\langle\delta\lambda^{\mu}(t)\rangle_{{\rm env}}=0$,
and the two-point time correlation function
\begin{equation}
\left\langle \delta\lambda^{\mu}(t)\delta\lambda^{\nu}(t')\right\rangle _{{\rm env}}=\tilde{\sigma}^{\mu\nu}(t-t')=\sigma^{\mu\nu}\delta_{\tau_{c}}(t-t').\label{noisedef}
\end{equation}
The last step in Eq.~\eqref{noisedef} introduces the real positive
$d\times d$ noise amplitude matrix with $\sigma^{\mu\nu}=\sigma^{\nu\mu}$.
The time dependence of the correlator in Eq.~\eqref{noisedef} is
modeled by a $\delta$-function broadened by the noise autocorrelation
time $\tau_{c}$. For example, in experiments designed for detecting
Abelian geometric dephasing, see Ref.~\citep{Berger2015}, tunable
artificial classical noise with $\delta_{\tau_{c}}(t)=e^{-|t|/\tau_{c}}/(2\tau_{c})$
was used. We denote the typical size of the $\sigma^{\mu\nu}$ matrix
elements (e.g., the largest eigenvalue) by $\sigma$. The typical
size of fluctuations $\delta\lambda^{\mu}(t)$ is therefore of order
${\cal O}\left(\sqrt{\sigma/\tau_{c}}\right)$. Without loss of generality,
all $\lambda^{\mu}$ parameters, and therefore also the $\sigma^{\mu\nu}$
coefficients, are assumed to have units of energy.

When the system is coupled to a quantum bath, $\delta{\bm{\lambda}}(t)$
represents operators acting on the bath Hilbert space. By treating
typical system-bath couplings within the standard Born-Markov approximation
\citep{WeissBook}, one arrives at Gaussian fluctuations described
by the measure \eqref{measure}, where $\sigma^{\mu\nu}$ depends
on details such as the bath temperature or the system-bath coupling
strength. In some cases, the system-bath coupling Hamiltonian may
also include terms not present in the $d$-dimensional parameter set
${\bm{\lambda}}$. One can then enlarge the parameter space to include
a deterministic part with additional components $\lambda^{\mu}(t)=0$,
at the same time allowing for Gaussian fluctuations $\delta\lambda^{\mu}(t)\ne0$.
Our approach therefore covers the general case. However, we assume
that the system-bath coupling is not able to lift the $N$-fold degeneracy
of the level of interest.

To illustrate the above discussion, let us consider
the Hamiltonian $H_{\mathrm{tot}}=H_{\mathrm{s}}({\bm{\lambda}})+H_{\mathrm{int}}+H_{\mathrm{b}}$,
where the system Hamiltonian,
\begin{equation}
H_{\mathrm{s}}({\bm{\lambda}})=\sum_{\mu}\lambda^{\mu}(t)O_{\mu},
\end{equation}
is expressed in terms of system operators $O_{\mu}$, and
$H_{\mathrm{b}}$ is the bath Hamiltonian. The system-bath interaction is encoded by
\be
H_{\mathrm{int}}=g\hat{X}O_{1},
\ee
with a bath operator $\hat{X}$  and a coupling constant $g$. We now assume that the bath state and the $H_{\mathrm{b}}$-generated
dynamics imply  $\langle\hat{X}(t)\rangle_{\rm b}=0$ and $\langle\hat{X}(t)\hat{X}(t')\rangle_{\rm b}=x^2\delta_{\tau_{c}}(t-t')$, with
some amplitude $x$.
For the system density matrix, $\rho_{\rm s}=\mathrm{tr}_{\mathrm{b}}\rho_{\mathrm{tot}}$,
the standard Born-Markov approximation then leads to a Lindbladian master equation,
\begin{equation}
\partial_{t}\rho_{\rm s}=-i\left[H_{\mathrm{s}},\rho_{s}\right]
-\frac{g^{2}x^{2}}{2}
\left(O_{1}^{2}\rho_{s}+\rho_{s}O_{1}^{2}-2O_{1}\rho_{s}O_{1}\right).\label{eq:system_master_equation}
\end{equation}
Alternatively,  the time evolution of the system can be described by a stochastic
Schr\"odinger equation,
\begin{equation}
i\partial_{t}\ket{\psi_{\mathrm{s}}}=\left(H_{\mathrm{s}}+\delta\lambda^{1}O_{1}\right)\ket{\psi_{\mathrm{s}}},\label{eq:system_stoch_schrod}
\end{equation}
where $\delta\lambda^{1}$ is a stochastic variable with zero average
and the correlation function $\left\langle \delta\lambda^{1}(t)\delta\lambda^{1}(t')\right\rangle_{\rm env}=g^{2}\langle\hat{X}(t)\hat{X}(t')\rangle_{\rm b}=\sigma^{11}\delta_{\tau_{c}}(t-t')$,
with $\sigma^{11}=g^{2}x^{2}$. Expanding the evolution of $\ket{\psi_{\mathrm{s}}}\bra{\psi_{\mathrm{s}}}$
to second order in $\delta\lambda^{1}$, and subsequently averaging over the
fluctuations, we find that $\rho_{\rm s}=\left\langle \ket{\psi_{\mathrm{s}}}\bra{\psi_{\mathrm{s}}}\right\rangle_{\rm env}$
obeys Eq.~(\ref{eq:system_master_equation}).
The stochastic Hamiltonian evolution in Eq.~\eqref{eq:system_stoch_schrod} thus represents
an unraveling of the master equation  (\ref{eq:system_master_equation}).

These insights enable us to
treat classical control parameter fluctuations and fluctuations due to the interaction of the system with a bath on the
same footing \cite{foot_Whitney}.  Nonetheless, a subtle distinction between both cases remains.
For classical
fluctuations of the control parameters, each individual run is fully coherent,
and dephasing effects arise only upon averaging over fluctuations. When the system is coupled to a bath,  system-bath entanglement implies a dephasing of the system dynamics
 even in a single experimental run.
 In the latter case, averaging over $\delta{\bm{\lambda}}(t)$
is merely a mathematical trick to describe the system-bath entanglement.
However, since in practice one obtains the environmental averages discussed below by repeating the experiment many times (for the same reference trajectory), this subtle distinction is of no importance in what follows.

\subsubsection{Parameter regime of interest}

\label{sec2c2}

In order to define the parameter regime covered by our theory below,
we introduce three dimensionless parameters,
\begin{equation}
\mathfrak{T}={\cal E}T,\quad\mathfrak{t_{c}}={\cal E}\tau_{c},\quad\mathfrak{s}=\sigma/{\cal E},\label{expansion_parameters}
\end{equation}
with the energy scale ${\cal E}$ in Eq.~\eqref{defE}. Below we
shall require validity of the inequality chain
\begin{equation}
\mathfrak{T}\gg\mathfrak{t_{c}}\gg1\gg\mathfrak{s}.\label{conditions}
\end{equation}
The first inequality implies that noise is local on the time scale
$T$, while the second inequality reinstates that we assume the adiabatic
limit for each realization of a fluctuating trajectory. We note that
for $\mathfrak{t_{c}}\alt1$, non-adiabatic processes may become important
where, in particular, noise fluctuations introduce corrections to
adiabatic eigenstates via block-off-diagonal terms in the Hamiltonian.
Such processes may affect both dynamic and geometric dephasing. However,
we leave studies of non-adiabatic corrections to future work and here
focus on the limit $\mathfrak{t_{c}}\gg1$. Finally, the inequality
$\mathfrak{s}\ll\mathfrak{t_{c}}$ implements the weak system-bath
coupling assumption behind the Born approximation \citep{WeissBook}.
While this leaves room for the intermediate regime $1\alt\mathfrak{s}\ll\mathfrak{t_{c}}$,
for simplicity, we here assume $\mathfrak{s}\ll1$.\\

\subsubsection{Averaged Berry matrix}

We now average the time evolution operator ${\cal U}$ for a perturbed
trajectory, see Eqs.~\eqref{Uperturbed} and \eqref{eq:corrected_evolution_first_order},
over the Gaussian fluctuations $\delta{\bm{\lambda}}(t)$. Expanding
the energy $E_{1}({\bm{\lambda}}+\delta{\bm{\lambda}})=E_{1}({\bm{\lambda}})+\delta\lambda^{\mu}\partial_{\mu}E_{1}({\bm{\lambda}})$
to lowest order in $\delta{\bm{\lambda}}$, we obtain
\begin{equation}
\bar{\mathcal{U}}=e^{-i\int_{0}^{T}dtE_{1}}\left\langle \mathcal{T}e^{-\int_{0}^{T}dt\left[\dot{\lambda}^{\mu}A_{\mu}+\delta\lambda^{\mu}\left(i\partial_{\mu}E_{1}+\dot{\lambda}^{\nu}F_{\mu\nu}\right)\right]}\right\rangle _{\mathrm{env}}.\label{aux1}
\end{equation}
Next we make use of the standard representation for time-ordered exponentials,
\begin{equation}
\mathcal{T}\exp\left(\int_{0}^{T}dt\ O(t)\right)=\sum_{n=0}^{\infty}\frac{1}{n!}\mathcal{T}\left(\int_{0}^{T}dt\ O(t)\right)^{n}.
\end{equation}
With the auxiliary functions
\begin{equation}
f(t)=-\dot{\lambda}^{\mu}A_{\mu},\quad g_{\mu}(t)=-i\partial_{\mu}E_{1}-\dot{\lambda}^{\nu}F_{\mu\nu},
\end{equation}
\\
 we can express the time-ordered exponential as
\begin{widetext}
\begin{eqnarray}
\mathcal{T}e^{\int_{0}^{T}dt\left[f(t)+\delta\lambda^{\mu}g_{\mu}(t)\right]} & = & \sum_{n=0}^{\infty}\frac{1}{n!}\sum_{k=0}^{n}C_{n}^{k}\mathcal{T}\left(\int_{0}^{T}dt\ f(t)\right)^{k}\left(\int_{0}^{T}dt\ \delta\lambda^{\mu}g_{\mu}(t)\right)^{n-k}\\
 & = & \sum_{k=0}^{\infty}\sum_{l=0}^{\infty}\frac{1}{k!l!}\mathcal{T}\left(\int_{0}^{T}dt\ f(t)\right)^{k}\left(\int_{0}^{T}dt\ \delta\lambda^{\mu}g_{\mu}(t)\right)^{l},\nonumber
\end{eqnarray}
where $C_{n}^{k}=n!/\left[k!(n-k)!\right]$ is the binomial coefficient.
Note that the time ordering prescription ensures the correct order
of operators. The fluctuation average in Eq.~\eqref{aux1} thus follows
as
\begin{equation}
\left\langle \mathcal{T}e^{-\int_{0}^{T}dt\left[\dot{\lambda}^{\mu}A_{\mu}+\delta\lambda^{\mu}\left(i\partial_{\mu}E_{1}+\dot{\lambda}^{\nu}F_{\mu\nu}\right)\right]}\right\rangle _{\mathrm{env}}=\sum_{k=0}^{\infty}\frac{1}{k!}\mathcal{T}\left[\left(\int_{0}^{T}dt\ f(t)\right)^{k}\sum_{l=0}^{\infty}\frac{1}{l!}\left\langle \mathcal{T}\left(\int_{0}^{T}dt\ \delta\lambda^{\mu}g_{\mu}(t)\right)^{l}\right\rangle _{\mathrm{env}}\right].
\end{equation}
Using the Gaussian fluctuation measure in Eq.~(\ref{measure}), we
then obtain
\begin{eqnarray}
\sum_{l=0}^{\infty}\frac{1}{l!}\left\langle \mathcal{T}\left(\int_{0}^{T}dt\ \delta\lambda^{\mu}g_{\mu}(t)\right)^{l}\right\rangle _{\mathrm{env}} & = & \sum_{l=0}^{\infty}\frac{(l-1)!!}{l!}\delta_{l,\mathrm{even}}\left\langle \mathcal{T}\left(\int_{0}^{T}dt\int_{0}^{T}dt'\ \tilde{\sigma}^{\mu\nu}(t-t')g_{\mu}(t)g_{\nu}(t)\right)^{l/2}\right\rangle _{\mathrm{env}}\nonumber \\
 & = & \sum_{m=0}^{\infty}\frac{1}{2^{m}m!}\left\langle \mathcal{T}\left(\int_{0}^{T}dt\int_{0}^{T}dt'\ \tilde{\sigma}^{\mu\nu}(t-t')g_{\mu}(t)g_{\nu}(t')\right)^{m}\right\rangle _{\mathrm{env}},\label{eq:NA_averaging}
\end{eqnarray}
where $(l-1)!!=(l-1)(l-3)\cdots1$ denotes the semifactorial. Using
$\tilde{\sigma}^{\mu\nu}(t)=\sigma^{\mu\nu}\delta(t)$, where the
effect of finite noise correlation times $\tau_{c}$ will be discussed
in a moment, we find
\begin{equation}
\int_{0}^{T}dt\int_{0}^{T}dt'\ \tilde{\sigma}^{\mu\nu}(t-t')g_{\mu}(t)g_{\nu}(t')=\sigma^{\mu\nu}\int_{0}^{T}dt\left[-\partial_{\mu}E_{1}\partial_{\nu}E_{1}+2i\dot{\lambda}^{\rho}F_{\mu\rho}\partial_{\nu}E_{1}+\dot{\lambda}^{\rho}\dot{\lambda}^{\eta}F_{\mu\rho}F_{\nu\eta}\right],\label{eq:NA_correlation}
\end{equation}
and thus
\begin{equation}
\left\langle \mathcal{T}e^{-\int_{0}^{T}dt\left[\dot{\lambda}^{\mu}A_{\mu}+\delta\lambda^{\mu}\left(i\partial_{\mu}E_{1}+\dot{\lambda}^{\nu}F_{\mu\nu}\right)\right]}\right\rangle _{\mathrm{env}}=\mathcal{T}e^{\int_{0}^{T}dt\left[-\dot{\lambda}^{\mu}A_{\mu}-\frac{1}{2}\sigma^{\mu\nu}\partial_{\mu}E_{1}\partial_{\nu}E_{1}+i\sigma^{\mu\nu}\dot{\lambda}^{\rho}F_{\mu\rho}\partial_{\nu}E_{1}\right]},\label{eq:NA_average}
\end{equation}
where the term $\sim\dot{\lambda}^{\rho}\dot{\lambda}^{\eta}={\cal O}\left(T^{-2}\right)$
has been dropped since it vanishes in the adiabatic limit.
\end{widetext}

A finite noise correlation time $\tau_{c}$, and thus the finite range
of $\tilde{\sigma}^{\mu\nu}(t-t')$, will introduce corrections to
Eq.~\eqref{eq:NA_average}. First, corrections may arise from $g_{\nu}(t')\neq g_{\nu}(t)$
in Eq.~(\ref{eq:NA_correlation}). Taylor expanding $g_{\nu}(t')$
around $t$ and using $\tilde{\sigma}^{\mu\nu}(t-t')=\tilde{\sigma}^{\mu\nu}(t'-t)$,
we find that Eq.~(\ref{eq:NA_correlation}) is accurate up to $\mathcal{O}\left(\mathfrak{t_{c}}^{2}/\mathfrak{T}\right)$
with the expansion parameters in Eq.~\eqref{expansion_parameters}.
Second, additional corrections may come from the fact that the $g_{\mu}(t)$
are matrices which do not commute at different times. For finite values
of $\tau_{c}$, this causes issues when going from Eq.~(\ref{eq:NA_correlation})
to Eq.~(\ref{eq:NA_average}) since the time ordering in Eq.~(\ref{eq:NA_averaging})
mixes the order of $F_{\mu\nu}$ belonging to different integrals.
We find that these corrections scale as $\mathcal{O}\left(\mathfrak{t_{c}}/\mathfrak{T}\right)$.
Therefore, all corrections to Eq.~\eqref{eq:NA_average} coming from
the finite noise correlation time $\tau_{c}$ will vanish in the adiabatic
limit.

We thereby obtain the averaged evolution operator as
\begin{equation}
\bar{{\cal U}}=e^{-i\int_{0}^{T}dtE_{1}}e^{-\frac{1}{2}\sigma^{\mu\nu}\int_{0}^{T}dt\ \partial_{\mu}E_{1}\partial_{\nu}E_{1}}\ \bar{{\cal U}}_{B},\label{eq:NAGP_averaged}
\end{equation}
where the averaged Berry matrix is expressed as a path-ordered exponential,
\begin{equation}
\bar{{\cal U}}_{B}=\mathcal{P}\exp\oint d\lambda^{\mu}\left(-A_{\mu}+i\sigma^{\nu\rho}F_{\nu\mu}\partial_{\rho}E_{1}\right),\label{avU}
\end{equation}
which evidently is a purely geometric contribution. The term $\sim E_{1}$
in Eq.~\eqref{eq:NAGP_averaged} contains the dynamic phase of the
deterministic base path. The second term has a trivial matrix structure
in the $N$-dimensional Hilbert space of the degenerate block
and encodes the conventional dynamic dephasing rate $\Gamma_{{\rm dyn}}$.
Writing this term as $e^{-\Gamma_{{\rm dyn}}T}$, we have
\begin{equation}
\Gamma_{\mathrm{dyn}}=\frac{\sigma^{\mu\nu}}{2T}\int_{0}^{T}dt\ \partial_{\mu}E_{1}\partial_{\nu}E_{1}.\label{dynamic_dephasing}
\end{equation}
The rate $\Gamma_{\mathrm{dyn}}\sim\sigma$ does not change when the
orientation sense of the protocol is reversed, i.e., when replacing
${\bm{\lambda}}(t)\rightarrow{\bm{\lambda}}'(t)\equiv{\bm{\lambda}}(T-t)$.

The averaged Berry matrix, $\bar{{\cal U}}_{B}$ in Eq.~\eqref{avU},
contains the non-Abelian Berry phase of the reference path (the term
$\sim A_{\mu}$) as well as the NAGD contribution $\sim F_{\nu\mu}$.
This term obeys $\left(i\sigma^{\nu\rho}F_{\nu\mu}\partial_{\rho}E_{1}\right)^{\dagger}=i\sigma^{\nu\rho}F_{\nu\mu}\partial_{\rho}E_{1}$,
implying that $\bar{{\cal U}}_{B}$ is not unitary and thus contains
dephasing contributions. We analyze its effect in more detail in Sec.~\ref{sec2e}.
As in the Abelian case \citep{Whitney2005}, geometric dephasing requires
the simultaneous presence of dynamic phase fluctuations, $\sim\partial_{\rho}E_{1}$,
and geometric phase fluctuations, $\sim F_{\nu\mu}$.

In contrast to the dynamic dephasing in Eq.~\eqref{dynamic_dephasing},
the NAGD term is sensitive to the orientation sense of the protocol.
In fact, we infer from Eq.~\eqref{avU} that a reversal of the orientation
sense simply amounts to the replacement $\bar{{\cal U}}_{B}\rightarrow\bar{{\cal U}}_{B}^{-1}$.
Indeed, consider the discretization of the time-ordered exponential,
\begin{eqnarray}
\bar{{\cal U}}_{B} & = & \mathcal{T}e^{\int_{0}^{T}dt\dot{\lambda}^{\mu}\left[-A_{\mu}+i\sigma^{\nu\rho}F_{\nu\mu}\partial_{\rho}E_{1}\right]}\equiv\mathcal{T}e^{\int_{0}^{T}dt\dot{\lambda}^{\mu}O_{\mu}(t)}\nonumber \\
 & = & \lim_{\delta_{t}\to0}\prod_{k=0}^{T/\delta_{t}}e^{\delta_{t}\dot{\lambda}^{\mu}(t_{k})O_{\mu}(\bm{\lambda}(t_{k}))},\label{eq:UB_discretization}
\end{eqnarray}
where $t_{k}=k\delta_{t}$ with a short-time discretization parameter
$\delta_{t}$. The time ordering over $k$ is implied in the product.
For the reversed protocol, ${\bm{\lambda}}(t)\rightarrow{\bm{\lambda}}'(t)\equiv{\bm{\lambda}}(T-t)$,
we thus have the averaged Berry matrix (with $\delta_{t}\to0$)
\begin{eqnarray}
\bar{{\cal U}}_{B}' & = & \prod_{k=0}^{T/\delta_{t}}e^{\delta_{t}\dot{\lambda}'^{\mu}(t_{k})O_{\mu}(\bm{\lambda}'(t_{k}))}\nonumber \\
 & = & \prod_{k=0}^{T/\delta_{t}}e^{-\delta_{t}\dot{\lambda}{}^{\mu}(T-t_{k})O_{\mu}(\bm{\lambda}(T-t_{k}))}\label{inverse}\\
 & = & \prod_{k=T/\delta_{t}}^{0}e^{-\delta_{t}\dot{\lambda}{}^{\mu}(t_{k})O_{\mu}(\bm{\lambda}(t_{k}))}=\bar{{\cal U}}_{B}^{-1}.\nonumber
\end{eqnarray}
We note that Eq.~(\ref{avU}) was derived by expanding the Hamiltonian
(\ref{eq:H_eff_11_second_order}) up to linear order in $\delta\bm{\lambda}(t)$.
Taking further orders of $\delta\bm{\lambda}(t)$ into account will
result in corrections to Eqs.~(\ref{eq:NAGP_averaged}) and (\ref{avU}).
In fact, both the dynamic and the geometric terms acquire extra contributions
which are suppressed by powers of $\mathfrak{s}$ and/or $\mathfrak{t_{c}}^{-1}$
compared to the terms kept above. Nevertheless, the relation $\bar{{\cal U}}_{B}'=\bar{{\cal U}}_{B}^{-1}$
in Eq.~\eqref{inverse} holds to any order of the expansion as it
only relies on the path ordering and on the geometric term being proportional
to $\dot{\lambda}^{\mu}$.

\subsection{Gauge covariance}

\label{sec2d}

Next we show that Eqs.~\eqref{eq:corrected_evolution_first_order}
and \eqref{avU} are gauge covariant, i.e., that the averaged Berry
matrix does not depend on the particular basis used for describing
the system. Consider first the transformation of the unperturbed Berry
matrix (\ref{udef}). Under a gauge transformation $\Omega(\bm{\lambda})$,
$\mathcal{U}_{B}$ transforms according to Eq.~(\ref{eq:T-exp_gauge_transform})
so that $U(\bm{\lambda}(T))\mathcal{U}_{B}U^{\dagger}(\bm{\lambda}(0))$
is gauge invariant.

Consider now the time-discretized version of $\bar{\mathcal{U}}_{B}$
in Eq.~(\ref{eq:UB_discretization}). With $\delta_{t}\rightarrow0$,
we have
\begin{equation}
\bar{\mathcal{U}}_{B}=\prod_{k=0}^{T/\delta_{t}}e^{\delta_{t}\dot{\lambda}^{\mu}O_{\mu}(\bm{\lambda}(t_{k}))}=\prod_{k=0}^{T/\delta_{t}}\left(1+\delta_{t}\dot{\lambda}^{\mu}O_{\mu}(\bm{\lambda}(t_{k}))\right).\label{auxx}
\end{equation}
Using the gauge transformation rules for $A_{\mu}$, Eq.~(\ref{eq:A_gauge_transform}),
and for $F_{\mu\nu}$, Eq.~(\ref{eq:F_gauge_transform}), we obtain
\begin{eqnarray}
O_{\mu}(\bm{\lambda}(t_{k})) & \rightarrow & \Omega^{\dagger}(\bm{\lambda}(t_{k}))\,O_{\mu}(\bm{\lambda}(t_{k}))\,\Omega(\bm{\lambda}(t_{k}))\nonumber \\
 & + & P_{1}\Omega^{\dagger}(\bm{\lambda}(t_{k}))\,\partial_{\mu}\Omega(\bm{\lambda}(t_{k})),
\end{eqnarray}
which implies that
\begin{eqnarray}
 &  & \left(1+\delta_{t}\dot{\lambda}^{\mu}O_{\mu}(\bm{\lambda}(t_{k}))\right)P_{1}\rightarrow\Omega^{\dagger}(\bm{\lambda}(t_{k}+\delta_{t}))\times\nonumber \\
 &  & \times\left(1+\delta_{t}\dot{\lambda}^{\mu}O_{\mu}(\bm{\lambda}(t_{k}))\right)\Omega(\bm{\lambda}(t_{k}))P_{1}+{\cal O}(\delta_{t}^{2}).
\end{eqnarray}
Hence we conclude from Eq.~\eqref{auxx} that $\bar{{\cal U}}_{B}$
is gauge covariant,
\begin{equation}
\bar{\mathcal{U}}_{B}\rightarrow\Omega^{\dagger}(\bm{\lambda}(T))\,\bar{\mathcal{U}}_{B}\,\Omega(\bm{\lambda}(0)),
\end{equation}
where it is implied that $\bar{\mathcal{U}}_{B}$ should only be applied
to states within block 1. A similar consideration proves the gauge
covariance of Eq.~\eqref{eq:corrected_evolution_first_order}.

\subsection{Polar decomposition and NAGD}

\label{sec2e}

We next discuss the physics of NAGD by considering the singular value
decomposition (SVD) and the matrix polar decomposition \citep{DLMF}
of the averaged Berry matrix $\bar{{\cal U}}_{B}$ in Eq.~\eqref{avU}.
We further explain the role of the Berry curvature ($F_{\mu\nu}$)
term in Eq.~\eqref{avU} and perform the polar decomposition of $\bar{\mathcal{U}}_{B}$
to leading order in the noise amplitudes $\sigma^{\mu\nu}$.

\subsubsection{General discussion}

We start with the SVD representation of the averaged Berry matrix,
$\bar{{\cal U}}_{B}=u\Lambda v^{\dagger}$, with the unitaries $u$
and $v$ and the diagonal matrix $\Lambda={\rm diag}(\Lambda_{1},\ldots,\Lambda_{N})$.
The real-valued parameters
\begin{equation}
\Lambda_{k}=e^{-\Gamma_{k}},\qquad k=1,\ldots,N,
\end{equation}
encode the real-valued geometric dephasing ``eigenrates'' $\Gamma_{k}$
for the $N$-fold degenerate subspace of the open system. One may
view this result as follows. There are two orthonormal bases, $u$
and $v$, encoded by the columns of the respective unitaries. The
action of $\bar{{\cal U}}_{B}$ takes a basis vector from $v$ and
converts it to the respective vector in the $u$ basis, while changing
its norm (which here represents coherence) by a factor of $\Lambda_{k}$.
Note that one has to be careful with this interpretation since only
density matrices (and not pure states) should be averaged over fluctuations,
cf.~Sec.~\ref{sec4f}. However, as we show in Sec.~\ref{sec4},
many experimentally observable quantities of interest for NAGD detection
can be expressed through the matrix $\bar{{\cal U}}_{B}$ which acts
on the initial state. Importantly, we will see below that in contrast
to dynamic dephasing, where one always has $\Gamma_{\mathrm{dyn}}T>0$,
physical settings with negative geometric dephasing rates, $\Gamma_{k}<0$,
are easily realizable. Thus, the action of $\Lambda$ may suppress
or \emph{enhance} state weights. Of course, the coherence represented
by the norm cannot exceed unity which is ensured by the factor $e^{-\Gamma_{{\rm dyn}}T}$
in Eq.~(\ref{eq:NAGP_averaged}). In particular, $\Gamma_{\mathrm{dyn}}T\gg\abs{\Gamma_{k}}$
holds for long time duration $T$ of the adiabatic evolution protocol.

It is also useful to perform the polar decomposition \citep{DLMF}
$\bar{{\cal U}}_{B}=VR$, where the unitary matrix $V$ and the Hermitian
positive-semidefinite matrix $R$ are uniquely defined. Comparing
with the SVD representation, we see that the matrices $V$ and $R$
are given by
\begin{equation}
\bar{{\cal U}}_{B}=VR,\quad V=uv^{\dagger},\quad R=v\Lambda v^{\dagger}=R^{\dagger}.\label{polardecomp}
\end{equation}
Thus the average Berry matrix acts as the composition of a Hermitian
matrix $R$ --- which encodes geometric dephasing in the particular
``dephasing eigenbasis'' defined by $v$ --- followed by a unitary
rotation, $V=uv^{\dagger}$.

When reversing the orientation sense of the protocol, the averaged
Berry matrix is replaced by its inverse, $\bar{{\cal U}}_{B}\rightarrow\bar{{\cal U}}_{B}^{-1}$,
see Sec.~\ref{sec2c}. The polar decomposition is now given by
\begin{equation}
\bar{{\cal U}}_{B}^{-1}=R^{-1}V^{\dagger}=\tilde{V}\tilde{R},\quad\tilde{V}=V^{\dagger},\quad\tilde{R}=u\Lambda^{-1}u^{\dagger},\label{polarback}
\end{equation}
where the unitary rotation is simply given by $\tilde{V}=V^{\dagger}$.
The Hermitian matrix $\tilde{R}$ encodes NAGD for the reversed protocol
and contains the diagonal matrix $\Lambda^{-1}$. One can therefore
change the sign of all $\Gamma_{k}$ simultaneously by changing the
orientation sense of the base loop path. However, $\Lambda^{-1}$
now acts \emph{in a different basis}, namely the one corresponding
to the unitary $u$, as compared to the original protocol, see Eq.~\eqref{polardecomp}.

As discussed in Sec.~\ref{sec1}, it is useful to distinguish ``poor
man's'' from ``truly'' non-Abelian systems. The poor man's case
refers to degenerate systems where Berry matrices $\mathcal{U}_{B}$
for different trajectory realizations are mutually commuting, and
hence $[V,R]=0$. We discuss such systems in Secs.~\ref{sec4a} and
\ref{sec5d}. By contrast, for a truly non-Abelian system, one has
$[V,R]\ne0$. This distinction has important consequences for the
physical manifestations of NAGD. We first note that the unitary Berry
matrix $\mathcal{U}_{B}$ for a given trajectory $\bm{\lambda}(t)$
can always be diagonalized in an orthonormal basis, with the eigenvalues
being phase factors. A different trajectory $\bm{\lambda}'(t)$ will
correspond to a different Berry matrix $\mathcal{U}_{B}'$. For the
poor man's case, we have $\left[\mathcal{U}_{B},\mathcal{U}_{B}'\right]=0$
for arbitrary ${\bm{\lambda}}$ and ${\bm{\lambda}}'$. This implies
that there is a common basis diagonalizing the Berry matrices for
all fluctuating trajectory realizations, and the environmental average
thus only affects the eigenvalues but not the eigenbasis. With a unitary
matrix $w$ encoding this fixed eigenbasis, we arrive at the averaged
Berry matrix $\bar{{\cal U}}_{B}^{\text{pm}}=w\Phi\Lambda w^{\dagger}$
for a poor man's non-Abelian system, where the diagonal matrices $\Phi$
and $\Lambda$ contain the average geometric phases and the geometric
dephasing eigenrates, respectively. The polar decomposition (\ref{polardecomp})
is then given by $V=w\Phi w^{\dagger}$ and $R=w\Lambda w^{\dagger}$,
and we indeed have $\left[V,R\right]=0$. For truly non-Abelian systems,
on the other hand, the eigenbasis of $\mathcal{U}_{B}$ also fluctuates.
One then generically finds $\left[V,R\right]\neq0$. This distinction
has experimentally observable consequences, see Secs.~\ref{sec4d}
and \ref{sec5e}.

\subsubsection{Role of Berry connection and curvature}

With the infinitesimal time step $\delta_{t}$, let us consider the
contribution $\exp\left[\delta_{t}\dot{\lambda}^{\mu}(t_{k})\left(-A_{\mu}+i\sigma^{\nu\rho}F_{\nu\mu}\partial_{\rho}E_{1}\right)\right]$
in Eq.~(\ref{eq:UB_discretization}). Up to corrections of $\mathcal{O}(\delta_{t}^{2})$,
this term can be written as the product $V_{k}R_{k}$ with
\begin{equation}
V_{k}=\exp\left(-\delta_{t}\dot{\lambda}^{\mu}(t_{k})A_{\mu}(\bm{\lambda}(t_{k}))\right)\label{uni1}
\end{equation}
and
\begin{equation}
R_{k}=\exp\left(\delta_{t}\dot{\lambda}^{\mu}(t_{k})i\sigma^{\nu\rho}F_{\nu\mu}(\bm{\lambda}(t_{k}))\partial_{\rho}E_{1}(\bm{\lambda}(t_{k}))\right).\label{herm1}
\end{equation}
The properties $A_{\mu}^{\dagger}=-A_{\mu}$ and $F_{\mu\nu}^{\dagger}=-F_{\mu\nu}$
imply that $V_{k}$ is unitary and $R_{k}$ is Hermitian positive-definite,
thus implementing the polar decomposition \eqref{polardecomp} for
the infinitesimal contribution. We observe that geometric dephasing
involves only the Berry curvature, see Eq.~\eqref{herm1}, while
the unitary part is solely due to the Berry connection, see Eq.~\eqref{uni1}.
This statement also applies to the full averaged Berry matrix for
Abelian and poor man's non-Abelian systems, where a common eigenbasis
exists. However, it is generally not valid for truly non-Abelian systems
as we show next.

Using Eq.~(\ref{eq:UB_discretization}), the averaged Berry matrix
has the discretized form
\begin{eqnarray}
\bar{{\cal U}}_{B} & = & \prod_{k=0}^{T/\delta_{t}}V_{k}R_{k}=\mathcal{U}_{B}\prod_{k=0}^{T/\delta_{t}}R_{k}'=\mathcal{U}_{B}\times\label{eq:UB_averaged_a-la_Stokes}\\
 & \times & \mathcal{T}\exp\left(i\sigma^{\rho\nu}\int_{0}^{T}dt\ \dot{\lambda}^{\mu}(t)\mathcal{V}^{\dagger}(t)F_{\nu\mu}(t)\mathcal{V}(t)\partial_{\rho}E_{1}(t)\right),\nonumber
\end{eqnarray}
where we define $R_{k}'=\mathcal{V}^{\dagger}(t_{k})R_{k}\mathcal{V}(t_{k})$
with
\begin{equation}
\mathcal{V}(t)=\prod_{k=0}^{t/\delta_{t}}V_{k}=\mathcal{T}\exp\left(-\int_{0}^{t}dt'\ \dot{\lambda}^{\mu}(t')A_{\mu}(t')\right).\label{eq:V(t)}
\end{equation}
Note that $\mathcal{V}(T)=\mathcal{U}_{B}$ is the unperturbed Berry
matrix. We first observe that the operators $R_{k}'$ involve not
only the Berry curvature but also the Berry connection unless $\left[F_{\mu\nu}(t),A_{\rho}(t')\right]=0$
for all times $t$ and $t'$. Second, even though each $R_{k}'$ operator
is Hermitian, their product in Eq.~\eqref{eq:UB_averaged_a-la_Stokes}
is generally non-Hermitian since different $R_{k}'$ may not commute.
We conclude that, in general, the Berry connection also contributes
to NAGD, and similarly the Berry curvature contributes to the unitary
part.

\subsubsection{Weak system-bath coupling}

\label{sec2e3}

We now show that for very weak system-bath coupling amplitude $\sigma$,
the polar decomposition of $\bar{{\cal U}}_{B}$ can be performed
analytically. To that end, we expand Eq.~(\ref{eq:UB_averaged_a-la_Stokes})
in powers of $\mathfrak{s}=\sigma/{\cal E}$, see Eq.~\eqref{expansion_parameters},
\begin{eqnarray}
\bar{{\cal U}}_{B} & = & \mathcal{U}_{B}\Biggl(\mathtt{1}+i\sigma^{\rho\nu}\int_{0}^{T}dt\ \dot{\lambda}^{\mu}(t)\times\label{eq:TexpF_linear_in_fluctuations}\\
 & \times & \mathcal{V}^{\dagger}(t)F_{\nu\mu}(t)\mathcal{V}(t)\partial_{\rho}E_{1}(t)+{\cal O}(\mathfrak{s}^{2})\Biggr).\nonumber
\end{eqnarray}
To leading order in $\mathfrak{s},$ this implements the polar decomposition,
$\bar{\mathcal{U}}_{B}=VR$, with
\begin{equation}
V=\mathcal{U}_{B}+{\cal O}(\mathfrak{s}^{2}),\label{eq:polar_approx_V}
\end{equation}
and the Hermitian operator
\begin{equation}
R=\mathtt{1}+i\sigma^{\rho\nu}\int_{0}^{T}dt\ \dot{\lambda}^{\mu}(t)\mathcal{V}^{\dagger}(t)F_{\nu\mu}(t)\mathcal{V}(t)\partial_{\rho}E_{1}(t)+{\cal O}(\mathfrak{s}^{2}).\label{eq:eq:polar_approx_R}
\end{equation}
The fact that $V$ does not receive corrections $\sim\sigma$ plays
a crucial role in one of our protocols for NAGD detection in Sec.~\ref{sec4}.

\section{Topological protection vs geometric dephasing}

\label{sec3}

We next discuss a non-Abelian generalization of the Stokes theorem,
see Sec.~\ref{sec3a}, and summarize its relation to the results
of Sec.~\ref{sec2}. In Sec.~\ref{sec3b}, we then discuss the connection
between NAGD and the topologically protected braiding of anyonic quasiparticles.

\subsection{Non-Abelian Stokes theorem}

\label{sec3a}

In the Abelian case, the Stokes theorem expresses the Berry phase
as a surface integral over the Berry curvature,
\begin{eqnarray}
 &  & \oint_{\partial S}dt\,\dot{\lambda}^{\mu}(t)A_{\mu}^{(\mathrm{abel})}({\bm{\lambda}}(t))=\label{AbelianBerry}\\
 &  & =\iint_{S}dsdt\ \partial_{s}\lambda^{\mu}(s,t)\partial_{t}\lambda^{\nu}(s,t)F_{\mu\nu}^{(\mathrm{abel})}({\bm{\lambda}}(s,t)),\nonumber
\end{eqnarray}
where $F_{\mu\nu}^{(\mathrm{abel})}$ follows from Eq.~\eqref{eq:F_munu_def}
with $A_{\mu}\rightarrow A_{\mu}^{({\rm abel)}}$, ${\bm{\lambda}}(t)$
parametrizes the closed path $\partial S$ in parameter space, and
$\partial S$ is the boundary of a simply connected surface $S$ which
in turn is parametrized by ${\bm{\lambda}}(s,t)$. In the non-Abelian
case, the Stokes theorem, and hence Eq.~\eqref{AbelianBerry}, does
not apply anymore since we are dealing with path-ordered exponentials.
Nonetheless, considerable progress has been made towards generalizing
Eq.~\eqref{AbelianBerry} to the non-Abelian case. Since the corresponding
works are perhaps not widely known, we briefly summarize them below.
In particular, a global ``non-Abelian Stokes theorem'' can be formulated
as \citep{Schlesinger1928,Klimo1973,Arefeva1980,Bralic1980,Smolin1989,Simonov1989,Shevchenko1998,Hirayama1998,Hirayama1998a,Karp2000,Hirayama2000}
\begin{widetext}
\begin{eqnarray}
 &  & \mathcal{P}\exp\left(-\oint_{\partial S}d\lambda^{\mu}A_{\mu}\right)=\mathcal{T}_{t}\exp\left(-\int_{t_{1}}^{t_{2}}dt\int_{s_{1}}^{s_{2}}ds\ \partial_{s}\lambda^{\mu}(s,t)\partial_{t}\lambda^{\nu}(s,t)\mathcal{V}^{\dagger}(s,t)F_{\mu\nu}\left({\bm{\lambda}}(s,t)\right)\mathcal{V}(s,t)\right),\label{eq:NA_Stokes_theorem}\\
 &  & \mathcal{V}(s,t)=\mathcal{T}_{\sigma}\exp\left(-\int_{s_{1}}^{s}d\sigma\ \partial_{\sigma}\lambda^{\mu}(\sigma,t)A_{\mu}\left({\bm{\lambda}}(\sigma,t)\right)\right)\mathcal{T}_{\tau}\exp\left(-\int_{t_{1}}^{t}d\tau\ \partial_{\tau}\lambda^{\mu}(s_{1},\tau)A_{\mu}\left({\bm{\lambda}}(s_{1},\tau)\right)\right),\nonumber
\end{eqnarray}
\end{widetext}
where $\{(t,s)|t_{1}\leq t\leq t_{2},s_{1}\leq s\leq s_{2}\}$ parametrizes
a simply connected surface $S$ with boundary $\partial S$ and the
operator $\mathcal{T}_{t}$ denotes time ordering with respect to
only $t$ (but not $s$). While Eq.~(\ref{eq:NA_Stokes_theorem})
certainly looks complicated enough, we emphasize that it is a nontrivial
simplification that time ordering is needed for a single parameter
only. To our knowledge, the first proof of Eq.~\eqref{eq:NA_Stokes_theorem}
has been given in the context of general relativity in 1928 \citep{Schlesinger1928}.
For later derivations, see Refs.~\citep{Klimo1973,Bralic1980,Arefeva1980,Smolin1989,Simonov1989,Shevchenko1998,Hirayama1998,Hirayama1998a,Karp2000,Hirayama2000}.
The essential idea of the proof is to split the surface area $S$
into infinitesimal plaquettes, where the original contour integral
over $\partial S$ is represented as an integral that starts at the
initial point of the contour, goes to a plaquette, encircles it, returns
to the initial point, then moves on to the next plaquette, encircles
it, returns to the initial point, and so on. The equivalence of the
two contours can be traced back to a cancellation between the ``return''
path from one plaquette and the ``forward'' path to the next plaquette
\citep{Arefeva1980}. Encircling a plaquette then gives an exponential
of $F_{\mu\nu}({\bm{\lambda}}(s,t))$, while the contribution of the
forward (return) path is expressed by $\mathcal{V}(s,t)$ ($\mathcal{V}^{\dagger}(s,t)$).
These operators essentially parallel transport $F_{\mu\nu}$ to the
initial point of the contour. We also note that our Berry matrix expression
for a perturbed path in Eq.~\eqref{eq:corrected_evolution_first_order}
appears natural in view of the non-Abelian Stokes theorem.

The result in Eq.~\eqref{eq:NA_Stokes_theorem} has several interesting
implications. First, in the Abelian case, the Berry curvature fully
determines the Berry phase, cf.~Eq.~\eqref{AbelianBerry}. However,
in the non-Abelian case, different Berry connections $A_{\mu}$ and
$A_{\mu}'$ may yield the same Berry curvature, $F_{\mu\nu}=F_{\mu\nu}'$,
yet give different Berry matrices due to the contribution of $\mathcal{V}(s,t)$.
For an explicit example, see Ref.~\citep{Zee1988}. Second, notwithstanding
this point, Eq.~\eqref{eq:NA_Stokes_theorem} shows that if the Berry
curvature vanishes throughout some simply connected region in parameter
space, $F_{\mu\nu}({\bm{\lambda}})=0$, the non-Abelian Berry phase
will vanish for any loop path within that region. This in turn implies
that the Berry connection is a pure gauge in that region, i.e., some
$\Omega({\bm{\lambda}})$ exists such that $A_{\mu}=\Omega^{\dagger}\partial_{\mu}\Omega$.

\subsection{Topological protection vs NAGD}

\label{sec3b}

\begin{figure}
\begin{centering}
\includegraphics[width=1\columnwidth]{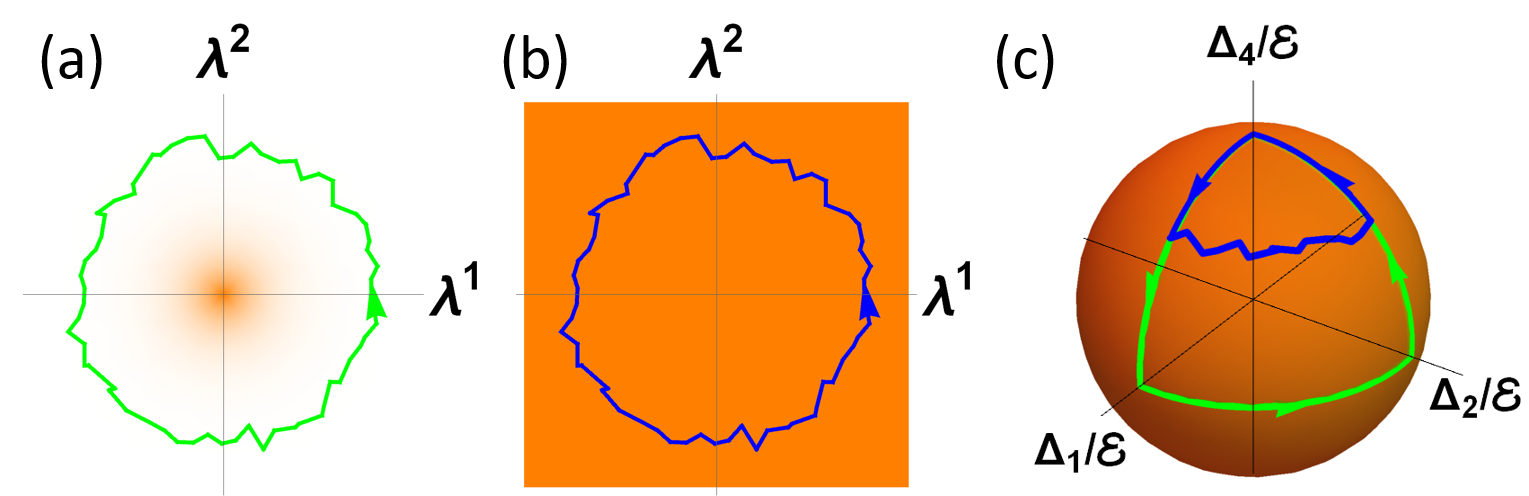}
\par\end{centering}
\caption{\protect Different mechanisms of topological protection. \textbf{(a):}
The Berry curvature (orange) is localized at isolated points in parameter
space. Fluctuations of the parameter trajectory (green) do not change
the Berry matrix as long as the trajectory stays in the region of
vanishing Berry curvature. This scenario is typical for the topological
protection of anyon braiding. \textbf{(b):} Same as before but the
Berry curvature remains finite in the relevant parameter space. The
fluctuating trajectory (blue) feels the Berry curvature and thus
the Berry matrix is sensitive to  fluctuations.
Hence there is no topological protection in this case. \textbf{(c):}
Also for this example --- corresponding to Majorana braiding setups
with control parameters $\Delta_{j}/{\cal E}$, see Sec.~\ref{sec5}
--- the Berry curvature does not vanish. However, for system-specific
reasons, it may happen that physically realizable fluctuations cannot deform certain trajectories
 (green). This results
in another topological protection mechanism. However,
more general trajectories
do exhibit fluctuations (blue), and the corresponding
Berry matrices are then unprotected. As explained in the main text,
topologically protected systems do not exhibit NAGD.}
\label{fig1}
\end{figure}

Topologically ordered phases of matter \citep{Wen2017} represent
an active topic of research in condensed matter physics. They support
anyonic excitations which could be used for quantum information processing
purposes \citep{Nayak2008}. Non-Abelian anyons give rise to a degenerate
subspace. In particular, consider an adiabatic evolution where one
moves an anyon along a real-space trajectory encircling other anyons
and returning back to the original position. This process results
in a non-trivial unitary transformation in the degenerate subspace.
One finds non-Abelian statistics due to the non-Abelian Berry connection
obtained by treating the anyon real-space coordinates as the parameters
${\bm{\lambda}}(t)$ \citep{Nayak2008}. Importantly, the resulting
Berry matrix does not depend on details of the trajectory as long
as the anyon stays away from all other anyons, see~Fig.~\ref{fig1}
for a schematic illustration. In fact, one may view this feature as
the defining property of the topological protection of braiding. Considering
small perturbations of the trajectory ${\bm{\lambda}}(t)$, we next
observe from Eq.~\eqref{eq:corrected_evolution_first_order} that
the Berry matrix will not depend on such perturbations if the Berry
curvature vanishes, $F_{\mu\nu}({\bm{\lambda}})=0$, for all ${\bm{\lambda}}$
configurations where the anyon remains far away from other anyons.
The non-Abelian Stokes theorem \eqref{eq:NA_Stokes_theorem} shows
that this condition is both necessary and sufficient. One sees, therefore,
that systems exhibiting topological protection will not show NAGD
since $F_{\mu\nu}=0$, cf.~Eq.~\eqref{avU}.

It is interesting to note that there is also another kind of topological
protection. In the standard three-star Majorana setup \citep{Alicea2012,Sau2011,VanHeck2012,Karzig2016,Rahmani2017,Cheng2011,Knapp2016}
that supports localized Majorana bound states, a unitary braiding
operation in the degenerate subspace can be performed by manipulating
the tunneling amplitudes $\Delta_{j}(t)$ between different Majorana
states according to a specific protocol, see Sec.~\ref{sec5d} for
details. (Here the parameters $\Delta_{j}/{\cal E}$ with ${\cal E}$
in Eq.~\eqref{Majenerg} correspond to the control parameters ${\bm{\lambda}}$.)
This operation generates the same unitary braiding transformation
as the one obtained by exchanging (in real space) two Majorana-hosting
vortices in a $p$-wave superconductor \citep{Nayak2008}, and it
is also insensitive to fluctuations $\delta\lambda^{\mu}(t)$ ---
albeit for a different reason than above. In fact, as discussed in
detail in Sec.~\ref{sec5}, the Berry curvature generally does not
vanish in topologically protected Majorana setups of the type shown
in Fig.~\ref{fig3} below. However, physically realizable fluctuations
are not able to change the geometric shape of the parameter trajectory,
see Fig.~\ref{fig1}(c). Alternatively, one can say that the realizable
fluctuations are such that no contribution to the unitary braiding
transformation arises, i.e., $\delta\lambda^{\nu}F_{\mu\nu}=0$ in
Eq.~\eqref{eq:corrected_evolution_first_order}. Therefore, also
this mechanism of topological protection is incompatible with NAGD.
In Sec.~\ref{sec5}, by considering an extended five-star Majorana
setup, we show how this type of topological protection emerges in
our formalism. In addition, we design protocols that break the protection
at will, and thus can provide clear NAGD signatures.

\section{Detection of non-Abelian geometric dephasing}

\label{sec4}

We now introduce protocols that could allow experimentalists to observe
NAGD. In Sec.~\ref{sec4a}, we briefly discuss how the matrix elements
of the averaged Berry matrix $\bar{{\cal U}}_{B}$ in Eq.~\eqref{avU}
may be inferred by averaging the interference signal obtained from
a Mach-Zehnder interferometer, see also Ref.~\citep{ourprl}. Since
for most non-Abelian systems of present interest, e.g., for the Majorana
setups in Sec.~\ref{sec5}, such schemes are very difficult or even
impossible to realize, we continue in Sec.~\ref{sec4b} with the
simplest type of one-block interference protocol where interference
between different components of the final state is probed. Unfortunately,
this protocol does not provide access to $\bar{{\cal U}}_{B}$. However,
we show in Sec.~\ref{sec4c} that a more general two-block interference
protocol can be used for NAGD detection. A particularly convenient
spin-echo type variant of this protocol is presented in Sec.~\ref{sec4d}.
In Sec.~\ref{sec4e}, we comment on the absence of relations between
the Berry connections for different blocks, and in Sec.~\ref{sec4f}
we close this section with a discussion of how to average the density
matrix of the system for the general multi-block case.

\subsection{Interference between initial and final state}

\label{sec4a}

\begin{figure}
\centering \includegraphics[width=0.95\columnwidth]{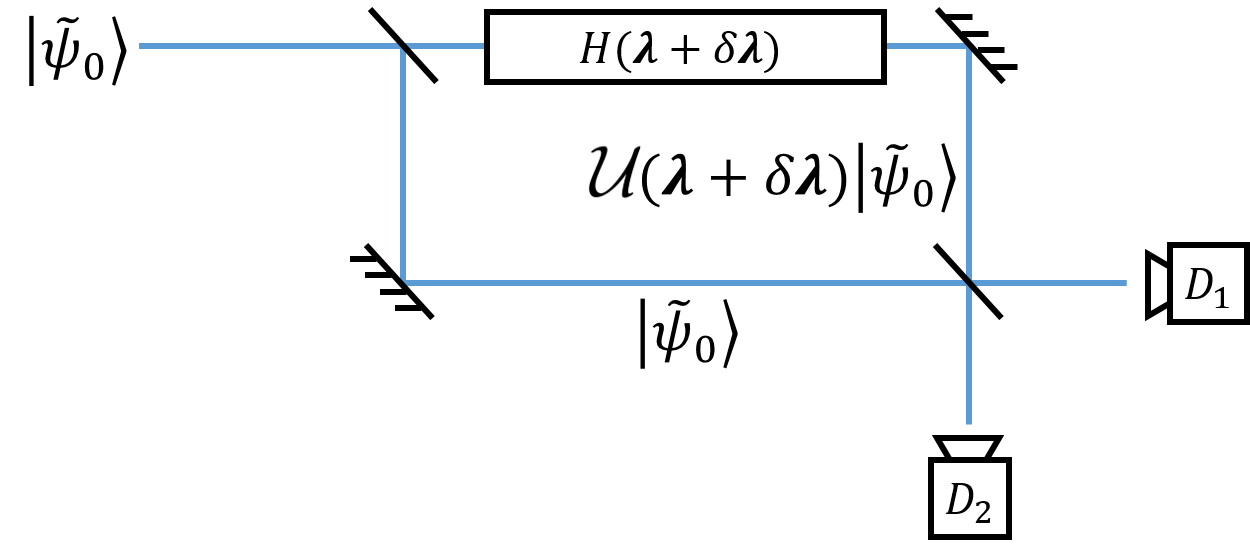}
\caption{\label{fig2} An interferometric setup for observing NAGD. For details, see main
text.}
\end{figure}

The conceptually simplest setup for probing the averaged Berry matrix
$\bar{{\cal U}}_{B}$ employs a Mach-Zehnder interferometer where
the system is split into a superposition state, cf.~Fig.~\ref{fig2}.
Starting from the initial state $|\psi_{{\rm in}}\rangle$, the state
evolves in one arm of the interferometer in the presence of the fluctuating
control parameters such that the final state $|\psi_{{\rm f}}\rangle={\cal U}_{B}|\psi_{{\rm in}}\rangle$
is generated for a given trajectory realization. The setup is prepared
such that these perturbations are absent in the other arm, where the
final state is simply $|\psi_{{\rm in}}\rangle$ up to a trivial phase
factor. The Mach-Zehnder interference signal thus effectively probes
the interference between $|\psi_{{\rm f}}\rangle$ and $|\psi_{{\rm in}}\rangle.$
By averaging the result over fluctuations, i.e., by repeating the
experiment many times, one gains access to the averaged Berry matrix
$\bar{{\cal U}}_{B}$. In order to extract the full matrix structure,
one has to repeat this experiment for different initial states.

To illustrate this idea, let us consider a particle with an internal
degree of freedom, say, a spin of size $S\geq1$ such that there is
at least one degenerate subspace while the system can have a non-trivial
Hamiltonian. The particle is initially prepared in the state $|\psi(0)\rangle=|\psi_{{\rm in}}\rangle=U_{0}|\tilde{\psi}_{0}\rangle$
within block 1, i.e., $P_{1}\ket{\tilde{\psi}_{0}}=\ket{\tilde{\psi}_{0}}$,
where we define
\begin{equation}
U_{0}=U({\bm{\lambda}}(0))=U({\bm{\lambda}}(T)).\label{u0def}
\end{equation}
The state $|\psi_{{\rm in}}\rangle$ should be identically prepared
for every trajectory realization such that one can perform the average
over noise fluctuations (we recall that $\delta{\bm{\lambda}}(0)=\delta{\bm{\lambda}}(T)=0$)
starting from precisely the same initial state. In one arm of the
interferometer, the spin dynamics evolves according to the Hamiltonian
$H({\bm{\lambda}}(t)+\delta{\bm{\lambda}}(t))$, arriving at the final
state $|\psi_{{\rm f}}\rangle=\ket{\psi(T)}=U_{0}\,\mathcal{U}\,U_{0}^{\dagger}\ket{\psi_{{\rm in}}}$,
with $\mathcal{U}$ in Eq.~(\ref{Uperturbed}). In the other arm,
no perturbations are present, $H=0$, and the state does not change.
The probabilities for the particle to appear at the respective detector
exit of the Mach-Zehnder interferometer are given by
\begin{equation}
P_{1,2}=\frac{1}{2}\left(1\pm\mathrm{Re}\langle\tilde{\psi}_{0}|\mathcal{U}|\tilde{\psi}_{0}\rangle\right).
\end{equation}
Averaging over the fluctuations $\delta\bm{\lambda}$, i.e., over
many experimental runs, one obtains the averaged probabilities
\begin{equation}
\bar{P}_{1,2}=\frac{1}{2}\left(1\pm\mathrm{Re}\langle\tilde{\psi}_{0}|\bar{\mathcal{U}}|\tilde{\psi}_{0}\rangle\right),
\end{equation}
with $\bar{\mathcal{U}}$ in Eqs.~(\ref{eq:NAGP_averaged}) and (\ref{avU}).
Repeating the experiment for different initial states, one can recover
the full matrix structure of $\bar{\mathcal{U}}$, and thus of the
averaged Berry matrix $\bar{{\cal U}}_{B}$.

As natural candidate for such an experiment, let us consider a nucleus
with half-integer spin $S=l-\frac{1}{2}\geq\frac{3}{2}$ subject to
nuclear quadrupole resonance. The effective Hamiltonian is given by
\citep{Zee1988}
\begin{equation}
H=(\bm{S}\cdot\bm{B})^{2}=B^{2}e^{-i\varphi S_{z}}e^{-i\theta S_{y}}S_{z}^{2}e^{i\theta S_{y}}e^{i\varphi S_{z}},\label{eq:H_quadrupole_resonance}
\end{equation}
with a ``magnetic'' field $\bm{B}=B(\sin\theta\cos\varphi,\sin\theta\sin\varphi,\cos\theta)$.
The system has $l$ two-fold degenerate blocks spanned by the eigenstates
of $\bm{S}\cdot\bm{B}$ with eigenvalues $\pm mB$ for half-integer
$m=\frac{1}{2},\ldots,S$. Defining the control parameters $\lambda^{\mu}$
as $\theta$ and $\varphi$, the Berry connection and curvature components
in a block with $m>1/2$ are given by \citep{Zee1988}
\begin{eqnarray}
A_{\theta}^{(m>1/2)} & = & 0,\quad A_{\varphi}^{(m>1/2)}=-im\sigma_{z}^{(m)}\cos\theta,\nonumber \\
F_{\theta\varphi}^{(m>1/2)} & = & im\sigma_{z}^{(m)}\sin\theta,
\end{eqnarray}
while for $m=1/2$, one finds
\begin{eqnarray}
A_{\theta}^{(m=1/2)} & = & -\frac{i}{2}l\sigma_{y}^{(m)},\nonumber \\
A_{\varphi}^{(m=1/2)} & = & -\frac{i}{2}\left(\sigma_{z}^{(m)}\cos\theta-l\sigma_{x}^{(m)}\sin\theta\right),\nonumber \\
F_{\theta\varphi}^{(m=1/2)} & = & -\frac{i}{2}(l^{2}-1)\sigma_{z}^{(m)}\sin\theta.\label{eq:F_NA_quadrupole_resonance}
\end{eqnarray}
Here the $\sigma_{j}^{(m)}$ are Pauli matrices acting in the respective
degenerate subspace. Note that for $m>1/2$, the Berry connection
and curvature are expressed via $\sigma_{z}^{(m)}$ only, and thus
are always diagonal in the eigenbasis of $\sigma_{z}^{(m)}$. Blocks
with $m>1/2$ therefore realize the poor man's non-Abelian case. By
contrast, for $m=1/2$, different components of the Berry connection
and curvature do not commute, and we encounter a truly non-Abelian
system. In either case, fluctuations of $\bm{B}$ will produce NAGD,
see Eqs.~\eqref{polardecomp}, \eqref{polarback}, and \eqref{avU}.
We note that the Berry connection has been probed by nuclear quadrupole
resonance experiments for $S=3/2$ nuclei \citep{Tycko1987,Zwanziger1990}.
However, macroscopic samples such as those used in Refs.~\citep{Tycko1987,Zwanziger1990}
are not suitable for the interference experiment outlined above.

\subsection{One-block interference protocol}

\label{sec4b}

The interferometer discussed in Sec.~\ref{sec4a} may be difficult
to implement in practice for many systems of present interest. In
particular, setups with superconducting qubits involving Abelian \citep{Leek2007,Berger2015}
or non-Abelian \citep{AbdumalikovJr2013} Berry phases, and condensed-matter
systems characterized by a non-Abelian Berry connection \citep{Hasan2010,Yang2014,Murakami2003},
do not allow one to prepare spatial superposition states. We thus
turn to alternative protocols for NAGD detection.

In fact, one could measure interference signals between different
components of the final state, and subsequently average the result
over the fluctuating paths. In practice, this strategy amounts to
measuring suitable Hermitian operators, $\tilde{M}^{(1)}$, which
are for now assumed to act only within the $N$-fold degenerate block
1,
\begin{equation}
\tilde{M}^{(1)}=U_{0}P_{1}U_{0}^{\dagger}M^{(1)}U_{0}P_{1}U_{0}^{\dagger},
\end{equation}
with $U_{0}$ in Eq.~\eqref{u0def}. After performing the average
over fluctuations, we obtain the final-state expectation value
\begin{equation}
\left\langle \langle\psi(T)|\tilde{M}^{(1)}|\psi(T)\rangle\right\rangle _{{\rm env}}=\left\langle \psi(0)\right|\bar{M}^{(1)}\left|\psi(0)\right\rangle ,\label{eq:avgM1}
\end{equation}
where the averaged operator $\bar{M}^{(1)}$ is given by
\begin{equation}
\bar{M}^{(1)}=\left\langle U_{0}\,\mathcal{U}^{\dagger}U_{0}^{\dagger}M^{(1)}U_{0}\,\mathcal{U}U_{0}^{\dagger}\right\rangle _{{\rm env}}.\label{avgM1}
\end{equation}
The evolution operator ${\cal U}$ for a fluctuating path has been
specified in Eq.~\eqref{Uperturbed}. Importantly, after averaging
over the bath-induced fluctuations, the expectation value of $\tilde{M}^{(1)}$
in the final state is expressed via Eq.~\eqref{eq:avgM1} as an expectation
value of the averaged operator $\bar{M}^{(1)}$ in Eq.~\eqref{avgM1}
with respect to the \emph{known} initial state. Here $\bar{M}^{(1)}$
does not depend on the initial state. Performing experiments with
different initial states and for different operators $\tilde{M}^{(1)}$,
one can acquire information about the averaged products of arbitrary
matrix elements, $\left\langle \mathcal{U}_{ij}^{*}\mathcal{U}_{kl}\right\rangle _{{\rm env}}$.
These averages differ from products of matrix elements of $\bar{\mathcal{U}}=\left\langle \mathcal{U}\right\rangle _{{\rm env}}$
in Eq.~\eqref{eq:NAGP_averaged}. Nevertheless, in the adiabatic
limit, by following a similar approach as in Sec.~\ref{sec2c}, one
can calculate $\bar{M}^{(1)}$.

To that end, let us first define a general time evolution operator,
\begin{equation}
\mathcal{U}(t,T)=e^{-i\int_{t}^{T}d\tau E_{1}\left({\bm{\lambda}}(\tau)+\delta{\bm{\lambda}}(\tau)\right)}\mathcal{T}e^{\int_{t}^{T}d\tau\dot{\lambda}^{\mu}(-A_{\mu}+\delta\lambda^{\nu}F_{\mu\nu})}.\label{U1def}
\end{equation}
Evidently we have $\mathcal{U}=\mathcal{U}(0,T)$, see Eq.~\eqref{Uperturbed}.
Similarly, we define the averaged time-dependent operator
\begin{equation}
\bar{M}(t,T)=\left\langle U_{0}\,\mathcal{U}^{\dagger}(t,T)U_{0}^{\dagger}M^{(1)}U_{0}\,\mathcal{U}(t,T)U_{0}^{\dagger}\right\rangle _{{\rm env}},\label{M1def}
\end{equation}
where we have $\bar{M}^{(1)}=\bar{M}(0,T)$, see Eq.~\eqref{avgM1}.
We now observe that all dynamic contributions $\sim E_{1}({\bm{\lambda}}(t)+\delta{\bm{\lambda}}(t))$
cancel out in Eq.~\eqref{M1def},
\begin{eqnarray}
\bar{M}(t,T) & = & \Bigl\langle U_{0}\bar{\mathcal{T}}e^{-\int_{t}^{T}d\tau\dot{\lambda}^{\mu}(-A_{\mu}+\delta\lambda^{\nu}F_{\mu\nu})}U_{0}^{\dagger}M^{(1)}\nonumber \\
 & \times & U_{0}\mathcal{T}e^{\int_{t}^{T}d\tau\dot{\lambda}^{\mu}(-A_{\mu}+\delta\lambda^{\nu}F_{\mu\nu})}U_{0}^{\dagger}\Bigr\rangle_{{\rm env}},
\end{eqnarray}
where the operator $\bar{\mathcal{T}}$ denotes anti-time ordering.
Since NAGD is caused by the interplay of dynamic and geometric phase
fluctuations, we can already anticipate the absence of NAGD under
such a single-block interference protocol. This expectation is confirmed
by an explicit calculation as we show next.

To proceed, we first write
\begin{equation}
-\frac{d\bar{M}(t,T)}{dt}=\left[\bar{M}(t-\delta_{t},T)-\bar{M}(t,T)\right]/\delta_{t},
\end{equation}
with the infinitesimal time step $\delta_{t}\rightarrow0$. We then
have
\begin{eqnarray}
 &  & \bar{M}(t-\delta_{t},T)=\Bigr\langle e^{-\delta_{t}\dot{\lambda}^{\mu}[-A_{\mu}+\delta\lambda^{\nu}(t)F_{\mu\nu}]}\bar{M}(t,T)\nonumber \\
 &  & \qquad\qquad\qquad\times\ e^{\delta_{t}\dot{\lambda}^{\mu}[-A_{\mu}+\delta\lambda^{\nu}(t)F_{\mu\nu}]}\Bigr\rangle_{{\rm env}}\\
 &  & =\bar{M}(t,T)+\delta_{t}\dot{\lambda}^{\mu}(t)\left[A_{\mu},\bar{M}(t,T)\right]+\frac{\delta_{t}}{2}\sigma^{\mu\nu}\dot{\lambda}^{\rho}\dot{\lambda}^{\chi}\nonumber \\
 &  & \times\Bigl(\left\{ F_{\rho\mu}F_{\chi\nu},\bar{M}(t,T)\right\} -2F_{\rho\mu}\bar{M}(t,T)F_{\chi\nu}\Bigr)+{\cal O}\left(\delta_{t}^{3/2}\right).\nonumber
\end{eqnarray}
Therefore we arrive at
\begin{eqnarray}
 &  & -\frac{d\bar{M}(t,T)}{dt}=\dot{\lambda}^{\mu}(t)\left[A_{\mu},\bar{M}(t,T)\right]+\frac{1}{2}\sigma^{\mu\nu}\dot{\lambda}^{\rho}\dot{\lambda}^{\chi}\nonumber \\
 &  & \times\ \Bigl(\left\{ F_{\rho\mu}F_{\chi\nu},\bar{M}(t,T)\right\} -2F_{\rho\mu}\bar{M}(t,T)F_{\chi\nu}\Bigr),\label{MMdef}
\end{eqnarray}
where the last term has the form of a Lindbladian. However, this term
can be neglected because it scales $\sim\dot{\lambda}^{2}={\cal O}(T^{-2})$.
We conclude that, in the adiabatic limit, $\bar{M}(t,T)$ and thus
also $\bar{M}^{(1)}$ in Eq.~\eqref{avgM1} will not depend on the
$\sigma^{\mu\nu}$ noise amplitudes at all. Indeed, Eq.~\eqref{MMdef}
implies that
\begin{eqnarray}
\bar{M}^{(1)} & = & U_{0}\bar{{\cal T}}e^{\int_{0}^{T}dt\dot{\lambda}^{\mu}A_{\mu}}U_{0}^{\dagger}M^{(1)}U_{0}{\cal T}e^{-\int_{0}^{T}dt\dot{\lambda}^{\mu}A_{\mu}}U_{0}^{\dagger}\nonumber \\
 & = & U_{0}\,\mathcal{U}_{B}^{\dagger}U_{0}^{\dagger}M^{(1)}U_{0}\,\mathcal{U}_{B}U_{0}^{\dagger}.
\end{eqnarray}
This single-block interference scheme is therefore insensitive to
fluctuations in the adiabatic limit. Of course, this also implies
that one cannot detect NAGD in this way. Note that this result disagrees with the results of Refs.~\citep{Solinas2004,Fuentes-Guridi2005,Sarandy2006}
in which noise is found to affect single-block observables. The origin
of this discrepancy is twofold. Unlike us, Ref.~\citep{Solinas2004}
investigated non-adiabatic noise, while Refs.~\citep{Fuentes-Guridi2005,Sarandy2006}
investigated the effect of noise that lifts the degeneracy in the
block.

\subsection{Two-block interference protocol}

\label{sec4c}

\begin{figure}
\begin{centering}
\includegraphics[width=1\columnwidth]{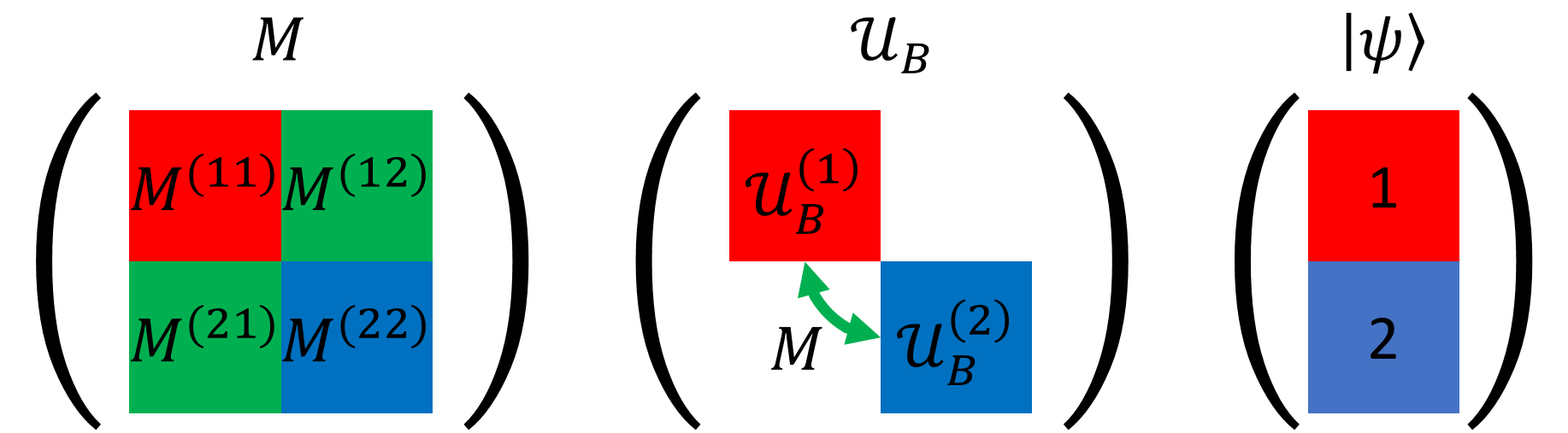}
\par\end{centering}
\caption{\protect The essence of two-block interference. The adiabatic evolution
changes the system state $\protect\ket{\psi}$ by the action of the
Berry matrix $\mathcal{U}_{B}$ which acts differently in the two
blocks (red, blue). The measured operator $M$ can have block-diagonal
and block-off-diagonal components. The block-diagonal components (red,
blue) are not sensitive to energy fluctuations and thus to NAGD. The
block-off-diagonal component (green) probes the interference between
the two blocks and is sensitive to NAGD.}
\label{fig3}
\end{figure}

The simplest possibility for an interference protocol of the type
considered in Sec.~\ref{sec4b} but with sensitivity to fluctuations
is to consider an initial state that also has weight in another subspace,
$j=2$. This block is $N_{2}$-fold degenerate and has the energy
$E_{2}({\bm{\lambda}})$, where $N_{2}=1$ is possible. Note that the protocol proposed here can be implemented in nuclear
quadrupole resonance experiments similar to those of Refs.~\citep{Tycko1987,Zwanziger1990}. We now show that a two-block interferometric scheme indeed provides
direct access to NAGD, see Fig.~\ref{fig3} for an illustration.

Choosing an initial state subject to the condition $(P_{1}+P_{2})|\tilde{\psi}(0)\rangle=|\tilde{\psi}(0)\rangle$,
one executes the noisy parameter loop protocol as described in Sec.~\ref{sec4b}.
At the final time, $t=T$, one measures the expectation value of an
arbitrary Hermitian operator, $\tilde{M}$, acting only on blocks
1 and 2,
\begin{gather}
\tilde{M}=\sum_{i,j=1,2}\tilde{M}^{(ij)},\quad[\tilde{M}^{(ij)}]^{\dagger}=\tilde{M}^{(ji)},\nonumber \\
\tilde{M}^{(ij)}=U_{0}P_{i}U_{0}^{\dagger}M^{(ij)}U_{0}P_{j}U_{0}^{\dagger}.\label{genM}
\end{gather}
For a given trajectory realization, we thereby obtain the expectation
value $\langle\psi(T)|\tilde{M}|\psi(T)\rangle$. By subsequently
averaging over the control parameter fluctuations, we arrive at a
similar expression as in Eq.~\eqref{eq:avgM1},
\begin{equation}
\bar{{\cal M}}=\Bigl\langle\langle\psi(T)|\tilde{M}|\psi(T)\rangle\Bigr\rangle_{{\rm env}}=\langle\psi(0)|\bar{M}|\psi(0)\rangle.\label{M2def}
\end{equation}
Below we find that the off-diagonal component, $\tilde{M}^{(21)}=[\tilde{M}^{(12)}]^{\dagger}$
in Eq.~\eqref{genM}, is responsible for noise-sensitive contributions
in the adiabatic limit. Hence this type of protocol can be used for
detecting NAGD. On the other hand, for the same reasons as discussed
for the single-block interferometry scheme in Sec.~\ref{sec4b},
all diagonal entries, $\tilde{M}^{(jj)}$, produce only expectation
values that are insensitive to fluctuations in the adiabatic limit.
For simplicity, we henceforth assume $\tilde{M}^{(jj)}=0$. Writing
$\bar{M}=\bar{M}^{(12)}+\bar{M}^{(21)}$, due to the Hermiticity condition
in Eq.~\eqref{genM}, it suffices to compute the time-dependent averaged
function
\begin{equation}
\bar{M}^{(21)}(t,T)=\left\langle U_{0}\mathcal{U}^{(2)\dagger}(t,T)U_{0}^{\dagger}M^{(21)}U_{0}\mathcal{U}^{(1)}(t,T)U_{0}^{\dagger}\right\rangle _{{\rm env}}\label{M21def}
\end{equation}
with $\bar{M}^{(21)}=\bar{M}^{(21)}(0,T)$ and $U_{0}$ in Eq.~\eqref{u0def}.
The adiabatic time evolution operator, $\mathcal{U}^{(j)}(t,T)$,
which has been given in Eq.~\eqref{U1def} for block 1, follows for
block $j$ as
\begin{eqnarray}
 &  & \mathcal{U}^{(j)}(t,T)=e^{-i\int_{t}^{T}d\tau E_{j}\left({\bm{\lambda}}(\tau)+\delta{\bm{\lambda}}(\tau)\right)}\times\label{Ujdef}\\
 &  & \quad\times\mathcal{T}\exp{\int_{t}^{T}d\tau\dot{\lambda}^{\mu}\left(-A_{\mu}^{(j)}+\delta\lambda^{\nu}F_{\mu\nu}^{(j)}\right)}.\nonumber
\end{eqnarray}
We denote the evolution operator for the full perturbed loop trajectory
as $\mathcal{U}^{(j)}=\mathcal{U}^{(j)}(0,T)$. The Berry connection
$A_{\mu}^{(j)}$ in block $j=1,2$ is here defined precisely as in
Eq.~\eqref{eq:A_mu_def} but with the replacement $P_{1}\rightarrow P_{j}$.
Likewise, the corresponding field strength tensor $F_{\mu\nu}^{(j)}$
follows from Eq.~\eqref{eq:F_munu_def} with the replacement $A_{\mu}\rightarrow A_{\mu}^{(j)}$.
In contrast to what we found in Sec.~\ref{sec4b}, the dynamic contributions
$\sim E_{1,2}$ do not cancel out anymore as we show next.

The evolution equation for $\bar{M}^{(21)}(t,T)$ in Eq.~\eqref{M21def}
can be derived along the same lines as in Sec.~\ref{sec4b}. Defining
the energy difference
\begin{equation}
\tilde{E}_{1}({\bm{\lambda}})=-\tilde{E}_{2}({\bm{\lambda}})=E_{1}-E_{2},\label{tildeE}
\end{equation}
we find
\begin{widetext}
\begin{eqnarray}
-\frac{d\bar{M}^{(21)}(t,T)}{dt} & = & \left(i\tilde{E}_{2}({\bm{\lambda}}(t))-\frac{1}{2}\sigma^{\mu\nu}\partial_{\mu}\tilde{E}_{2}\partial_{\nu}\tilde{E}_{2}\right)\bar{M}^{(21)}(t,T)+\dot{\lambda}^{\mu}(t)\left(A_{\mu}^{(2)}\bar{M}^{(21)}(t,T)-\bar{M}^{(21)}(t,T)A_{\mu}^{(1)}\right)\nonumber \\
 & + & i\sigma^{\mu\nu}\dot{\lambda}^{\rho}\partial_{\mu}\tilde{E}_{2}\left(F_{\nu\rho}^{(2)}\bar{M}^{(21)}(t,T)-\bar{M}^{(21)}(t,T)F_{\nu\rho}^{(1)}\right)+{\cal O}\left(T^{-2}\right).
\end{eqnarray}
\end{widetext}

In the adiabatic limit, we then obtain $\bar{M}^{(21)}$ as
\begin{equation}
\bar{M}^{(21)}=U_{0}\bar{{\cal U}}^{(2)\dagger}\,U_{0}^{\dagger}M^{(21)}U_{0}\,\bar{{\cal U}}^{(1)}U_{0}^{\dagger},\label{eq:Two-block-interference_averaged}
\end{equation}
with the averaged evolution operators
\begin{eqnarray}
\bar{{\cal U}}^{(j)} & = & e^{-i\int_{0}^{T}dt\left(E_{j}-\frac{i}{2}\sigma^{\mu\nu}\partial_{\mu}E_{j}\partial_{\nu}\tilde{E}_{j}\right)}\,\bar{{\cal U}}_{B}^{(j)},\label{eq:Two-block-averaging}\\
\bar{{\cal U}}_{B}^{(j)} & = & \mathcal{P}\exp\oint d\lambda^{\mu}\left(-A_{\mu}^{(j)}+i\sigma^{\nu\rho}F_{\nu\mu}^{(j)}\partial_{\rho}\tilde{E}_{j}\right).\nonumber
\end{eqnarray}
The averaged Berry matrices, $\bar{{\cal U}}_{B}^{(j)}$, are defined
almost identically as in Eq.~\eqref{avU} but with the Berry connection
and curvature for the respective block and by replacing $E_{1}$ with
$E_{j}$ or $\tilde{E}_{j}$. In such two-block interferometric measurements,
the averaged evolution operators $\bar{{\cal U}}^{(j)}$ appearing
in $\bar{M}^{(21)}$, see Eq.~\eqref{eq:Two-block-interference_averaged},
know about the presence of the other sector via the energy difference
$E_{1}-E_{2}$. Measuring the expectation value $\bar{{\cal M}}$
in Eq.~(\ref{M2def}) for various operators $\tilde{M}$ and different
initial states $\ket{\psi(0)}$, one can therefore map out the matrices
$\bar{{\cal U}}_{B}^{(j)}$ and thus probe NAGD. An exception to this
conclusion concerns the Abelian component $\det\left(\bar{{\cal U}}_{B}^{(j)}\right)$,
where only $\det\left(\bar{{\cal U}}_{B}^{(2)\dagger}\bar{{\cal U}}_{B}^{(1)}\right)$
is accessible within the above scheme.

One may wonder how many different operators $M$ one has to repeatedly measure in order to reconstruct the averaged Berry matrix using the
 above protocol.
Given the degeneracies $N_1$ and $N_2$ of the two blocks,
the operator $M^{(21)}$ has $N_1N_2$ independent entries. Taking into account that putting a real number into one of the entries and zero into all others,
one gets access to the real part of $\left\langle \tilde\psi_0\right| U_B^{(2) \dagger} \left| j^{(2)} \right\rangle\left\langle k^{(1)} \right| U_B^{(1)}\left| \tilde \psi_0 \right\rangle$,
while inserting a purely imaginary number, the imaginary part of the same matrix element can be accessed.
Therefore, in order to reconstruct the above matrix
for all $\{ k^{(1)}, j^{(2)} \}$, one needs to measure
$2N_1N_2$ different operators $M$.
A similar consideration shows that accessing all
$\left\langle m^{(2)}\right| U_B^{(2)\dagger} \left | j^{(2)}\right \rangle\left\langle k^{(1)} \right| U_B^{(1)} \left| n^{(1)} \right\rangle$
matrix elements requires one to measure the results obtained for $2N_1N_2$ different initial states $|\tilde \psi_0\rangle$ for each $M$ operator.
However, mapping $U_B^{(1)}$ and $U_B^{(2)}$ does not require measuring all products of these matrix elements. In fact, it is sufficient to measure the product for all $k^{(1)}$
and $n^{(1)}$ for some fixed $m^{(2)}$ and $j^{(2)}$ in order to infer $U_B^{(1)}$ up to a common prefactor.
This requires measuring $2N_1$ different operators $M$ and $2N_1$
different states $|\tilde \psi_0\rangle$.
We thus conclude that one has to (repeatedly) perform $4N_j^2$ different measurements in order to map out $U_B^{(j)}$.

\subsection{Spin echo protocols}

\label{sec4d}

The expression for $\bar{M}^{(21)}$ in Eq.~(\ref{eq:Two-block-interference_averaged})
contains a deterministic dynamic phase factor $e^{i\int_{0}^{T}dt(E_{2}-E_{1})}$.
In view of the condition $\abs{E_{2}-E_{1}}T\gg1$, see Eq.~\eqref{conditions},
it is highly desirable to exclude such phase factors in order to simplify
the analysis of experimental data. This can be done using a spin echo
protocol inspired by nuclear magnetic resonance techniques, which
becomes particularly simple for $N_{2}=N_{1}=N$. Moreover, we assume
that one can implement a ``flip operator'' $\Sigma_{x}$ which exchanges
blocks 1 and 2 in a one-to-one correspondence and generalizes the
Pauli operator $\sigma_{x}$ (for $N=1$) to the case $N\ge2$. For
$N=2$, the matrix representation of $\Sigma_{x}$ within blocks 1
and 2 (all other matrix elements vanish) is given by
\begin{equation}
\Sigma_{x}=\left(\begin{array}{cc|cc}
0 & 0 & 1 & 0\\
0 & 0 & 0 & 1\\
\hline 1 & 0 & 0 & 0\\
0 & 1 & 0 & 0
\end{array}\right).\label{sigmaxs}
\end{equation}
We consider a spin echo protocol of total duration $2T$, where the
base trajectory ${\bm{\lambda}}(t)$ describes a loop path for $0\le t\le T$.
At $t=T$ both subspaces are swapped by applying $\Sigma_{x}$. We
then traverse the same loop again, ${\bm{\lambda}}(t)={\bm{\lambda}}(t-T)$
for $T\le t\le2T$, where we demand that the fluctuations $\delta{\bm{\lambda}}(t)$
for $t<T$ and $t>T$ are uncorrelated. At time $t=2T$, both blocks
are swapped once more by another application of the operator $\Sigma_{x}$.
Finally, one measures the expectation value of an operator $\tilde{M}$
as in Sec.~\ref{sec4c}.

For a specific fluctuating trajectory, the final state, $|\psi_{{\rm f}}\rangle=|\psi(t=2T)\rangle$,
is given by
\begin{equation}
|\psi_{{\rm f}}\rangle=U_{0}\Sigma_{x}\Bigl(\mathcal{U}^{(1)}\,\Sigma_{x}\ \mathcal{U}^{(2)}P_{2}+\mathcal{U}^{(2)}\,\Sigma_{x}\,\mathcal{U}^{(1)}\,P_{1}\Bigr)U_{0}^{\dagger}|\psi(0)\rangle,\label{finalspinecho-1}
\end{equation}
with $\mathcal{U}^{(j)}$ in Eq.~\eqref{Ujdef}. Defining the averaged
expectation value as
\begin{equation}
\bar{\mathcal{M}}_{\mathrm{se}}=\Bigl\langle\langle\psi_{{\rm f}}|\tilde{M}|\psi_{{\rm f}}\rangle\Bigr\rangle_{{\rm env}}=\langle\psi(0)|\bar{M}_{\mathrm{se}}|\psi(0)\rangle\label{eq:M_expectation_spin_echo}
\end{equation}
and repeating the calculation of Sec.~\ref{sec4c}, we obtain the
averaged operator
\begin{widetext}
\begin{equation}
\bar{M}_{\mathrm{se}}^{(21)}=U_{0}\bar{{\cal U}}^{(2)\dagger}\Sigma_{x}\bar{{\cal U}}^{(1)\dagger}\Sigma_{x}U_{0}^{\dagger}M^{(21)}U_{0}\Sigma_{x}\bar{{\cal U}}^{(2)}\Sigma_{x}\bar{{\cal U}}^{(1)}U_{0}^{\dagger}=e^{-2\Gamma_{\mathrm{dyn}}T}\,U_{0}\bar{{\cal U}}_{B}^{(2)\dagger}\Sigma_{x}\bar{{\cal U}}_{B}^{(1)\dagger}\Sigma_{x}U_{0}^{\dagger}M^{(21)}U_{0}\Sigma_{x}\bar{{\cal U}}_{B}^{(2)}\Sigma_{x}\bar{{\cal U}}_{B}^{(1)}U_{0}^{\dagger},\label{spinechoM}
\end{equation}
\end{widetext}

where all contributions from deterministic dynamic phase factors indeed
cancel out. We note that this cancellation can also be achieved by
using a time-reversed protocol for $T\le t\le2T$, i.e., ${\bm{\lambda}}(t)={\bm{\lambda}}(2T-t)$.
In this case, we find
\begin{eqnarray}
 &  & \bar{M}_{\mathrm{se}}^{(21)}=e^{-2\Gamma_{\mathrm{dyn}}T}\times\\
 &  & \quad\times\bar{{\cal U}}_{B}^{(2)\dagger}\Sigma_{x}\left(\bar{{\cal U}}_{B}^{(1)\dagger}\right)^{-1}\Sigma_{x}M^{(21)}\Sigma_{x}\left(\bar{{\cal U}}_{B}^{(2)}\right)^{-1}\Sigma_{x}\bar{{\cal U}}_{B}^{(1)},\nonumber
\end{eqnarray}
since the averaged Berry matrix is replaced by its inverse when the
trajectory is traversed in the opposite direction, cf.~Sec.~\ref{sec2c}.
In either case, dynamic dephasing is encoded by the non-universal
dimensionless parameter
\begin{eqnarray}
2\Gamma_{\mathrm{dyn}}T & = & \frac{1}{2}\int_{0}^{2T}dt\ \sigma^{\mu\nu}\partial_{\mu}\tilde{E_{1}}\partial_{\nu}\tilde{E_{1}}\nonumber \\
 & = & \int_{0}^{T}dt\ \sigma^{\mu\nu}\partial_{\mu}\tilde{E_{1}}\partial_{\nu}\tilde{E_{1}}\sim\sigma T.\label{dynamicdeph}
\end{eqnarray}
It stands to reason that by repeating the experiment for different
initial states and for different operators $\tilde{M}$, such a spin-echo
protocol allows one to map out $\Sigma_{x}\bar{{\cal U}}_{B}^{(2)}\Sigma_{x}\bar{{\cal U}}_{B}^{(1)}$.
This is, however, a somewhat tedious procedure. Fortunately, as we
demonstrate next, in some cases one can study NAGD without a meticulous
determination of the averaged Berry matrix.

Suppose that for the base trajectory, the Berry matrices of the two
blocks satisfy the relation
\begin{equation}
\Sigma_{x}\mathcal{U}_{B}^{(2)}\Sigma_{x}\mathcal{U}_{B}^{(1)}=\mathtt{1}.\label{eq:Berry_2blocks_special_property}
\end{equation}
We then consider the polar decomposition $\bar{{\cal U}}_{B}^{(j)}=V^{(j)}R^{(j)}$
in the weak-noise limit. The results of Sec.~\ref{sec2e3} together
with Eq.~\eqref{eq:Berry_2blocks_special_property} imply that
\begin{equation}
V^{(2)}=\Sigma_{x}V^{(1)\dagger}\Sigma_{x}+{\cal O}(\mathfrak{s}^{2}),
\end{equation}
while the NAGD part is given by
\begin{equation}
R^{(j)}=1+\mathfrak{s}r^{(j)}+{\cal O}(\mathfrak{s}^{2}),
\end{equation}
where $r^{(j)}$ can be read off from Eq.~\eqref{eq:eq:polar_approx_R}.
Up to ${\cal O}(\mathfrak{s}^{2})$ terms, we thereby find
\begin{equation}
\Sigma_{x}\bar{{\cal U}}_{B}^{(2)}\Sigma_{x}\bar{{\cal U}}_{B}^{(1)}=1+\mathfrak{s}\left(\Sigma_{x}V^{(2)}r^{(2)}V^{(2)\dagger}\Sigma_{x}+r^{(1)}\right).\label{sense1}
\end{equation}
Changing the orientation sense of the trajectory in the same spin
echo protocol, ${\bm{\lambda}}'(t)={\bm{\lambda}}(T-t)$ for $0\le t\le T$
and ${\bm{\lambda}}'(t)={\bm{\lambda}}(2T-t)$ for $T\le t\le2T$,
we have to replace $\bar{{\cal U}}_{B}^{(j)}\to\bigl(\bar{{\cal U}}_{B}^{(j)}\bigr)^{-1}$.
Instead of Eq.~\eqref{sense1}, we now obtain
\begin{multline}
\Sigma_{x}\bigl(\bar{{\cal U}}_{B}^{(2)}\bigr)^{-1}\Sigma_{x}\bigl(\bar{{\cal U}}_{B}^{(1)}\bigr)^{-1}\\
=1-\mathfrak{s}\left(\Sigma_{x}r^{(2)}\Sigma_{x}+V^{(1)}r^{(1)}V^{(1)\dagger}\right).\label{sense2}
\end{multline}

In the Abelian and poor man's non-Abelian cases, $V^{(j)}$ and $r^{(j)}$
commute with each other, implying that the unitaries $V^{(j)}$ effectively
disappear from Eqs.~\eqref{sense1} and \eqref{sense2}. Hence the
contributions $\sim\mathfrak{s}$ in these two equations differ by
a sign change. The terms $\sim\mathfrak{s}$ in $\bar{M}_{\mathrm{se}}^{(21)}$,
cf.~Eq.~\eqref{spinechoM}, and in $\bar{\mathcal{M}}_{\mathrm{se}}$,
cf.~Eq.~\eqref{eq:M_expectation_spin_echo}, thus have opposite
sign for opposite trajectory orientation. This feature represents
a clear hallmark of Abelian and poor man's non-Abelian systems. However,
in the truly non-Abelian case, $V^{(j)}$ and $r^{(j)}$ do not necessarily
commute, and hence contributions $\sim\mathfrak{s}$ to $\bar{{\cal M}}_{\mathrm{se}}$
may change arbitrarily (or even may not change at all) when the orientation
sense of the protocol is reversed. Detecting this feature amounts
to having a smoking gun signature for non-Abelian (as opposed to Abelian)
geometric dephasing. We give a specific example in Sec.~\ref{sec5e}
below.

\subsection{Berry matrices of different subblocks}

\label{sec4e}

The condition \eqref{eq:Berry_2blocks_special_property} for Berry
matrices in different blocks may seem natural for systems with only
two blocks. Indeed, it is well known that for the Abelian case realized
in a spin-$1/2$ system \citep{Berry1984}, the states with opposite
spin projections acquire opposite Berry phases. We show below that
in general, such a relation does not hold for non-Abelian Berry phases.
As a consequence, protocols respecting Eq.~\eqref{eq:Berry_2blocks_special_property}
have to be carefully designed.

Consider a two-block system with block degeneracies $N_{1}=N_{2}=N$.
The unitary matrix $U({\bm{\lambda}})$ describing the degenerate
spaces at different parameter values, see Eq.~\eqref{eq:degenerate_Ham_diag},
belongs to the group $U(2N)$. Since a common overall phase in $U({\bm{\lambda}})$
does not change the Hamiltonian, one can redefine $U({\bm{\lambda}})\rightarrow U({\bm{\lambda}})/\left[\det U({\bm{\lambda}})\right]^{1/2N}$,
and thus choose $U({\bm{\lambda}})\in SU(2N)$. Hence $\alpha_{\mu}=U^{\dagger}\partial_{\mu}U$
in Eq.~\eqref{eq:A_mu_def} is a traceless anti-Hermitian matrix
representing an element of the Lie algebra $su(2N)$. The Berry connection
in block $j$ is obtained by the projection $A_{\mu}^{(j)}=P_{j}\alpha_{\mu}P_{j}$.
As a consequence, $A_{\mu}^{(j)}$ is an element of the Lie algebra
$u(N)$. Note that the algebra is not restricted to $su(N)$ since
it does not have to be traceless. However, the tracelessness of $\alpha_{\mu}$
implies that we have
\begin{equation}
\mathrm{tr}A_{\mu}^{(1)}=-\mathrm{tr}A_{\mu}^{(2)}.\label{relation1}
\end{equation}
For a two-block Abelian ($N=1$) system, the Berry phases picked up
in different blocks therefore have opposite sign but equal absolute
value. However, for generic non-Abelian systems, Eq.~\eqref{relation1}
is not sufficient to uniquely relate $A_{\mu}^{(1)}$ and $A_{\mu}^{(2)}$.
Since there are no other restrictions on $\alpha_{\mu}$ even when
taking the gauge freedom into account, we conclude that for generic
non-Abelian systems, no specific relation between $A_{\mu}^{(1)}$
and $A_{\mu}^{(2)}$ exists. Moreover, even if it were to exist, it
would not imply a relation between ${\cal U}_{B}^{(1)}$ and ${\cal U}_{B}^{(2)}$
since the connection components $A_{\mu}^{(j)}({\bm{\lambda}})$ for
different ${\bm{\lambda}}$ do not commute.

An explicit example for Berry matrices ${\cal U}_{B}^{(1)}$ and ${\cal U}_{B}^{(2)}$
that are not connected by a simple relation can be constructed for
a spin $S=3/2$ system experiencing nuclear quadrupole resonance,
cf.~Eqs.~\eqref{eq:H_quadrupole_resonance} to \eqref{eq:F_NA_quadrupole_resonance}.
Defining block $1$ from the states with spin projection $\pm3/2$,
we see that this block represents a poor man's non-Abelian system
since for any parameter trajectory, ${\cal U}_{B}^{(1)}$ is diagonal
in the basis of $\ket{\pm3/2}$ states. However, block 2 is spanned
by the states $\ket{\pm1/2}$ and represents a truly non-Abelian system.
Indeed, for any state in this block, there is a trajectory such that
${\cal U}_{B}^{(2)}$ does not leave the state invariant.

Another example follows from the Majorana five-star geometry in Sec.~\ref{sec5},
where the Berry curvature $F_{\mu\nu}^{(s)}$ of the two blocks (denoted
by $s=\pm1$ below) is given by Eq.~\eqref{eq:F_23_Majoranas}. Here
one finds that the $F_{\mu\nu}^{(s)}$ components involving one index
$\mu,\nu=\theta_{4}$ are $\sim s$ while the three other curvature
components are independent of $s$. We have checked that it is not
possible to make all six components independent of $s$ (or all $\sim s$)
by gauge transformations, see Eq.~\eqref{eq:F_gauge_transform}.
In fact, all three Pauli matrices appear in the $s$-independent components
and in those $\sim s$, and a gauge transformation flipping the sign
of any Pauli matrix affects both groups. Since $F_{\mu\nu}^{(s)}$
can be viewed as the leading non-trivial contribution to the Berry
matrix ${\cal U}_{B}^{(s)}$ over an infinitesimal contour in the
$(\lambda^{\mu},\lambda^{\nu})$ plane \citep{AltlandBook}, we have
another example for two-block Berry matrices not connected by a simple
relation.

\subsection{Averaged density matrices}

\label{sec4f}

The observables of interest in systems supporting non-Abelian Berry
phases may also involve more than one or two blocks as studied so
far, see Secs.~\ref{sec4b} to \ref{sec4d}. We next briefly discuss
the averaged density matrix of a general system subject to NAGD. One
can calculate the expectation values of arbitrary observables as follows.
The initial system state $\ket{\psi(0)}$ corresponds to the density
matrix $\rho(0)=\ket{\psi(0)}\bra{\psi(0)}$. The density matrix at
time $t$ is then given by
\begin{equation}
\rho(t)=\sum_{j,k}U_{0}\mathcal{U}^{(j)}(t)P_{j}U_{0}^{\dagger}\rho(0)U_{0}P_{k}\mathcal{U}^{(k)\dagger}(t)U_{0}^{\dagger},
\end{equation}
with $U_{0}$ in Eq.~(\ref{u0def}) and
\begin{eqnarray}
\mathcal{U}^{(j)}(t) & = & e^{-i\int_{0}^{t}d\tau E_{j}\left({\bm{\lambda}}(\tau)+\delta{\bm{\lambda}}(\tau)\right)}\label{Ujdef-1}\\
 & \times & \mathcal{T}\exp{\int_{0}^{t}d\tau\dot{\lambda}^{\mu}\left(-A_{\mu}^{(j)}+\delta\lambda^{\nu}F_{\mu\nu}^{(j)}\right)}.\nonumber
\end{eqnarray}
The average $\bar{\rho}(t)=\left\langle \rho(t)\right\rangle _{{\rm env}}$
can be computed along the lines of the calculation of $\bar{M}^{(21)}(t,T)$
in Sec.~\ref{sec4c}, i.e., one evaluates $d\bar{\rho}/dt$, performs
the fluctuation average, neglects terms ${\cal O}(T^{-2})$, and solves
the resulting differential equation. We thereby find
\begin{equation}
\bar{\rho}(t)=\sum_{j,k}U_{0}\bar{\mathcal{U}}^{(j;k)}(t)P_{j}U_{0}^{\dagger}\rho(0)U_{0}P_{k}\bar{\mathcal{U}}^{(k;j)\dagger}(t)U_{0}^{\dagger},\label{aux54}
\end{equation}
with the averaged evolution operators
\begin{eqnarray}
 &  & \bar{{\cal U}}^{(j;k)}(t)=e^{-i\int_{0}^{t}dt\left(E_{j}-\frac{i}{2}\sigma^{\mu\nu}\partial_{\mu}E_{j}\partial_{\nu}\left[E_{j}-E_{k}\right]\right)}\times\label{eq:multi-block-averaging}\\
 &  & \times\mathcal{P}\exp\int_{0}^{t}dt\dot{\lambda}^{\mu}\left(-A_{\mu}^{(j)}+i\sigma^{\nu\rho}F_{\nu\mu}^{(j)}\partial_{\rho}\left[E_{j}-E_{k}\right]\right).\nonumber
\end{eqnarray}
The final result for $\bar{\rho}(T)$ immediately follows from Eq.~\eqref{aux54}.
In agreement with the results of Sec.~\ref{sec4b}, we observe that
both dynamic and geometric dephasing terms vanish in a one-block ($j=k$)
system.

\section{Application: Modified Majorana braiding protocols}

\label{sec5}

\begin{figure}
\centering{}\includegraphics[width=1\columnwidth]{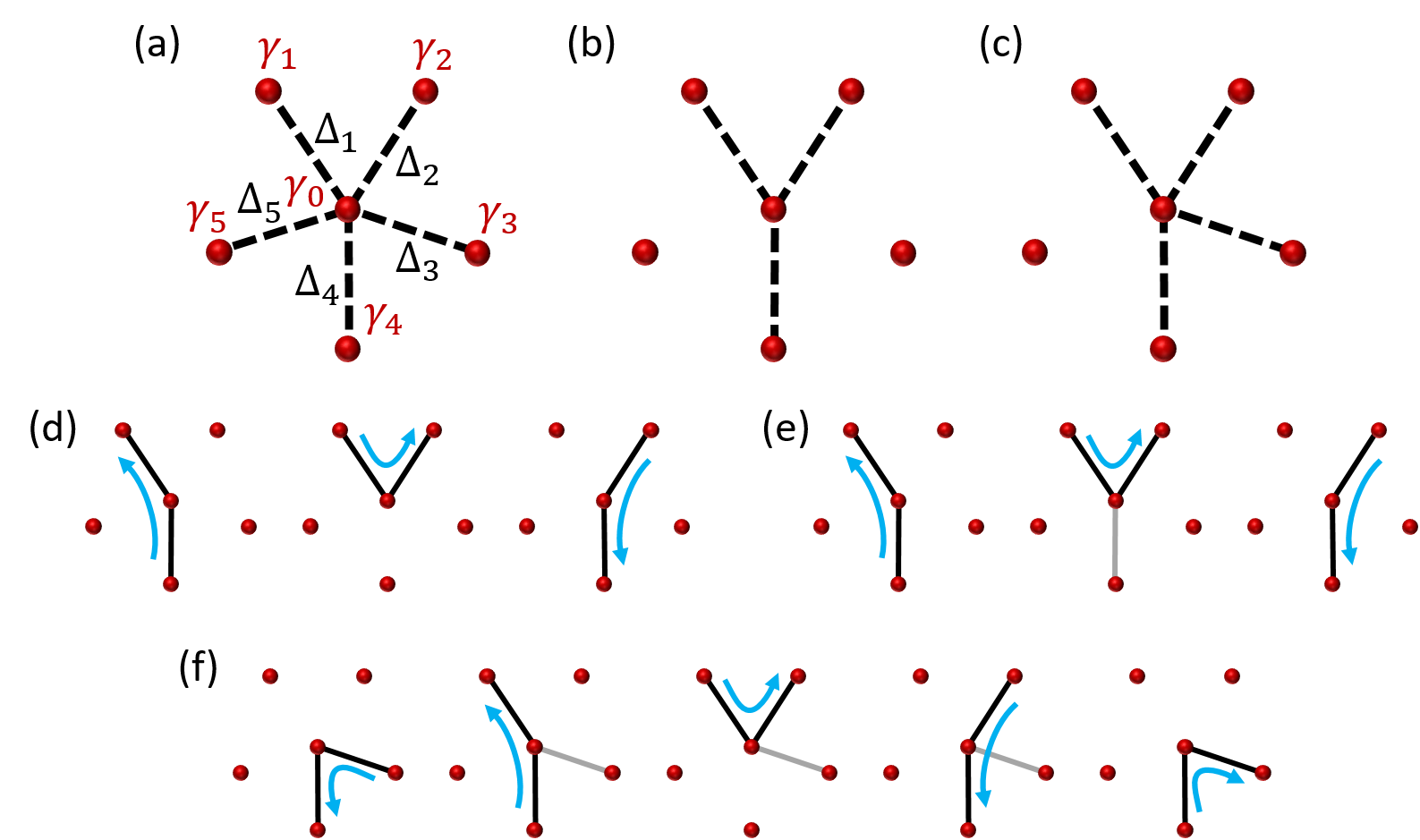} \caption{\protect Star-like Majorana setups and protocols for NAGD detection.
\textbf{(a):} Five-star Majorana setup with tunnel couplings $\Delta_{j=1,\ldots,5}$
between Majorana operators $\gamma_{0}$ and $\gamma_{j}$, see Eq.~\eqref{eq:H_5-star}.
\textbf{(b):} Three-star setup with $\Delta_{3}=\Delta_{5}=0$ (dashed
black lines indicate $\Delta_{j}\protect\ne0$). This is a poor man's
non-Abelian setup --- while the Hamiltonian has a degeneracy, all
components of the Berry connection (and of the Berry curvature) commute
with each other. \textbf{(c):} Five-star Majorana setup used for the
NAGD detection protocol with $\Delta_{5}=0$. This represents a truly
non-Abelian setup. \textbf{(d):} Topologically protected Majorana
braiding protocol in the three-star setup, cf.~Ref.~\citep{Alicea2012,Sau2011,VanHeck2012}.
Solid black lines indicate the non-zero couplings, $\Delta_{j}(t)\protect\ne0$,
during the respective time interval. The shown sequence leads to an
exchange of $\gamma_{1}$ and $\gamma_{2}$. In the first step, one
starts with $\Delta_{4}>0$ and all other $\Delta_{j}=0$. The blue
arrow indicates that $\Delta_{4}(t)$ is slowly reduced to zero while
$\Delta_{1}(t)$ is simultaneously ramped up, and similarly for the
other steps. \textbf{(e):} Elementary steps for the NAGD detection
protocol in the poor man's non-Abelian case using the three-star setup
in panel (b). The grey solid line indicates an additional constant
coupling $\Delta_{4}'\protect\neq0$ which breaks topological protection
and implies NAGD. For a full description of this protocol, see Sec.~\ref{sec5d}.
\textbf{(f):} Same as in panel (e) but for the truly non-Abelian case,
where $\Delta_{3}'\protect\ne0$ denotes a constant additional coupling
(grey) present during the respective time intervals. For details,
see Sec.~\ref{sec5e}.}
\label{fig4}
\end{figure}

In this section, we consider a concrete application to illustrate
how one may experimentally detect NAGD features using a specific spin
echo protocol, cf.~Sec.~\ref{sec4d}. From a theoretical point of
view, the arguably simplest system that would allow one to observe
NAGD signatures in an unambiguous manner is defined by the star-like
Majorana bound state (MBS) structures illustrated in Fig.~\ref{fig4}.
In particular, for this application we can easily identify observables
that satisfy the conditions specified in Secs.~\ref{sec4c} and \ref{sec4d}.
Related setups have been theoretically studied in the context of non-Abelian
braiding \citep{Alicea2012,Sau2011,VanHeck2012,Karzig2016,Rahmani2017,Cheng2011,Knapp2016}.
We emphasize that here we study modified Majorana braiding setups
with the aim of detecting NAGD, and not for characterizing errors
in Majorana qubit operations. The latter errors concern only one degenerate
block and, as follows from the considerations in Sec.~\ref{sec4b},
are due to non-adiabatic corrections \citep{Rahmani2017}.

We introduce the Hamiltonian used for modeling this setup in Sec.~\ref{sec5a}.
Next, techniques for reading out and/or manipulating the system are
described. Albeit these proposals can be found in the literature \citep{Lutchyn2017,Alicea2012,Fu2010,Flensberg2011,Sau2011,VanHeck2012,Aasen2016,Landau2016,Plugge2016,Plugge2017,Karzig2017},
for the convenience of the reader, Sec.~\ref{sec5b} provides a brief
summary of these techniques. We then discuss in Sec.~\ref{sec5c}
how fluctuating parameter trajectories can be generated in this system.

Protocols for NAGD detection in the poor man's non-Abelian case, cf.~Sec.~\ref{sec2e},
are addressed in Sec.~\ref{sec5d}. By poor man's non-Abelian we
mean cases where the matrices $V$ and $R$ in the polar decomposition
\eqref{polardecomp} always commute. We note that the conventional
three-star Majorana setup in Fig.~\ref{fig4}(b) yields commuting
$V$ and $R$ matrices independently of the protocol employed. The
protected braiding protocol is shown in Fig.~\ref{fig4}(d), see
Refs.~\citep{Alicea2012,Sau2011,VanHeck2012,Karzig2016,Rahmani2017,Cheng2011,Knapp2016}.
To break topological protection and obtain NAGD contributions, see
Sec.~\ref{sec3b}, we allow for a non-zero coupling $\Delta_{4}'\ne0$
as shown for the protocol in Fig.~\ref{fig4}(e). This protocol allows
one to detect NAGD for the poor man's non-Abelian case, as we discuss
in Sec.~\ref{sec5d}.

To obtain NAGD features in a truly non-Abelian system with $[V,R]\ne0$,
we study a five-star Majorana setup, cf.~Fig.~\ref{fig4}(c), using
the protocol shown in Fig.~\ref{fig4}(f). In that case, an additional
coupling $\Delta_{3}'\ne0$ is switched on during the indicated time
intervals and serves to break topological protection. We show in detail
how to implement the corresponding NAGD detection protocol in Sec.~\ref{sec5e}.
Importantly, the relative simplicity of this problem admits an analytical
solution for the adiabatic quantum dynamics in both the poor man's
and the truly non-Abelian case. This allows us to devise protocols
for observing simple signatures of NAGD.

\subsection{Model}

\label{sec5a}

\subsubsection{Five-star Majorana setup}

The setup in Fig.~\ref{fig3}(a) is modeled by the Hamiltonian
\begin{equation}
H_{M}(t)=i\gamma_{0}\sum_{j=1}^{5}\Delta_{j}(t)\gamma_{j},\label{eq:H_5-star}
\end{equation}
where the Majorana operators $\gamma_{k}^{\thinspace}=\gamma_{k}^{\dagger}$
satisfy the anticommutator algebra $\{\gamma_{k},\gamma_{l}\}=2\delta_{kl}$
\citep{Alicea2012}. The $d=5$ real-valued tunnel couplings, $\Delta_{j}(t)$,
correspond to our general control parameters ${\bm{\lambda}}(t)$.
Note that Eq.~\eqref{eq:H_5-star} represents an effective low-energy
theory, where all relevant energy scales are assumed to be well below
the proximity gap of the topological superconductor hosting the MBSs.
(At higher energies, one also has to include effects caused by above-gap
quasi-particle excitations.) The standard three-star setup follows
from Eq.~\eqref{eq:H_5-star} by simply putting some tunnel couplings
to zero, see Fig.~\ref{fig4}(b).

Next we observe that $H_{M}^{2}(t)=\sum_{j=1}^{5}\Delta_{j}^{2}$
and that the combination $\sum_{j=1}^{5}\Delta_{j}(t)\gamma_{j}$
defines an effective Majorana fermion (up to a real-valued overall
prefactor). Therefore $H_{M}$ has exactly two eigenenergies,
\begin{equation}
E_{\pm}=\pm{\cal E}/2,\quad{\cal E}=2\Biggl(\sum_{j}\Delta_{j}^{2}\Biggr)^{1/2}.\label{Majenerg}
\end{equation}
The energies $E_{-}$ and $E_{+}$ correspond to the energies $E_{1}$
and $E_{2}$ in Sec.~\ref{sec4}, respectively, and ${\cal E}$ represents
the energy scale appearing in the expansion parameters listed in Eq.~\eqref{expansion_parameters}.

Importantly, the total fermion number parity is conserved by the Hamiltonian
\eqref{eq:H_5-star}. We thus can focus on a Hilbert space sector
with fixed total parity. For concreteness, we choose even total parity
from now on, but the results for the odd parity sector follow accordingly.
The Hilbert space is then four-dimensional, and each of the levels
$E_{\pm}=\pm{\cal E}/2$ is two-fold degenerate ($N=2)$ for arbitrary
choice of the tunnel couplings $\{\Delta_{j}\}$. Below we use the
four basis states
\begin{equation}
\Bigl\{|0_{12}0_{34}0_{05}\rangle,|1_{12}0_{34}1_{05}\rangle,|0_{12}1_{34}1_{05}\rangle,|1_{12}1_{34}0_{05}\rangle\Bigr\}\label{basis1}
\end{equation}
to span the even-parity sector of the Hilbert space, where $n_{jk}=0,1$
contains the Majorana parity eigenvalue $i\gamma_{j}\gamma_{k}=2n_{jk}-1=\pm1$.
In what follows, it is convenient to parametrize the tunnel couplings
in terms of hyper-spherical coordinates,
\begin{eqnarray}
\Delta_{5} & = & \frac{{\cal E}}{2}\cos\theta_{1},\nonumber \\
\Delta_{3} & = & \frac{{\cal E}}{2}\sin\theta_{1}\cos\theta_{2},\nonumber \\
\Delta_{4} & = & \frac{{\cal E}}{2}\sin\theta_{1}\sin\theta_{2}\cos\theta_{3},\label{eq:spherical_coords_for_Delta}\\
\Delta_{1} & = & \frac{{\cal E}}{2}\sin\theta_{1}\sin\theta_{2}\sin\theta_{3}\cos\theta_{4},\nonumber \\
\Delta_{2} & = & \frac{{\cal E}}{2}\sin\theta_{1}\sin\theta_{2}\sin\theta_{3}\sin\theta_{4},\nonumber
\end{eqnarray}
with the angles $\theta_{i=1,2,3}\in[0,\pi]$ and $\theta_{4}\in[0,2\pi]$.
Writing $\theta_{0}={\cal E}/2$, the parameters $\theta_{\mu}$ with
$\mu=0,\ldots,4$ can be identified with the control parameters $\lambda^{\mu}$
(instead of the tunnel couplings $\Delta_{j}$).

\subsubsection{Eigenstates}

Using Eq.~\eqref{eq:spherical_coords_for_Delta}, we can express
$H_{M}$ in Eq.~(\ref{eq:H_5-star}) as
\begin{equation}
H_{M}=U_{4}U_{3}U_{2}U_{1}\left(\frac{i{\cal E}}{2}\gamma_{0}\gamma_{5}\right)U_{1}^{\dagger}U_{2}^{\dagger}U_{3}^{\dagger}U_{4}^{\dagger},\label{aux4}
\end{equation}
with the parameter-dependent unitary matrices
\begin{eqnarray}
U_{1} & = & e^{-\theta_{1}\gamma_{5}\gamma_{3}/2},\quad U_{2}=e^{-\theta_{2}\gamma_{3}\gamma_{4}/2},\nonumber \\
U_{3} & = & e^{-\theta_{3}\gamma_{4}\gamma_{1}/2},\quad U_{4}=e^{-\theta_{4}\gamma_{1}\gamma_{2}/2}.\label{eq:U1-4}
\end{eqnarray}
Note that $U_{4}U_{3}U_{2}U_{1}$ corresponds to $U({\bm{\lambda}})$
in Eq.~\eqref{eq:degenerate_Ham_diag}. Denoting the two eigenstates
for energy $E_{s=\pm}=s{\cal E}/2$ by $|a,s\rangle$ (with $a=1,2$),
Eq.~\eqref{aux4} directly gives
\begin{equation}
|a,s\rangle=U_{4}U_{3}U_{2}U_{1}|a,s\rangle_{0}^ {},\label{eq:Basis_rot_m}
\end{equation}
where $|a,s\rangle_{0}^ {}$ is the corresponding eigenstate of $(i{\cal E}/2)\gamma_{0}\gamma_{5}$.
In the basis (\ref{basis1}), we have
\begin{eqnarray}
|1,-\rangle_{0}^ {} & = & \left(\begin{array}{c}
1\\
0\\
0\\
0
\end{array}\right),\quad|2,-\rangle_{0}^ {}=\left(\begin{array}{c}
0\\
0\\
0\\
1
\end{array}\right),\nonumber \\
|1,+\rangle_{0}^ {} & = & \left(\begin{array}{c}
0\\
1\\
0\\
0
\end{array}\right),\quad|2,+\rangle_{0}^ {}=\left(\begin{array}{c}
0\\
0\\
1\\
0
\end{array}\right).\label{barestates}
\end{eqnarray}
Clearly, all four states in Eq.~\eqref{eq:Basis_rot_m} are orthonormal.

\subsubsection{Berry connection}

\label{sec5a3}

Next we compute the Berry connection, $A_{\mu}^{(s)}$, for the respective
block $s=\pm$. Instead of the tunnel couplings, we use the parameters
in Eq.~\eqref{eq:spherical_coords_for_Delta}, i.e., the four angles
$\theta_{i=1,\ldots,4}$ and $\theta_{0}={\cal E}/2$. According to
Eq.~\eqref{eq:A_mu_def}, the $2\times2$ matrix $A_{\theta_{\mu}}^{(s)}$
has the elements $\left(A_{\theta_{\mu}}^{(s)}\right)_{aa'}=\langle a,s|\partial_{\theta_{\mu}}|a',s\rangle.$
Given the above expressions, it is straightforward to compute the
Berry connection. We find
\begin{eqnarray}
A_{\theta_{0}}^{(s)} & = & A_{\theta_{1}}^{(s)}=0,\quad A_{\theta_{2}}^{(s)}=-\frac{i}{2}\cos\theta_{1}\,\sigma_{z}^{(s)},\label{eq:A_2_Majoranas}\\
A_{\theta_{3}}^{(s)} & = & -\frac{i}{2}\left(\cos\theta_{2}\,\sigma_{x}^{(s)}-\sin\theta_{2}\,\cos\theta_{1}\,\sigma_{y}^{(s)}\right),\nonumber \\
A_{\theta_{4}}^{(s)} & = & \frac{is}{2}\Bigl(\cos\theta_{3}\,\sigma_{z}^{(s)}+\sin\theta_{3}\cos\theta_{2}\,\sigma_{y}^{(s)}+\nonumber \\
 &  & \quad+\sin\theta_{3}\,\sin\theta_{2}\,\cos\theta_{1}\,\sigma_{x}^{(s)}\Bigr),\nonumber
\end{eqnarray}
where we define Pauli matrices acting in the subspace with fixed $s=\pm$
in Eq.~\eqref{basis1}. For instance, $\sigma_{x}^{(s)}$ has the
matrix elements
\begin{equation}
_{0}^ {}\langle a,s'|\sigma_{x}^{(s)}|a',s'\rangle_{0}^ {}=\delta_{ss'}\left(\begin{array}{cc}
0 & 1\\
1 & 0
\end{array}\right)_{aa'},
\end{equation}
and similarly for $\sigma_{y,z}^{(s)}$.

We note that the Berry connection components $A_{\theta_{\mu}}^{(s)}$
can become singular when some of the parameters reach $\theta_{\mu}=0$
or $\pi$. For example, for $\theta_{1}=0$, the values of $\theta_{2}$,
$\theta_{3}$, and $\theta_{4}$ are undefined since they do not affect
the tunnel couplings $\Delta_{j}$ in Eq.~\eqref{eq:spherical_coords_for_Delta}.
Nonetheless, $A_{\theta_{4}}^{(s)}$ will depend on the values of
$\theta_{2}$ and $\theta_{3}$. This singularity is caused by the
fact that the $U_{j}$ in Eq.~\eqref{eq:U1-4} depend on all the
angles. The ambiguity corresponds to choosing different bases in the
degenerate blocks when $\theta_{\mu}=0$, i.e., to performing a gauge
transformation. In general, one has to define several charts (coordinate
systems) covering the parameter space, calculate $A_{\mu}^{(s)}$
in each of them, split the evolution trajectory into pieces that do
not approach singular points in the appropriate charts, calculate
the respective pieces of the Berry matrix, and glue them together
using appropriate gauge transformations. In our case, however, we
can resolve this issue in a simpler manner. In fact, the protocols
below will only start and end at singular points but never cross them.
The adiabatic evolution of $\ket{\psi(t)}$ in Eq.~(\ref{schrodinger})
yields
\begin{equation}
\ket{\psi(T)}=U(\bm{\lambda}(T))\mathcal{U}U^{\dagger}(\bm{\lambda}(0))\ket{\psi(0)}\label{eq:UUU}
\end{equation}
with $\mathcal{U}$ in Eq.~(\ref{adiabatic_evolution}). Hence one
simply has to replace $U_{0}$ in Sec.~\ref{sec4} by either $U(\bm{\lambda}(T))$
or $U(\bm{\lambda}(0))$ in the appropriate places. We can thereby
account for the fact that the bases at the start and at the end of
the loop trajectory are different.

\subsubsection{Berry curvature}

The components of the Berry curvature tensor $F_{\mu\nu}^{(s)}$,
see Eq.~(\ref{eq:F_munu_def}), then follow as
\begin{eqnarray}
F_{\theta_{1}\theta_{2}}^{(s)} & = & \frac{i}{2}\sin\theta_{1}\,\sigma_{z}^{(s)},\quad F_{\theta_{1}\theta_{3}}^{(s)}=-\frac{i}{2}\sin\theta_{1}\,\sin\theta_{2}\,\sigma_{y}^{(s)},\nonumber \\
F_{\theta_{1}\theta_{4}}^{(s)} & = & \frac{is}{2}\sin\theta_{1}\,\sin\theta_{2}\,\sin\theta_{3}\,\sigma_{x}^{(s)},\nonumber \\
F_{\theta_{2}\theta_{3}}^{(s)} & = & \frac{i}{2}\sin^{2}\theta_{1}\,\sin\theta_{2}\,\sigma_{x}^{(s)},\label{eq:F_23_Majoranas}\\
F_{\theta_{2}\theta_{4}}^{(s)} & = & -is\cos\theta_{1}\,\cos\theta_{2}\,\sin\theta_{3}\,\sigma_{x}^{(s)}\nonumber \\
 &  & -\frac{is}{2}(1+\cos^{2}\theta_{1})\sin\theta_{2}\,\sin\theta_{3}\,\sigma_{y}^{(s)},\nonumber \\
F_{\theta_{3}\theta_{4}}^{(s)} & = & -is\cos\theta_{1}\,\sin\theta_{2}\,\cos\theta_{3}\,\sigma_{x}^{(s)}\nonumber \\
 &  & -\frac{is}{2}(1+\cos^{2}\theta_{1})\sin^{2}\theta_{2}\,\sin\theta_{3}\,\sigma_{z}^{(s)}.\nonumber
\end{eqnarray}
All other matrix elements follow by antisymmetry, $F_{\nu\mu}^{(s)}=-F_{\mu\nu}^{(s)}$,
or vanish identically.

\subsection{Readout and manipulation of Majorana parities}

\label{sec5b}

\begin{figure}
\centering{}\includegraphics[width=0.8\columnwidth]{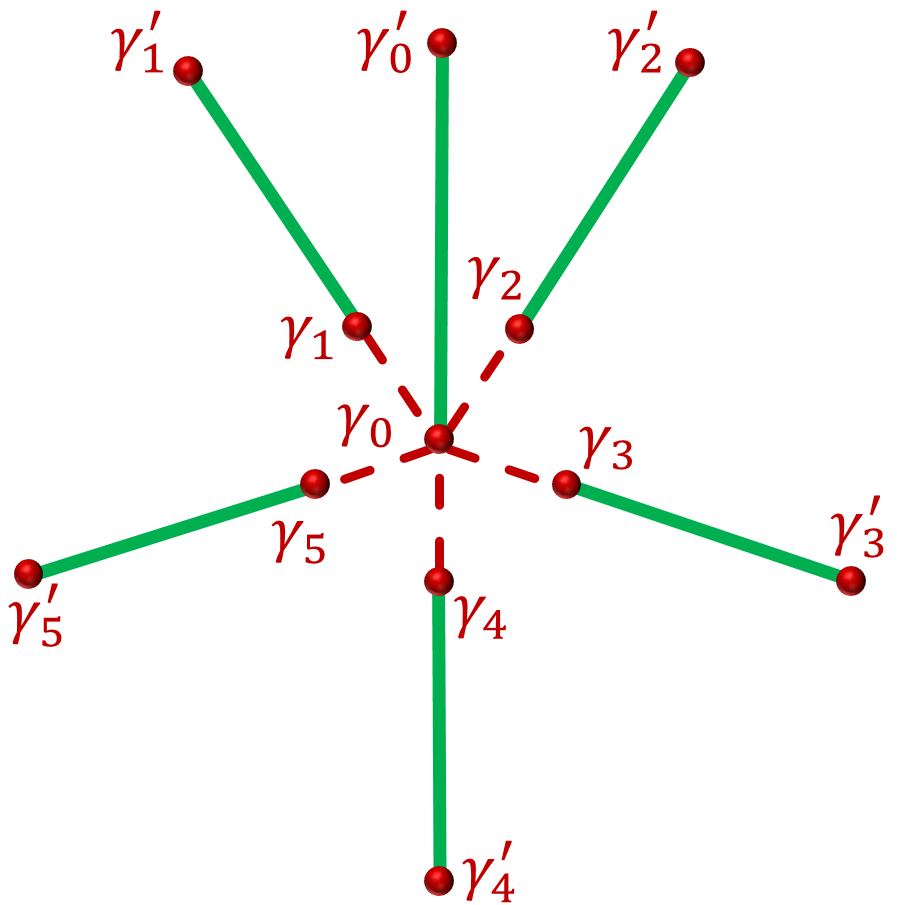}
\caption{\protect Schematic setup for a possible implementation of the Majorana five-star device.
Six long Majorana nanowires (green) are connected to a common floating superconductor and thus form a superconducting island with charging energy $E_C$.
The outer MBSs described by the $\gamma_j^\prime$ operators are spectator modes, i.e., their presence does not affect our protocols.
By arranging six Majorana nanowires ($j=0,\ldots,5$) as shown, the five Majorana operators $\gamma_{j=1,...,5}$ are tunnel-coupled only to the central Majorana operator $\gamma_0$. One can thereby realize the setup in Fig.~\ref{fig4}(a).}
\label{fig5}
\end{figure}

Now the inner Majoranas form a 5-star as such, and the tunnel couplings can be directly controlled by
gates. The price is that now we have 6 outer Majoranas that are inert and do not affect our protocols (we will have to explain
this in a sentence).
The difference between the two options is that one includes the six wires and the common proximitizing superconductor (the
green ring) in the picture, while the other one only includes six wires, and the common proximitizing superconductor should be
stated in words only.

In the NAGD detection protocols discussed below for the Majorana setups
in Fig.~\ref{fig4}, it is necessary to have experimental access
to the local Majorana parities $\hat{p}_{j,k}=i\gamma_{j}\gamma_{k}$
with eigenvalues $p_{j,k}=\pm1$. In this subsection, in order to
keep the paper self-contained, we shall summarize some of the recent
proposals \citep{Fu2010,Flensberg2011,Aasen2016,Landau2016,Plugge2016,Plugge2017,Karzig2017}
on (i) how to perform a projective readout of Majorana parities, (ii)
how to apply the ``flip operator'' $\hat{p}_{j,k}$ to the system
state (note that $\hat{p}_{j,k}$ anticommutes with $\hat{p}_{j,n\ne k}$
and $\hat{p}_{m\ne j,k}$, and thus flips those parities), and (iii)
how to initialize the system in a specific superposition state $|\psi(0)\rangle$.

For simplicity, we here assume that the five-star setup in Fig.~\ref{fig4}
is realized by using a single floating superconducting island with
the large charging energy $E_{C}$ (such that the tunnel couplings
$\Delta_{j}\ll E_{C}$) \citep{Plugge2017}. One possible implementation realizing the five-star device is sketched in Fig.~\ref{fig5}.
We also assume that the system is operated under Coulomb valley conditions,
i.e., the total charge on the island is quantized. While similar techniques
can be used for the grounded case with $E_{C}\rightarrow0$, see Refs.~\citep{Flensberg2011,Aasen2016},
working with a large charging energy has several key advantages. On
one hand, it protects the system against accidental electron tunneling
processes involving the measurement and/or manipulation devices. On
the other hand, it also provides a partial protection against quasi-particle
poisoning processes, see Refs.~\citep{Plugge2016,Karzig2017} for
details. In effect, the Hamiltonian $H_{M}$ in Eq.~\eqref{eq:H_5-star}
then provides an accurate description of the low-energy physics taking
place within the quantized ground-state charge sector. In particular,
the couplings $\Delta_{j}(t)$ in Eq.~\eqref{eq:H_5-star}
describe the tunneling amplitudes connecting $\gamma_{0}$ and
$\gamma_{j}$. In practice, they can be varied
by changing the voltage on a finger-gate electrode near those two
MBSs \citep{Lutchyn2017}. By depleting the corresponding region,
one can completely switch off the respective coupling, $\Delta_{j}\rightarrow0$.

Let us first describe how one can projectively measure the Majorana
parity eigenvalue $p_{j,k}=\pm1$ for a given system state. This task
can be achieved by interferometric readout techniques \citep{Fu2010,Landau2016,Plugge2016,Plugge2017,Karzig2017}.
Consider, for instance, two normal-conducting leads that are tunnel-coupled
with amplitudes $t_{j}$ and $t_{k}$ to the respective Majorana operators
$\gamma_{j}$ and $\gamma_{k}$. For large $E_{C}$, transport between
those two leads is only possible through cotunneling processes. The
transmission amplitude is then given by $it_{{\rm cot}}\gamma_{j}\gamma_{k}$
with $t_{{\rm cot}}\approx t_{j}t_{k}^{\ast}/E_{C}$. To enable an
interferometric readout of $i\gamma_{j}\gamma_{k}$, one needs a phase-coherent
reference arm of tunnel amplitude $t_{0}$ which provides a direct
tunneling path between both leads not involving thev Majorana island.
The total transmission probability, $|t_{0}+t_{{\rm cot}}p_{j,k}|^{2}$,
is then sensitive to the eigenvalue $p_{j,k}$, where we exploit that
the measurement is projective \citep{Landau2016,Plugge2017}. Since
the total transmission probability directly determines the linear
conductance between both leads, measuring the latter will also provide
an interferometric readout of the Majorana parity. This idea can also
be implemented with quantum dots instead of leads, see Refs.~\citep{Plugge2017,Karzig2017}.

Similarly, one can apply the operator $\hat{p}_{j,k}$ to a given
state $|\psi\rangle$ by forcing a single electron to tunnel between
both leads (in the absence of the reference arm). Since it is difficult
to control the leads in that manner, a more practical variant is to
instead employ a pair of single-level quantum dots and to vary their
energy levels such that an electron is transferred across the island
with high probability. While this is always possible by using a sufficiently
slow (adiabatic) change of the dot energy levels, one could also perform
this operation in a non-adiabatic way by confirming the outcome, i.e.,
by measuring the dot occupations after the electron pumping protocol.
In any case, once we know that a single electron has been transferred,
the above arguments imply that the operator $\hat{p}_{j,k}$ has been
applied to the system state. For further details, see Refs.~\citep{Plugge2016,Plugge2017,Karzig2017}.

Finally, one can prepare a specific initial state $|\psi(0)\rangle$
of the system by the readout of suitable Majorana parity operators.
For instance, suppose we want to prepare the state
\begin{equation}
|\psi(0)\rangle=\left(|0_{12}0_{03}0_{45}\rangle+|1_{12}1_{03}0_{45}\rangle\right)/\sqrt{2}.
\end{equation}
This state can be obtained from $|0_{12}0_{03}0_{45}\rangle$ by measuring
$\hat{p}_{0,2}$ with outcome $p_{0,2}=-1$, cf.~Ref.~\citep{Plugge2017}.
In order to prepare the product state $|0_{12}0_{03}0_{45}\rangle$,
it suffices to measure the local parity eigenvalues $p_{1,2}=p_{0,3}=p_{4,5}=+1$
by using the above readout techniques.

\subsection{Coupling to the environment}

\label{sec5c}

The independently fluctuating parameters for our Majorana setup are
the tunnel couplings, $\Delta_{j}(t)\rightarrow\Delta_{j}(t)+\delta\Delta_{j}(t)$,
where the Gaussian fluctuations $\delta\Delta_{j}(t)$ have zero mean,
$\langle\delta\Delta_{j}(t)\rangle_{{\rm env}}=0$. We assume for
simplicity that fluctuations of different tunnel couplings are uncorrelated.
To rationalize this assumption, let us give an example for how such
fluctuations could be generated in practice. We consider the effects
of charge fluctuations in the gate electrode regulating the respective
coupling. The electrostatic potential thereby fluctuates around some
average value, where the magnitude of the fluctuations can be controlled
by the circuit parameters defining the electromagnetic environment
of the setup \citep{Munk2019}. Since charge fluctuations on different
gates are independent, we see that the fluctuations $\delta\Delta_{j}$
connecting different Majorana pairs are indeed uncorrelated. Moroever,
since the tunnel couplings $\Delta_{j}$ are exponentially sensitive
to changes in the electrostatic potential, or to changes in the distance
between the two MBSs coupled by $\Delta_{j}$, fluctuations generally
act in a multiplicative way for this type of system, i.e., $\delta\Delta_{j}(t)\sim\Delta_{j}(t)$,
see Refs.~\citep{Karzig2016,Rahmani2017} for a detailed discussion
of this point. Equation \eqref{noisedef} thus yields the noise correlation
function
\begin{equation}
\langle\delta\Delta_{j}(t)\delta\Delta_{k}(t')\rangle_{{\rm env}}=\sigma_{j}^{\Delta}(t)\,\delta_{jk}\,\delta_{\tau_{c}}(t-t'),\label{noiseM}
\end{equation}
where the multiplicative fluctuation law $\delta\Delta_{j}\sim\Delta_{j}$
implies that the noise amplitude parameters $\sigma_{j}^{\Delta}$
can be written as
\begin{equation}
\sigma_{j}^{\Delta}(t)=\kappa_{j}\Delta_{j}^{2}(t).\label{multiplicative}
\end{equation}
We demand that the constants $\kappa_{j}$ satisfy $\kappa_{j}{\cal E}\ll1$
such that the conditions \eqref{conditions} are met. Let us emphasize
that whenever some tunnel coupling $\Delta_{j}$ is switched off,
$\Delta_{j}(t)=0$, the corresponding fluctuations $\delta\Delta_{j}$
are fully quenched by virtue of Eq.~\eqref{multiplicative}. The
absence of fluctuations for $\Delta_{j}=0$ is behind the topological
protection of the Majorana braiding protocol shown in Fig.~\ref{fig4}(d),
see Refs.~\citep{Karzig2016,Rahmani2017} as well as Secs.~\ref{sec3b}
and \ref{sec5d}.

However, as we have seen in Sec.~\ref{sec5a}, the Berry connection
and curvature are more conveniently computed in terms of the $\theta_{\mu}$
variables in Eq.~\eqref{eq:spherical_coords_for_Delta}. We thus
express the $\delta\theta_{\mu}$ fluctuations in terms of the $\delta\Delta_{j}$.
Using $\delta\Delta_{j}=\sum_{\mu}(\partial_{\theta_{\mu}}\Delta_{j})\,\delta\theta_{\mu}$,
calculating the derivatives, and inverting the relation, we find $\delta\theta_{\mu}=\sum_{j}{\bm{T}}_{\mu,j}\delta\Delta_{j}$.
The explicit form of the transformation matrix ${\bm{T}}$ is specified
in the Appendix, see Eq.~\eqref{TransfMatrix}. We thereby arrive
at the equivalent noise correlation function
\begin{equation}
\langle\delta\theta_{\mu}(t)\delta\theta_{\nu}(t')\rangle_{{\rm env}}=\sigma^{\mu\nu}(t)\,\delta_{\tau_{c}}(t-t'),
\end{equation}
where the noise amplitude matrix for the $\theta_{\mu}$ parameters
is with Eq.~\eqref{multiplicative} given by
\begin{equation}
\sigma^{\mu\nu}(t)=\sum_{j}\kappa_{j}\Delta_{j}^{2}(t){\bm{T}}_{\mu,j}(t){\bm{T}}_{\nu,j}(t).\label{sigmamunu}
\end{equation}

\subsection{Poor man's non-Abelian case}

\label{sec5d}

The poor man's non-Abelian case arises in a degenerate system where
always $[V,R]=0$ in the polar decomposition \eqref{polardecomp}.
This case can be realized for the three-star setup in Fig.~\ref{fig4}(b),
as we show in this subsection.

\subsubsection{Topologically protected braiding protocol}

We start with the standard Majorana braiding protocol \citep{Alicea2012,Cheng2011,Karzig2016,Knapp2016,Rahmani2017}
shown in Fig.~\ref{fig4}(d). As discussed in Sec.~\ref{sec3b},
one does not expect NAGD in this case since there are no fluctuations
of the geometric phase. Indeed, at any given time no more than two
$\Delta_{j}(t)\ne0$. In view of Eq.~\eqref{multiplicative}, only
those two couplings can fluctuate and, as a consequence, fluctuations
can only move one back and forth along the trajectory. The geometric
shape of the path in parameter space therefore remains unchanged.
Since NAGD is caused by cross-correlations of energy fluctuations
and geometric phase fluctuations, this braiding protocol is protected
against geometric dephasing.

In order to explicitly verify the absence of geometric phase fluctuations
within our formalism in Sec.~\ref{sec5a}, let us write out the braiding
protocol in Fig.~\ref{fig4}(d) in terms of the $\theta_{\mu}$ parameters.
(1) The protocol starts from a configuration with only $\Delta_{4}\ne0$
such that $\theta_{1}=\theta_{2}=\pi/2$ but $\theta_{3}=\theta_{4}=0$,
see Eq.~\eqref{eq:spherical_coords_for_Delta}. (2) One then changes
$\theta_{3}(t)$ adiabatically from $\theta_{3}=0$ to $\theta_{3}=\pi/2$.
At the end of this step, we have only $\Delta_{1}\ne0$. The relevant
Berry connection vanishes during this step since $\theta_{1}=\theta_{2}=\pi/2$,
and hence $A_{\theta_{3}}^{(\pm)}=0$ from Eq.~\eqref{eq:A_2_Majoranas}.
The corresponding NAGD contribution \eqref{avU} also vanishes because
of $\theta_{4}=0$,
\begin{eqnarray}
 &  & \dot{\theta}_{3}\sigma^{\theta_{0}\nu}F_{\nu\theta_{3}}^{(s)}=\dot{\theta}_{3}\frac{is{\cal E}}{4}\sigma_{z}^{(s)}\sin^{3}\theta_{3}\times\\
 &  & \times\sin\theta_{4}\cos\theta_{4}\left(\kappa_{2}\sin^{2}\theta_{4}-\kappa_{1}\cos^{2}\theta_{4}\right)=0,\nonumber
\end{eqnarray}
where the needed Berry curvature components follow from Eq.~\eqref{eq:F_23_Majoranas}.
(3) Next one varies $\theta_{4}$ from $0$ to $\pi/2$. Again, the
relevant connection component vanishes, $A_{\theta_{4}}^{(s)}=0$,
since $\theta_{1}=\theta_{2}=\theta_{3}=\pi/2$. The NAGD contribution
then vanishes because of $\theta_{3}=\pi/2$,
\begin{eqnarray}
 &  & \sigma^{\theta_{0}\nu}F_{\nu\theta_{4}}^{(s)}=-\frac{is{\cal E}}{4}\sigma_{z}^{(s)}\sin^{2}\theta_{3}\cos\theta_{3}\\
 &  & \times\left(\kappa_{1}\sin^{2}\theta_{3}\cos^{4}\theta_{4}+\kappa_{2}\sin^{2}\theta_{3}\sin^{4}\theta_{4}-\kappa_{4}\cos^{2}\theta_{3}\right)=0.\nonumber
\end{eqnarray}
(4) Finally, one changes $\theta_{3}(t)$ from $\pi/2$ to $0$ to
arrive back at only $\Delta_{4}\ne0$. Again, as during the second
step, $A_{\theta_{3}}=\sigma^{\theta_{0}\nu}F_{\nu\theta_{3}}^{(s)}=0$,
now due to $\theta_{4}=\pi/2$. We conclude that no phase or dephasing
terms are accumulated along the braiding trajectory, and $\mathcal{U}=\mathtt{1}$.

One may ask how braiding then arises. To answer this question, we
have to account for a basis change between the start and end points
of the protocol. Using Eq.~\eqref{eq:UUU}, while the states $|\tilde{\psi}(t)\rangle$
in the local ${\bm{\lambda}}$-dependent basis are transformed by
$\mathcal{U}=\mathtt{1}$ when the braiding protocol is executed,
one finds that the physical states $|\psi(t)\rangle$ are transformed
by $U({\bm{\lambda}}(T))\mathcal{U}U^{\dagger}({\bm{\lambda}}(0))=e^{-\pi\gamma_{1}\gamma_{2}/4}$.
We thereby recover the correct braiding operator, cf.~Ref.~\citep{Alicea2012}.

\subsubsection{NAGD in the poor man's non-Abelian setup}

\label{sec5d2}

For an arbitrary parameter trajectory in the three-star setup with
$\Delta_{3}=\Delta_{5}=0$, see Fig.~\ref{fig4}(b), NAGD contributions
will arise once all three remaining couplings ($\Delta_{1}$, $\Delta_{2}$,
and $\Delta_{4}$) are simultaneously switched on during a part of
the protocol. For example, consider the specific NAGD detection protocol
in Fig.~\ref{fig4}(e). Here we include a tunnel coupling $\Delta_{4}'\ne0$
during intermediate steps when, at the same time, also fluctuating
tunnel couplings $\Delta_{1}(t)$ and $\Delta_{2}(t)$ are present.

Let us first show that a setup with $\Delta_{3}=\Delta_{5}=0$ represents
the poor man's non-Abelian case. Using the parametrization in Eq.~\eqref{eq:spherical_coords_for_Delta},
we observe that arbitrary values of $\theta_{0},\theta_{3}$, and
$\theta_{4}$ are possible but we have $\theta_{1}=\theta_{2}=\pi/2$.
The relevant components of the Berry connection in Eq.~\eqref{eq:A_2_Majoranas}
are then given by
\begin{equation}
A_{\theta_{3}}^{(s)}=0,\quad A_{\theta_{4}}^{(s)}=\frac{is}{2}\cos\theta_{3}\,\sigma_{z}^{(s)},\label{eq:almostAbelianBerry}
\end{equation}
and the NAGD contributions follow from
\begin{eqnarray}
 &  & \sigma^{\theta_{0}\nu}F_{\nu\theta_{3}}^{(s)}=\frac{is{\cal E}}{4}\sigma_{z}^{(s)}\,\sin^{3}\theta_{3}\,\sin\theta_{4}\,\cos\theta_{4}\times\label{NAGD:almost}\\
 &  & \qquad\times\left[\kappa_{2}\sin^{2}\theta_{4}-\kappa_{1}\cos^{2}\theta_{4}\right],\nonumber \\
 &  & \sigma^{\theta_{0}\nu}F_{\nu\theta_{4}}^{(s)}=-\frac{is{\cal E}}{4}\sigma_{z}^{(s)}\,\sin^{2}\theta_{3}\,\cos\theta_{3}\times\nonumber \\
 &  & \qquad\times\left[\sin^{2}\theta_{3}\left(\kappa_{1}\cos^{4}\theta_{4}+\kappa_{2}\sin^{4}\theta_{4}\right)-\kappa_{4}\cos^{2}\theta_{3}\right].\nonumber
\end{eqnarray}
Importantly, all terms in Eq.~\eqref{NAGD:almost} involve only the
$\sigma_{z}^{(s)}$ Pauli matrix acting in the $s=\pm$ degenerate
subspace. Therefore, the two states in that space are not mixed, and
there is a protocol-independent basis diagonalizing both $V$ and
$R$ in Eq.~\eqref{polardecomp} simultaneously. Hence we have $[V,R]=0$,
i.e., the three-star setup defines a poor man's non-Abelian system.
By contrast, for the truly non-Abelian case studied in Sec.~\ref{sec5e},
the NAGD terms corresponding to Eq.~\eqref{NAGD:almost} will involve
all three Pauli matrices. As a consequence, there one has $[V,R]\ne0$
and the full non-Abelian matrix structure of NAGD becomes crucial.

Let us then analyze the specific NAGD detection protocol illustrated
in Fig.~\ref{fig4}(e). In the first part of the protocol, one adiabatically
increases $\theta_{3}$ from $0$ to $\alpha$ (while keeping $\theta_{4}=0$),
where the angle $\alpha$ encodes $\Delta_{4}'\ne0$,
\begin{equation}
\cos{\alpha}=2\Delta_{4}'/{\cal E}.\label{alphadef}
\end{equation}
Clearly, the corresponding NAGD term \eqref{NAGD:almost} vanishes
during this part. Next, one changes $\theta_{4}$ from $0$ to $\pi/2$,
where Eq.~\eqref{NAGD:almost} yields a finite contribution for $\alpha<\pi/2$.
In the final step, we vary $\theta_{3}$ from $\alpha$ to zero, where
no NAGD contributions are generated.

From the above expressions, we then obtain the averaged evolution
operators, $\bar{{\cal U}}^{(s)}$, cf.~Eq.~\eqref{eq:Two-block-averaging},
in analytical form. The dynamic dephasing rate follows as $\Gamma_{{\rm dyn}}=T^{-1}\int_{0}^{T}dt\,\sigma^{\theta_{0}\theta_{0}}$,
cf.~Eq.~\eqref{dynamic_dephasing}. This rate will depend on the
precise time dependence of the $\{\theta_{\mu}\}$ protocol. The analytical
result for $\Gamma_{{\rm dyn}}$ can be explicitly computed from Eq.~\eqref{sigmamunu}
but is of no immediate interest here. The averaged Berry matrices
follow as $\bar{{\cal U}}_{B}^{(s)}=e^{-\frac{i\pi s}{4}\cos\alpha\sigma_{z}^{(s)}}e^{\zeta\sigma_{z}^{(s)}}$,
with the dimensionless constant
\begin{equation}
\zeta=\frac{\pi}{4}{\cal E}\sin^{2}\alpha\,\cos\alpha\left[\frac{3}{8}\left(\kappa_{1}+\kappa_{2}\right)\sin^{2}\alpha-\kappa_{4}\cos^{2}\alpha\right].\label{zetadef}
\end{equation}
Using explicit expressions for how the $i\gamma_{j}\gamma_{k}$ operators
act on the basis states $\ket{a,s}$ in Eq.~(\ref{eq:Basis_rot_m}),
and including the unitaries at the initial and the final time, see
Eq.~\eqref{eq:UUU}, we obtain
\begin{eqnarray}
U({\bm{\lambda}}(T))\,\bar{{\cal U}}^{(s)}\,U^{\dagger}({\bm{\lambda}}(0)) & = & e^{-is\frac{{\cal E}}{2}T-\Gamma_{{\rm dyn}}T}\times\label{evolution_poor_man's_NA}\\
 & \times & e^{-\varphi_{\alpha}\gamma_{1}\gamma_{2}}e^{-\zeta\gamma_{0}\gamma_{1}\gamma_{2}\gamma_{4}},\nonumber
\end{eqnarray}
where the angle
\begin{equation}
\varphi_{\alpha}=\frac{\pi}{4}\left(1-\cos\alpha\right)\label{varphidef}
\end{equation}
captures the correction to the ideal braiding phase $\pi/4$ due to
$\Delta_{4}^{\prime}\ne0$, see Eq.~\eqref{alphadef}, i.e., due
to the deviation of the protocol in Fig.~\ref{fig4}(e) from the
protected braiding protocol in Fig.~\ref{fig4}(d). We note that
Eq.~\eqref{evolution_poor_man's_NA} is also valid for the odd total
fermion parity sector. In particular, a relative Berry phase $2\varphi_{\alpha}$
is accumulated between states with different local fermion parity
$i\gamma_{1}\gamma_{2}=\pm1$. The last exponential in Eq.~\eqref{evolution_poor_man's_NA}
describes the NAGD-induced suppression or enhancement depending on
$(i\gamma_{0}\gamma_{4})(i\gamma_{1}\gamma_{2})=\pm1$.

Finally, we note that when the protocol is executed in the opposite
direction, according to the arguments in Sec.~\ref{sec2c}, the second
line in Eq.~\eqref{evolution_poor_man's_NA} has to be replaced with
its inverse. This implies that one simply has to replace $\varphi_{\alpha}\rightarrow-\varphi_{\alpha}$
and $\zeta\rightarrow-\zeta$.

\subsubsection{Spin echo protocols}

We now apply the spin echo protocol described in Sec.~\ref{sec4d}
to the Majorana setup. The protocol starts by preparing the system
in an initial state $\ket{\psi(0)}$. One then performs the evolution
described in Sec.~\ref{sec5d2}, applies a flip operator $\Sigma_{x}$,
runs the evolution protocol in Sec.~\ref{sec5d2} once again, applies
$\Sigma_{x}$, and finally measures a block-off-diagonal operator
$\tilde{M}$. We employ the Majorana parity operators $\hat{p}_{0,j}=i\gamma_{0}\gamma_{j}$
(with $j=1,2,3,5$) both to realize the $\tilde{M}$ operators and
to implement the flip operator $\Sigma_{x}$. We observe that all
$\hat{p}_{0,j}$ operators anticommute with the final Hamiltonian,
$H(T)={\cal E}\hat{p}_{0,4}/2$, see Eq.~\eqref{eq:H_5-star}. This
means that they indeed have the block off-diagonal structure required
in Secs.~\ref{sec4c} and \ref{sec4d}. We find the averaged final-state
expectation values of $\hat{p}_{0,j}$ in the form
\begin{equation}
\bar{p}_{0,j}=e^{-2\Gamma_{\mathrm{dyn}}T}\bra{\psi(0)}\bar{U}_{B}^{\eta\dagger}\Sigma_{x}\bar{U}_{B}^{\eta\dagger}\Sigma_{x}\hat{p}_{0,j}\Sigma_{x}\bar{U}_{B}^{\eta}\Sigma_{x}\bar{U}_{B}^{\eta}\ket{\psi(0)},\label{eqNN}
\end{equation}
where $\eta=\pm1$ is the orientation sense of the protocol in Sec.~\ref{sec5d2}
and
\begin{equation}
\bar{U}_{B}^{\eta}=e^{-\eta\varphi_{\alpha}\gamma_{1}\gamma_{2}}e^{-\eta\zeta\gamma_{0}\gamma_{1}\gamma_{2}\gamma_{4}}.
\end{equation}
Choosing $\Sigma_{x}=\hat{p}_{0,2}$, we then obtain
\begin{equation}
\Sigma_{x}\bar{U}_{B}^{\eta}\Sigma_{x}\bar{U}_{B}^{\eta}=e^{-2\eta\zeta\gamma_{0}\gamma_{1}\gamma_{2}\gamma_{4}}.
\end{equation}
With the initial state
\begin{equation}
\ket{\psi(0)}=\left(\ket{1_{12}1_{04}0_{35}}+\ket{0_{12}0_{04}0_{35}}\right)/{\sqrt{2}},
\end{equation}
we thus find from Eq.~\eqref{eqNN} the result
\begin{equation}
\bar{p}_{0,j}=-\delta_{j,2}e^{4\eta\zeta-2\Gamma_{{\rm dyn}}T}.\label{nophase}
\end{equation}
Remarkably, Eq.~\eqref{nophase} only depends on the (dynamic and
geometric) dephasing contributions but makes no reference to phase
terms. We note that the corresponding expectation values in the initial
state are given by $\bra{\psi(0)}\hat{p}_{0,j}\ket{\psi(0)}=-\delta_{j,2}$.
The suppression of the expectation value has two contributions: the
dynamic ($e^{-2\Gamma_{{\rm dyn}}T}$) and the geometric ($e^{4\eta\zeta}$)
dephasing factors. The latter, remarkably, can increase or decrease
$\bar{p}_{0,2}$ as compared to $-\delta_{j,2}e^{-2\Gamma_{{\rm dyn}}T}$
depending on the protocol orientation $\eta$. Since $\hat{p}_{0,2}$
flips $\ket{1_{12}1_{04}0_{35}}\leftrightarrow\ket{0_{12}0_{04}0_{35}}$,
its expectation value is exactly the interference term.
This is an example of what we refer to as suppression or amplification
of coherence by geometric dephasing.

Choosing instead the flip operator $\Sigma_{x}=\hat{p}_{0,5}$, one
has
\begin{equation}
\Sigma_{x}\bar{U}_{B}^{\eta}\Sigma_{x}\bar{U}_{B}^{\eta}=e^{-2\eta\varphi_{\alpha}\gamma_{1}\gamma_{2}}.
\end{equation}
With the same initial state as above, we now obtain
\begin{eqnarray}
\bar{p}_{0,1} & = & \eta e^{-2\Gamma_{{\rm dyn}}T}\sin(4\varphi_{\alpha}),\nonumber \\
\bar{p}_{0,2} & = & -e^{-2\Gamma_{{\rm dyn}}T}\cos(4\varphi_{\alpha}),\nonumber \\
\bar{p}_{0,3} & = & \bar{p}_{0,4}=0.\label{noNAGD}
\end{eqnarray}
Note that neither geometric dephasing nor dynamic phase terms appear
in Eq.~\eqref{noNAGD}. Finally, it is also possible to exclusively
probe dynamic dephasing by choosing $\Sigma_{x}=\hat{p}_{0,5}$ and
\begin{equation}
\ket{\psi(0)}=\left(\ket{0_{12}0_{04}0_{35}}+\ket{0_{12}1_{04}1_{35}}\right)/{\sqrt{2}}.
\end{equation}
In that case, we find $\bar{p}_{0,j}=-\delta_{j,2}e^{-2\Gamma_{{\rm dyn}}T}$.
We conclude that the poor man's non-Abelian Majorana setup provides
a unique opportunity to separate the effects of dynamic dephasing,
geometric dephasing, and of the geometric phase.

\subsection{Truly non-Abelian case}

\label{sec5e}

We finally turn to the truly non-Abelian case, where NAGD contributions
involve several Pauli matrices $\sigma_{x,y,z}^{(s)}$ and the full
matrix structure of NAGD becomes observable. To realize this case,
we consider the five-star setup in Fig.~\ref{fig4}(c) with $\Delta_{5}=0$
and hence $\theta_{1}=\pi/2$, see Eq.~\eqref{eq:spherical_coords_for_Delta}.
A specific protocol for NAGD detection in the truly non-Abelian case
is shown in Fig.~\ref{fig4}(f). This protocol also involves a (non-fluctuating)
coupling, $\Delta_{3}'\ne0$, during parts of the protocol. Similar
to the coupling $\Delta_{4}'\ne0$ in Sec.~\ref{sec5d}, such a coupling
is needed to break topological protection.

\subsubsection{NAGD in the truly non-Abelian setup}

\label{sec5e1}

The protocol in Fig.~\ref{fig4}(f) starts from a configuration where
only $\Delta_{3}\ne0$. In that case, Eq.~\eqref{eq:spherical_coords_for_Delta}
gives the initial values $\theta_{2}=\theta_{3}=\theta_{4}=0$ while
$\theta_{1}=\pi/2$ throughout. One then adiabatically increases $\theta_{2}$
from zero to $\theta_{2}=\alpha$, where we define the angle $\alpha$
in analogy to Eq.~\eqref{alphadef} but with $\Delta_{4}'\rightarrow\Delta_{3}'\ne0$.
For $\alpha<\pi/2$, topological protection is broken and one has
a chance to observe NAGD. At the end of this step, one has the non-zero
tunnel couplings $\Delta_{4}(t)$ and $\Delta_{3}'$. We find that
all Berry curvature components relevant for generating NAGD terms
during this step are identically zero. Next, one changes $\theta_{3}$
from zero to $\pi/2$, thereby switching off (on) $\Delta_{4}$ $(\Delta_{1})$.
During this part of the protocol, the Berry connection $A_{\theta_{3}}^{(s)}=-\frac{i}{2}\cos\alpha\,\sigma_{x}^{(s)}$
and the NAGD contribution
\begin{eqnarray}
\sigma^{\theta_{0}\nu}F_{\nu\theta_{3}}^{(s)} & = & \frac{i{\cal E}}{4}\sigma_{x}^{(s)}\sin^{2}\alpha\,\cos\alpha\,\times\label{comp1}\\
 & \times & \left[\sin^{2}\alpha\left(\kappa_{1}\sin^{4}\theta_{3}+\kappa_{4}\cos^{4}\theta_{3}\right)-\kappa_{3}\cos^{2}\alpha\right]\nonumber
\end{eqnarray}
are generated. In the next step, one switches off (on) $\Delta_{1}$
$(\Delta_{2})$ by changing $\theta_{4}$ from zero to $\pi/2$. During
this part of the protocol, we find $A_{\theta_{4}}^{(s)}=\frac{is}{2}\cos\alpha\,\sigma_{y}^{(s)}$
and the NAGD contribution
\begin{eqnarray}
\sigma^{\theta_{0}\nu}F_{\nu\theta_{4}}^{(s)} & = & -\frac{is{\cal E}}{4}\sigma_{y}^{(s)}\,\sin^{2}\alpha\,\cos\alpha\label{comp2}\\
 & \times & \left[\sin^{2}\alpha\left(\kappa_{1}\cos^{4}\theta_{4}+\kappa_{2}\sin^{4}\theta_{4}\right)-\kappa_{3}\cos^{2}\alpha\right].\nonumber
\end{eqnarray}
We now switch off $\Delta_{2}$ and simultaneously turn on $\Delta_{4}$
by changing $\theta_{3}$ from $\pi/2$ to zero. The corresponding
NAGD contribution is again given by Eq.~\eqref{comp1} while $A_{\theta_{3}}^{(s)}=-\frac{i}{2}\cos\alpha\,\sigma_{x}^{(s)}$.
Finally, one changes $\theta_{2}$ from $\alpha$ to zero, where no
Berry connection or NAGD contributions are generated. The loop protocol
is thereby completed.

We first observe that different Pauli matrices $\sigma_{x,y}^{(s)}$
appear in the NAGD terms \eqref{comp1} and \eqref{comp2}, reflecting
the truly non-Abelian character of the problem. However, during each
time segment only one of these Pauli matrices appears, and the calculation
of path-ordered exponentials therefore remains tractable. Using the
dynamic dephasing rate $\Gamma_{{\rm dyn}}$ as defined in Sec.~\ref{sec5d2},
the averaged evolution operators in Eq.~\eqref{eq:Two-block-averaging}
are given by
\begin{eqnarray}
\bar{{\cal U}}^{(s)} & = & e^{-i\frac{s{\cal E}}{2}T-\Gamma_{{\rm dyn}}T}\,e^{\vartheta_{1,4}^{(s)}\sigma_{x}^{(s)}}\,e^{-s\vartheta_{1,2}^{(s)}\sigma_{y}^{(s)}}e^{-\vartheta_{2,4}^{(s)}\sigma_{x}^{(s)}},\nonumber \\
\vartheta_{j,k}^{(s)} & = & -s\zeta_{j,k}+\frac{i\pi}{4}\cos\alpha.\label{eq:averaged_protocol3}
\end{eqnarray}
where we define a generalized version of the NAGD parameter $\zeta$
in Eq.~\eqref{zetadef} for the truly non-Abelian case in terms of
the quantities
\begin{equation}
\zeta_{j,k}=\frac{\pi}{4}{\cal E}\sin^{2}\alpha\,\cos\alpha\left[\frac{3}{8}\left(\kappa_{j}+\kappa_{k}\right)\sin^{2}\alpha-\kappa_{3}\cos^{2}\alpha\right].\label{zetadef2}
\end{equation}
Multiplying out the exponentials of Pauli matrices in Eq.~\eqref{eq:averaged_protocol3},
one finds that $\bar{{\cal U}}^{(s)}$ has the form
\begin{equation}
\bar{{\cal U}}_{B}^{(s)}=A_{0}^{(s)}\mathtt{1}+A_{x}^{(s)}\sigma_{x}^{(s)}+A_{y}^{(s)}\sigma_{y}^{(s)}+A_{z}^{(s)}\sigma_{z}^{(s)},
\end{equation}
where the analytical form of the coefficients $A_{0,x,y,z}^{(s)}$
is easily obtained but of no immediate interest here. We emphasize
that this is the most general form of a $2\times2$ matrix $\bar{{\cal U}}_{B}^{(s)}$.

\subsubsection{Toward spin echo protocols}

The flip operator $\Sigma_{x}$ should act on the states in Eq.~\eqref{eq:Basis_rot_m}
according to $\Sigma_{x}\ket{a,s}=\ket{a,-s}$, see Eq.~\eqref{sigmaxs}.
This property also implies that $\Sigma_{x}\sigma_{j}^{(s)}\Sigma_{x}=\sigma_{j}^{(-s)}$
(with $j=x,y,z$). In particular, for all $\kappa_{j}=0$, we observe
from Eq.~\eqref{eq:averaged_protocol3} that $\Sigma_{x}\bar{{\cal U}}^{(-s)}\Sigma_{x}\bar{{\cal U}}^{(s)}=\mathtt{1}$,
i.e., we recover Eq.~(\ref{eq:Berry_2blocks_special_property}).
As discussed in Sec.~\ref{sec4d}, this property allows for a particularly
simple approach to the detection of truly non-Abelian geometric dephasing.
However, due to the ambiguity of the basis choice, we actually have
to consider $U(\bm{\lambda}(T))\bar{{\cal U}}^{(s)}U^{\dagger}(\bm{\lambda}(0))$
instead of $\bar{{\cal U}}^{(s)}$, see~Sec.~\ref{sec5a3}. In order
to make use of the above identity, one needs to cancel the contributions
of $U_{f}\equiv U(\bm{\lambda}(T))$ and $U_{i}\equiv U^{\dagger}(\bm{\lambda}(0))$
which is possible by a slight modification of the protocol of Sec.~\ref{sec5e1}.

To that end, we consider the evolution as shown in Fig.~\ref{fig4}(f)
but performed in the opposite direction and with $\Delta_{3}'=0$.
The corresponding evolution operator is given by $U_{i}U_{f}^{\dagger}$.
Therefore, performing the two sequences back-to-back yields the evolution
operator
\begin{equation}
U_{i}U_{f}^{\dagger}U_{f}\bar{{\cal U}}^{(s)}U_{i}^{\dagger}=U_{i}\bar{{\cal U}}^{(s)}U_{i}^{\dagger},
\end{equation}
in direct correspondence with the operator $U_{0}\bar{\mathcal{U}}_{B}^{(j)}U_{0}^{\dagger}$
in Sec.~\ref{sec4}. This modification allows us to implement a spin
echo protocol for NAGD detection as outlined in Sec.~\ref{sec4d}.

\subsubsection{Spin echo protocol}

For simplicity, we now assume that only $\Delta_{4}(t)$ fluctuates,
i.e., we put $\kappa_{1}=\kappa_{2}=\kappa_{3}=0$. We then start
from the initial state
\begin{equation}
|\psi(0)\rangle=\left(\ket{0_{12}0_{45}0_{03}}+\ket{1_{12}0_{45}1_{03}}\right)/\sqrt{2}.
\end{equation}
Note that this state represents a superposition of states with different
energies $\pm{\cal E}/2$ when all $\Delta_{j}=0$ apart from $\Delta_{3}\ne0$.
One then executes the imperfect braiding protocol as detailed in Sec.~\ref{sec5e1},
followed by the same protocol but with $\Delta_{3}'=0$ executed in
the opposite direction. The latter part thus corresponds to a protected
braiding protocol. At this point, $t=2T$, one applies the flip operator
\begin{equation}
\Sigma_{x}=\hat{p}_{0,2}=i\gamma_{0}\gamma_{2},\label{flipop}
\end{equation}
which exchanges both degenerate spaces $s=\pm$. One then runs the
above protocol again, i.e., one first executes the imperfect braiding
protocol, followed by a perfect braiding protocol in the opposite
direction. Subsequently, one applies the flip operator \eqref{flipop}
once more. Finally, one measures (one of) the local Majorana parities
$\hat{p}_{0,j}=i\gamma_{0}\gamma_{j}$. Performing the Gaussian average
over the noise fluctuations, with the orientation sense $\eta=\pm1$
of the entire protocol, we obtain
\begin{eqnarray}
\bar{p}_{0,1} & = & e^{-4\Gamma_{{\rm dyn}}T}\times{\cal O}(\bar{\zeta}^{2}),\quad\bar{p}_{0,3}=0,\nonumber \\
\bar{p}_{0,2} & = & -e^{-4\Gamma_{{\rm dyn}}T}\left[1-2\eta\bar{\zeta}\sin(4\beta)+{\cal O}(\bar{\zeta}^{2})\right],\label{eq:proper_NAGD_expectations}\\
\bar{p}_{0,4} & = & -8\bar{\zeta}e^{-4\Gamma_{{\rm dyn}}T}\sin^{2}\beta+{\cal O}(\bar{\zeta}^{2}),\nonumber
\end{eqnarray}
with $\beta=(\pi/4)\cos{\alpha}$ and the dimensionless NAGD parameter
\begin{equation}
\bar{\zeta}=\frac{3\pi}{16}\kappa_{4}{\cal E}\,\sin^{4}\alpha\,\cos\alpha.\label{zeta}
\end{equation}
The dynamic dephasing rate scales as $\Gamma_{{\rm dyn}}\sim\kappa_{4}{\cal E}^{2}$
but will again depend on details of the time-dependent protocol. We
note that the doubling of the exponent in $e^{-4\Gamma_{{\rm dyn}}T}$
as compared to Sec.~\ref{sec5d} arises due to doubled time duration
of the protocol.

The experimental confirmation of the results in Eq.~\eqref{eq:proper_NAGD_expectations}
would provide unambiguous smoking gun evidence for NAGD. In particular,
measuring $\bar{p}_{0,4}$ will immediately confirm that there is
a non-trivial NAGD matrix structure where the unitary ($V$) and the
Hermitian ($R$) parts of the averaged Berry matrix do not commute.
In fact, the sign change under $\eta\to-\eta$ observed in Eq.~\eqref{eq:proper_NAGD_expectations}
for the contribution $\sim\bar{\zeta}$ to $\bar{p}_{0,2}$ is precisely
as expected for the poor man's non-Abelian --- and for the Abelian
--- case, see Sec.~\ref{sec5d}. Crucially, the absence of such
a sign change in the expectation value $\bar{p}_{0,4}$ is only possible
due to the truly non-Abelian nature of NAGD in this setup, cf. Sec.~\ref{sec4d}.
Moreover, it is also worth noting that $\bar{p}_{0,4}=0$ in the absence
of noise.

\section{Conclusions}

\label{sec6}

In this work, see also Ref.~\citep{ourprl}, we have presented a
theory of adiabatic quantum transport in degenerate systems coupled
to a bath. The presence of a system-bath coupling implies that the
control parameters, ${\bm{\lambda}}(t)$, driving the adiabatic dynamics
will acquire a fluctuating component, ${\bm{\lambda}}(t)\rightarrow{\bm{\lambda}}(t)+\delta{\bm{\lambda}}(t)$.
These fluctuations obey Gaussian statistics, where we assume a sufficiently
long noise correlation time to justify the quantum adiabatic theorem
also for fluctuating trajectories. For short noise correlation times,
non-adiabatic corrections may become important \citep{footSyzranov},
and we believe that this issue is an interesting topic for future
research.

Our analysis shows that general systems of this class will exhibit
non-Abelian geometric dephasing when the system is steered from an
initial state to a final state along a fluctuating closed loop trajectory.
In contrast to dynamic dephasing, NAGD is universal in the sense that
it yields geometric (independent of the precise time dependence of
the protocol) contributions that admit a model-independent expression.
NAGD arises due to the cross-correlations of dynamic phase fluctuations
and geometric phase fluctuations, and it is sensitive to the orientation
sense of the protocol. Here geometric phase fluctuations are encoded
by the Berry curvature tensor $F_{\mu\nu}$. Interestingly, in topologically
protected systems, one has $F_{\mu\nu}=0$ away from singular points,
and generic protocols employed, say, for braiding anyonic quasiparticles,
will then be free from NAGD. As soon as topological protection is
broken, however, one will have $F\ne0$, resulting in NAGD.

A key object in our approach is the averaged time evolution operator
$\bar{{\cal U}}$, which can be factored into a term containing dynamic
phase and dephasing contributions, and a gauge covariant averaged
Berry phase matrix, $\bar{{\cal U}}_{B}$. This non-unitary matrix
is purely geometric and contains both the non-Abelian Berry phase
--- a loop integral over the Berry connection --- as well as the
NAGD contribution. By performing a matrix polar decomposition of $\bar{{\cal U}}_{B}$,
one can obtain an intuitive understanding of the effects of NAGD.
Most importantly, objects like $\bar{{\cal U}}_{B}$ also appear in
the NAGD detection protocols proposed in Sec.~\ref{sec4}, which
specifically target the non-Abelian matrix structure of NAGD. The
latter is responsible for key differences to the previously studied
case of Abelian geometric dephasing \citep{Whitney2005}.

As an application, we have discussed noisy Majorana braiding protocols.
On one hand, systems that may host Majorana bound states are presently
in the focus of intense experimental efforts \citep{Lutchyn2017}.
On the other hand, we think that, on a conceptual level, this setup
provides the simplest test case for NAGD. (We stress that our intention
is not to propose a new way of braiding Majorana states nor to analyze
the effect of fluctuations onto braiding-based quantum gates but to
detect NAGD in a related setup.) The simplicity of the model in Sec.~\ref{sec5}
allowed us to obtain fully analytical results despite of the intricate
path-ordered exponentials appearing in the theory. For most other
applications, it will be necessary to perform a numerical analysis
of suitably discretized path-ordered exponentials.

Finally, in view of the wide range of applications where the non-Abelian
Berry connection or curvature is of decisive importance, see, e.g.,
Refs.~\citep{Wilczek1984,Zee1988,Wen1991,Nayak2008,Pachos1999,Jones2000,Hasan2010,Yang2014,Wen2017,Xiao2010,Osterloh2005,Li2016,Vozmediano2010,Carroll},
the concepts laid out in our work should be beneficial for an understanding
of what happens when the respective systems are weakly coupled to
an environment. To mention just one application where NAGD contributions
may potentially show up, consider the phenomenon of weak anti-localization
in a metal \citep{AltlandBook}. When a conduction electron is backscattered
along an impurity sequence forming either a clockwise or a counter-clockwise
loop, the corresponding quantum amplitudes will interfere constructively.
In the presence of a magnetic Aharonov-Bohm field, the two amplitudes
are no longer in phase and the amount of backscattering gets reduced.
In more involved settings where Abelian and/or non-Abelian Berry phases
influence the dynamics of the system \citep{Xiao2010} and the latter
is coupled to an environment, geometric dephasing contributions could
be present. We believe that there are many interesting applications
of our theory, and we are confident that experimental checks will
soon be available.

\begin{acknowledgements} We thank A. Altland for discussions. This
project has been funded by the Deutsche Forschungsgemeinschaft (DFG,
German Research Foundation), Projektnummer 277101999, TRR 183 (project
C01). \end{acknowledgements}

\appendix

\section{Transformation matrix}

We here quote the explicit form of the transformation matrix ${\bm{T}}$
in Sec.~\ref{sec5c}, which connects the parameter fluctuations $\delta\theta_{\mu}$
to those of the tunnel couplings, $\delta\Delta_{j}$. Using the notation
$\theta_{0}={\cal E}$ and writing the linear matrix relation as
\begin{equation}
\begin{pmatrix}\delta\theta_{0}\\
\delta\theta_{1}\\
\delta\theta_{2}\\
\delta\theta_{3}\\
\delta\theta_{4}
\end{pmatrix}={\bm{T}}\begin{pmatrix}\delta\Delta_{5}\\
\delta\Delta_{3}\\
\delta\Delta_{4}\\
\delta\Delta_{1}\\
\delta\Delta_{2}
\end{pmatrix},\label{TransfL}
\end{equation}
we obtain the explicit form
\begin{widetext}
\begin{equation}
{\bm{T}}=\frac{2}{{\cal E}}\begin{pmatrix}\frac{{\cal E}}{2}\cos\theta_{1} & \frac{{\cal E}}{2}\sin\theta_{1}\cos\theta_{2} & \frac{{\cal E}}{{2}}\sin\theta_{1}\sin\theta_{2}\cos\theta_{3} & \frac{{\cal E}}{2}\sin\theta_{1}\sin\theta_{2}\sin\theta_{3}\cos\theta_{4} & \frac{{\cal E}}{2}\sin\theta_{1}\sin\theta_{2}\sin\theta_{3}\sin\theta_{4}\\
-\sin\theta_{1} & \cos\theta_{1}\cos\theta_{2} & \cos\theta_{1}\sin\theta_{2}\cos\theta_{3} & \cos\theta_{1}\sin\theta_{2}\sin\theta_{3}\cos\theta_{4} & \cos\theta_{1}\sin\theta_{2}\sin\theta_{3}\sin\theta_{4}\\
0 & -\frac{\sin\theta_{2}}{\sin\theta_{1}} & \frac{\cos\theta_{2}\cos\theta_{3}}{\sin\theta_{1}} & \frac{\cos\theta_{2}\sin\theta_{3}\cos\theta_{4}}{\sin\theta_{1}} & \frac{\cos\theta_{2}\sin\theta_{3}\sin\theta_{4}}{\sin\theta_{1}}\\
0 & 0 & -\frac{\sin\theta_{3}}{\sin\theta_{1}\sin\theta_{2}} & \frac{\cos\theta_{3}\cos\theta_{4}}{\sin\theta_{1}\sin\theta_{2}} & \frac{\cos\theta_{3}\sin\theta_{4}}{\sin\theta_{1}\sin\theta_{2}}\\
0 & 0 & 0 & -\frac{\sin\theta_{4}}{\sin\theta_{1}\sin\theta_{2}\sin\theta_{3}} & \frac{\cos\theta_{4}}{\sin\theta_{1}\sin\theta_{2}\sin\theta_{3}}
\end{pmatrix}.\label{TransfMatrix}
\end{equation}
\end{widetext}


\begin{thebibliography}{10}
\bibitem{foot_fluctuations_at_0} Relaxing this assumption implies $U(\bm{\lambda}(0))\rightarrow U(\bm{\lambda}(0)+\delta\bm{\lambda}(0))$
and similarly for $U(\bm{\lambda}(T))$, which ultimately leads to
corrections of order  ${\cal O}(\mathfrak{s}/\mathfrak{t_{c}})$
to observable quantities. Such terms are negligible compared to the leading NAGD contribution ${\cal O}(\mathfrak{s})$,
see Sec.~\ref{sec2c2} for the definition of the expansion parameters
$\mathfrak{s}$ and $\mathfrak{t_{c}}$, and Sec.~\ref{sec2e3} for NAGD
estimates.
\end{thebibliography}
\end{document}